\documentclass{article}
\usepackage{amsmath}
\usepackage{amsfonts}

\input{tcilatex}

\begin{document}

\title{On $\mathcal{A}_{n-1}^{(1)},$ $\mathcal{B}_{n}^{(1)},$ $\mathcal{C}%
_{n}^{(1)},$ $\mathcal{D}_{n}^{(1)},$ $\mathcal{A}_{2n}^{(2)},$ $\mathcal{A}%
_{2n-1}^{(2)},$ and $\mathcal{D}_{n+1}^{(2)}$ Reflection $K$-Matrices}
\author{R. Malara and A. Lima-Santos \thanks{{\footnotesize E-mail
addresses: malara@df.ufscar.br (R. Malara), dals@df.ufscar.br (A.
Lima-Santos).}} \\
%EndAName
\textit{Universidade Federal de S\~{a}o Carlos, Departamento de F\'{\i}sica, 
}\\
\textit{Caixa Postal 676, CEP 13565-905 S\~{a}o Carlos, Brazil}}
\date{}
\maketitle

\begin{abstract}
We present the classification of the most general regular solutions to the
boundary Yang-Baxter equations for vertex models associated with
non-exceptional affine Lie algebras. Reduced solutions found by applying a
limit procedure to the general solutions are discussed. We also present the
list of diagonal $K$-matrices. Special cases are considered separately.
\end{abstract}

\section{Introduction}

The quest for solutions of the Yang-Baxter equation \cite{Bax1,Fa2,KI,AA}%
\begin{equation}
\mathcal{R}_{12}(u-v)\mathcal{R}_{13}(u)\mathcal{R}_{23}(v)=\mathcal{R}%
_{23}(v)\mathcal{R}_{13}(u)\mathcal{R}_{12}(u-v)  \label{YBE}
\end{equation}%
has been successfully accomplished through the quantum group approach \cite%
{Dri1,Ne8,Alt1}, reducing the problem to a linear one. The $\mathcal{R}$%
-matrices corresponding to vector representations of all non-exceptional
affine Lie algebras were determined in this way in \cite{Jim1}. In the study
of the two-dimensional integrable systems of quantum field theories and
statistical physics, the Yang-Baxter equation (\ref{YBE}) has played an
essential role in establishing the integrability of models without a
boundary.

In a pioneering paper from the middle of the eighties, Cherednik \cite{Ch1}
suggested a possible generalization of factorized scattering theory to
integrable models with reflecting boundary conditions which preserve
integrability. The theoretical framework of the problem in the context of
the quantum inverse scattering method was set up by Sklyanin in \cite{Sk1},
where a systematic approach to build quantum integrable models with
nontrivial boundary conditions is developed, reflection $K$-matrices are
introduced and the relations they must fulfill in systems invariant under $P$%
-symmetry, $T$-symmetry, unitarity and crossing unitarity are obtained.
These relations feature the boundary Yang-Baxter equations or reflection
equations%
\begin{equation}
\mathcal{R}_{12}(u-v)K_{1}^{-}(u)\mathcal{R}%
_{21}(u+v)K_{2}^{-}(v)=K_{2}^{-}(v)\mathcal{R}_{12}(u+v)K_{1}^{-}(u)\mathcal{%
R}_{21}(u-v)  \label{BYBE}
\end{equation}%
and%
\begin{eqnarray}
&&\mathcal{R}_{12}(-u+v)\left( K_{1}^{+}\right) ^{t_{1}}(u)M_{1}^{-1}%
\mathcal{R}_{21}(-u-v-2\rho )M_{1}\left( K_{2}^{+}\right) ^{t_{2}}(v)= 
\notag \\
&&\left( K_{2}^{+}\right) ^{t_{2}}(v)M_{1}\mathcal{R}_{12}(-u-v-2\rho
)M_{1}^{-1}\left( K_{1}^{+}\right) ^{t_{1}}(u)\mathcal{R}_{21}(-u+v),
\label{dualBYBE}
\end{eqnarray}%
extended to models invariant under the less restrictive condition of $PT$%
-symmetry by Mezincescu and Nepomechie in \cite{Ne1,Ne2}. We remark that
there exists the following useful isomorphism: given a solution $K^{-}(u)$
of (\ref{BYBE}), the quantity%
\begin{equation}
K^{+}(u)=\left( K^{-}\right) ^{t}(-u-\rho )M,\text{ \ \ \ }M\equiv
U^{t}U=M^{t},  \label{Isomorphism}
\end{equation}%
satisfies (\ref{dualBYBE}), where $t_{i}$ denotes transposition in the $i$th
vector space, $\rho $ is the crossing parameter and $U$ is the crossing
matrix, both of them being specific to each model \cite{Baz2,Baz1}.

Quantum integrable models with non-periodic boundary conditions have been
extensively studied both in lattice and continuum theories, where the
boundary interaction is specified by a reflection $K$-matrix for lattice
systems \cite{HY,Bat3,AK,Zh1} or by a boundary $S$-matrix for quantum field
theories \cite{FK,Ne9,MM,MS}.

Recently, much attention has been directed to the research of an independent
systematic method of constructing the boundary quantum group generators
which would enable us to find solutions of the reflection equation (\ref%
{BYBE}). Studies of boundary quantum groups were initiated in \cite{Ne7} and
have been carried out in order to uncover the basic algebraic structure of
their generators. In this context, the most prominent works are the
references \cite{Ne6}, in which new solutions emerged some years ago, and 
\cite{Ba1}, which ultimately states that the boundary quantum group
structure associated with the reflection equation is actually the
tridiagonal algebra (\textit{q}-deformed Dolan-Grady relations \cite{DG}),
invariant under the coproduct homomorphism of $\mathcal{U}_{q}(\widehat{%
sl_{2}})$. It should be emphasized however that only the models associated
with $\mathcal{U}_{q}(\widehat{sl_{2}})$ $\mathcal{R}$-matrices enjoy this
tridiagonal algebraic symmetry. For higher rank affine Lie algebras, the
analogue of the deformed relations of the boundary quantum algebra remains
an open question.\ Somewhat earlier, some $\mathcal{A}_{n-1}^{(1)}$
reflection $K$-matrices as well as the $\mathcal{A}_{2}^{(2)}$ case were
rederived in \cite{Ne5}. Since appropriate classical integrable boundary
conditions are not yet known, one cannot investigate boundary affine Toda
field theory \cite{BCD}. An immediate question would be whether it is
possible to reveal all the solutions of the reflection equation by employing
quantum group generators. According to Baseilhac, all known reflection
matrices for models associated to the $\mathcal{U}_{q}(\widehat{sl_{2}})$ $%
\mathcal{R}$-matrix of quantum field theories and lattice systems with
boundary can be derived using this tridiagonal algebraic structure through
intertwining relations involving generators of the tridiagonal algebra.
Further developments intended as a broad outline of the construction of
boundary quantum group generators by studying the asymptotic behavior of the
open transfer matrix are presented in \cite{Dk1} for the XXZ case, and
continued in \cite{Dk2} for open spin chains associated to the $sl(n)$ $%
\mathcal{R}$-matrix. The non-local conserved quantity found in these works
turns out to be a special case of a (in)finite set of non-local conserved
quantities possessed by the models, which are all constructed in \cite{Ba2},
leading to the conclusion that these models are superintegrable.

Although regarded as a difficult problem, solutions of the boundary
Yang-Baxter equation (\ref{BYBE}) have been exploited for some $\mathcal{R}$%
-matrices by means of direct computation. For instance, we mention the
solutions for two-component systems \cite{VGR2,JWWX,LJWW}, for $19$-vertex
models \cite{IOZ,Li1}, for A-D-E interaction-round-a-face models \cite{BP},
for Andrews-Baxter-Forrester models in the RSOS/SOS representation \cite{AY}%
, and for vector representations of Yangians and super-Yangians \cite%
{Mac2,AADF}. In addition, $K$-matrices have been obtained for $\mathcal{A}%
_{n-1}^{(1)}$ models \cite{VGR1,Ab1} and, more recently, for $\mathcal{D}%
_{n+1}^{(2)}$ models \cite{Ma2}. Diagonal $K$-matrices for the $\mathcal{R}$%
-matrix associated with the minimal representation of the exceptional affine
algebra $\mathcal{G}_{2}^{(1)}$ have been considered in \cite{Bat4}. For
this model, the complete collection of non-diagonal solutions has been
displayed in \cite{Li7}.

Such classifications of solutions to the reflection equation have been
extended to include supersymmetric models, which can also be encountered in
the literature concerning topics in condensed matter physics \cite%
{HPW,BGZZ,ZGLG,ZGG,HYZZ}. In statistical mechanics, the emphasis has been
laid on deriving all the solutions of the reflection equation because
different $K$-matrices lead to different universality classes of surface
critical behavior \cite{Bat2}, allowing the calculation of various surface
critical phenomena both at and away from criticality \cite{Bat5,Bat1,Bat6}.

Most of works devoted to the investigation of the boundary Yang-Baxter
equations usually concentrate on acquiring regular solutions. Nevertheless,
it turns out that non-regular $K$-matrices are also of recent interest for
the series $\mathcal{A}_{n-1}^{(1)}$ \cite{Gan1,Mac1} and $\mathcal{D}%
_{n}^{(1)}$ \cite{DG1}.

In spite of the \textit{c}-number solutions of the reflection equations,
considerable attention has been paid to nontrivial $K$-matrices that include
boundary degrees of freedom, derived for the sine-Gordon model in \cite%
{BG,BK}. The construction of integrable quantum field theories on the
half-line including degrees of freedom at the boundary was originally
suggested by Ghoshal and Zamolodchikov in \cite{GZ}.

The development of the theory involved a great deal of efforts and,
motivated by this theme, we decided to put an extra effort into it. Now we
focus in this article based on our early works \cite{Li2,Li3,Li4,LM} on the
most general reflection $K$-matrices for the quantum $\mathcal{R}$-matrices
associated with the affine Lie algebras $\mathcal{A}_{n-1}^{(1)},$ $\mathcal{%
B}_{n}^{(1)},$ $\mathcal{C}_{n}^{(1)},$ $\mathcal{D}_{n}^{(1)},$ $\mathcal{A}%
_{2n}^{(2)},$ $\mathcal{A}_{2n-1}^{(2)},$ and $\mathcal{D}_{n+1}^{(2)}$. Our
classification scheme also provides reduced solutions generated by applying
a limit procedure to our general solutions previously presented. The list of
diagonal $K$-matrices is included and the special cases which do not exhibit
all the properties usually featured by most of the reflection $K$-matrices
are treated separately.

We have organized this paper as follows. We begin the next section by
considering the reflection equations for the vertex models associated with
the non-exceptional affine Lie algebras. In Section 3 we derive their
general solutions and in Section 4 reduced $K$-matrices are discussed. The
diagonal solutions as well as the special cases are presented in Sections 5
and 6, respectively. The last section is reserved for the conclusion.

\section{Reflection Equations}

The quantum $\mathcal{R}$-matrices for the vertex models associated with
non-exceptional affine Lie algebras in the fundamental representation as
presented by Jimbo \cite{Jim1} have the form%
\begin{eqnarray}
\mathcal{R} &=&(\mathrm{e}^{u}-q^{2})\sum E_{ii}\otimes E_{ii}+q(\mathrm{e}%
^{u}-1)\sum_{i\neq j}E_{ii}\otimes E_{jj}  \notag \\
&&-(q^{2}-1)\left( \sum_{i<j}E_{ij}\otimes E_{ji}+\mathrm{e}%
^{u}\sum_{i>j}E_{ij}\otimes E_{ji}\right)  \label{A(n)R-matrices}
\end{eqnarray}%
for the $\mathcal{A}_{n-1}^{(1)}$ models $(n\geq 2)$ \cite{Ch2,VGV},%
\begin{eqnarray}
\mathcal{R} &=&a_{1}\sum_{i\neq i^{\prime }}E_{ii}\otimes
E_{ii}+a_{2}\sum_{i\neq j,j^{\prime }}E_{ii}\otimes
E_{jj}+a_{3}\sum_{i<j,i\neq j^{\prime }}E_{ij}\otimes E_{ji}  \notag \\
&&+a_{4}\sum_{i>j,i\neq j^{\prime }}E_{ij}\otimes
E_{ji}+\sum_{i,j}a_{ij}E_{ij}\otimes E_{i^{\prime }j^{\prime }},
\label{B(n),C(n),D(n)R-matrices}
\end{eqnarray}%
where the Boltzmann weights with functional dependence on the spectral
parameter $u$ are given by%
\begin{eqnarray}
a_{1}(u) &=&(\mathrm{e}^{u}-q^{2})(\mathrm{e}^{u}-\xi ),\text{\ \ \ \ }%
a_{2}(u)=q(\mathrm{e}^{u}-1)(\mathrm{e}^{u}-\xi ),  \notag \\
a_{3}(u) &=&-(q^{2}-1)(\mathrm{e}^{u}-\xi ),\text{\ \ \ }a_{4}(u)=\mathrm{e}%
^{u}a_{3}(u),  \notag \\
a_{ij}(u) &=&\left\{ 
\begin{array}{c}
(q^{2}\mathrm{e}^{u}-\xi )(\mathrm{e}^{u}-1)\text{ \ \ \ \ \ \ \ \ \ \ \ \ \
\ \ \ \ \ \ \ \ \ \ \ \ \ \ \ }(i=j,i\neq i^{\prime }), \\ 
q(\mathrm{e}^{u}-\xi )(\mathrm{e}^{u}-1)+(\xi -1)(q^{2}-1)\mathrm{e}^{u}%
\text{ \ \ \ }(i=j,i=i^{\prime }), \\ 
(q^{2}-1)(\varepsilon _{i}\varepsilon _{j}\xi q^{\bar{\imath}-\bar{j}}(%
\mathrm{e}^{u}-1)-\delta _{ij^{\prime }}(\mathrm{e}^{u}-\xi ))\text{ \ \ \ \
\ \ }(i<j), \\ 
(q^{2}-1)\mathrm{e}^{u}(\varepsilon _{i}\varepsilon _{j}q^{\bar{\imath}-\bar{%
j}}(\mathrm{e}^{u}-1)-\delta _{ij^{\prime }}(\mathrm{e}^{u}-\xi ))\text{ \ \
\ \ }(i>j),%
\end{array}%
\right.  \label{B(n),C(n),D(n)weights}
\end{eqnarray}%
for the models of types $\mathcal{B}_{n}^{(1)},$ $\mathcal{C}_{n}^{(1)},$ $%
\mathcal{D}_{n}^{(1)},$ $\mathcal{A}_{2n}^{(2)},$ $\mathcal{A}_{2n-1}^{(2)}$%
, and%
\begin{eqnarray}
\mathcal{R} &=&\sum_{i,j\neq n+1,n+2}a_{ij}E_{ij}\otimes E_{i^{\prime
}j^{\prime }}+a_{1}\sum_{i\neq n+1,n+2}E_{ii}\otimes E_{ii}  \notag \\
&&+a_{2}\sum_{\substack{ i\neq j,j^{\prime }  \\ i\text{ or }j\neq n+1,n+2}}%
E_{ii}\otimes E_{jj}  \notag \\
&&+a_{3}\sum_{\substack{ i<j,i\neq j^{\prime }  \\ i,j\neq n+1,n+2}}%
E_{ij}\otimes E_{ji}+a_{4}\sum_{\substack{ i>j,i\neq j^{\prime }  \\ i,j\neq
n+1,n+2}}E_{ij}\otimes E_{ji}  \notag \\
&&+a_{5}\sum_{\substack{ i<n+1  \\ j=n+1,n+2}}(E_{ij}\otimes
E_{ji}+E_{j^{\prime }i^{\prime }}\otimes E_{i^{\prime }j^{\prime }})  \notag
\\
&&+a_{6}\sum_{\substack{ i>n+2  \\ j=n+1,n+2}}(E_{ij}\otimes
E_{ji}+E_{j^{\prime }i^{\prime }}\otimes E_{i^{\prime }j^{\prime }})  \notag
\\
&&+a_{7}\sum_{\substack{ i<n+1  \\ j=n+1,n+2}}(E_{ij}\otimes E_{j^{\prime
}i}+E_{j^{\prime }i^{\prime }}\otimes E_{i^{\prime }j})  \notag \\
&&+a_{8}\sum_{\substack{ i>n+2  \\ j=n+1,n+2}}(E_{ij}\otimes E_{j^{\prime
}i}+E_{j^{\prime }i^{\prime }}\otimes E_{i^{\prime }j})  \notag \\
&&+\frac{1}{2}\sum_{\substack{ i\neq n+1,n+2  \\ j=n+1,n+2}}%
[b_{i}^{+}(E_{ij}\otimes E_{i^{\prime }j^{\prime }}+E_{j^{\prime }i^{\prime
}}\otimes E_{ji})+b_{i}^{-}(E_{ij}\otimes E_{i^{\prime }j}+E_{ji^{\prime
}}\otimes E_{ji})]  \notag \\
&&+\sum_{i=n+1,n+2}[c^{+}E_{ii}\otimes E_{i^{\prime }i^{\prime
}}+c^{-}E_{ii}\otimes E_{ii}+d^{+}E_{ii^{\prime }}\otimes E_{i^{\prime
}i}+d^{-}E_{ii^{\prime }}\otimes E_{ii^{\prime }}],  \notag \\
&&  \label{D(n+1)R-matrices}
\end{eqnarray}%
with corresponding Boltzmann weights given by%
\begin{eqnarray}
a_{1}(u) &=&(\mathrm{e}^{2u}-q^{2})(\mathrm{e}^{2u}-\xi ^{2}),\text{\ \ \ \ }%
a_{2}(u)=q(\mathrm{e}^{2u}-1)(\mathrm{e}^{2u}-\xi ^{2}),  \notag \\
a_{3}(u) &=&-(q^{2}-1)(\mathrm{e}^{2u}-\xi ^{2}),\text{\ \ \ \ }a_{4}(u)=%
\mathrm{e}^{2u}a_{3}(u),  \notag \\
a_{5}(u) &=&\frac{1}{2}(\mathrm{e}^{u}+1)a_{3}(u),\text{\ \ \ \ }a_{6}(u)=%
\frac{1}{2}(\mathrm{e}^{u}+1)\mathrm{e}^{u}a_{3}(u),  \notag \\
a_{7}(u) &=&-\frac{1}{2}(\mathrm{e}^{u}-1)a_{3}(u),\text{\ \ \ \ }a_{8}(u)=%
\frac{1}{2}(\mathrm{e}^{u}-1)\mathrm{e}^{u}a_{3}(u),  \label{D(n+1)weights1}
\end{eqnarray}%
and%
\begin{eqnarray}
a_{ij}(u) &=&\left\{ 
\begin{array}{c}
(q^{2}\mathrm{e}^{2u}-\xi ^{2})(\mathrm{e}^{2u}-1)\text{ \ \ \ \ \ \ \ \ \ \
\ \ \ \ \ \ \ \ \ \ \ \ \ \ \ \ \ \ \ \ }(i=j), \\ 
(q^{2}-1)(\xi ^{2}q^{\bar{\imath}-\bar{j}}(\mathrm{e}^{2u}-1)-\delta
_{ij^{\prime }}(\mathrm{e}^{2u}-\xi ^{2}))\text{ \ \ \ \ }(i<j), \\ 
(q^{2}-1)\mathrm{e}^{2u}(q^{\bar{\imath}-\bar{j}}(\mathrm{e}^{2u}-1)-\delta
_{ij^{\prime }}(\mathrm{e}^{2u}-\xi ^{2}))\text{\ \ \ \ }(i>j),%
\end{array}%
\right.  \notag \\
b_{i}^{\pm }(u) &=&\left\{ 
\begin{array}{c}
\pm q^{i-1/2}(q^{2}-1)(\mathrm{e}^{2u}-1)(\mathrm{e}^{u}\pm \xi )\text{ \ \
\ \ \ \ \ \ \ \ \ \ \ \ \ \ }(i<n+1), \\ 
q^{i-n-5/2}(q^{2}-1)(\mathrm{e}^{2u}-1)\mathrm{e}^{u}(\mathrm{e}^{u}\pm \xi )%
\text{ \ \ \ \ \ \ \ \ \ \ \ \ }(i>n+2),%
\end{array}%
\right.  \notag \\
c^{\pm }(u) &=&\pm \frac{1}{2}(q^{2}-1)(\xi +1)\mathrm{e}^{u}(\mathrm{e}%
^{u}\mp 1)(\mathrm{e}^{u}\pm \xi )+q(\mathrm{e}^{2u}-1)(\mathrm{e}^{2u}-\xi
^{2}),  \notag \\
d^{\pm }(u) &=&\pm \frac{1}{2}(q^{2}-1)(\xi -1)\mathrm{e}^{u}(\mathrm{e}%
^{u}\pm 1)(\mathrm{e}^{u}\pm \xi ),  \label{D(n+1)weights2}
\end{eqnarray}%
for the $\mathcal{D}_{n+1}^{(2)}$ models, where $q=\mathrm{e}^{-2\eta }$
denotes an arbitrary parameter for all models described above.

By convention, the indices $i$, $j$ range over $1,2,...,\mathbb{N}$, where $%
\mathbb{N}$ is the size of the matrix: $\mathbb{N}=n,2n+1,2n,2n,2n+1,2n,2n+2$
respectively for $\mathcal{A}_{n-1}^{(1)},$ $\mathcal{B}_{n}^{(1)},$ $%
\mathcal{C}_{n}^{(1)},$ $\mathcal{D}_{n}^{(1)},$ $\mathcal{A}_{2n}^{(2)},$ $%
\mathcal{A}_{2n-1}^{(2)},$ and $\mathcal{D}_{n+1}^{(2)}$. We set $i^{\prime
}=\mathbb{N}+1-i$ and $E_{ij}$ are the elementary matrices $%
((E_{ij})_{ab}=\delta _{ia}\delta _{jb})$. We further let $\varepsilon
_{i}=1 $ $(1\leq i\leq n)$, $=-1$ $(n+1\leq i\leq 2n)$ for the $\mathcal{C}%
_{n}^{(1)}$ models and $\varepsilon _{i}=1$ in the remaining cases.

Here we have $\xi =q^{2n-1},q^{2n+2},q^{2n-2},-q^{2n+1},-q^{2n},q^{n}$
respectively for $\mathcal{B}_{n}^{(1)},$ $\mathcal{C}_{n}^{(1)},$ $\mathcal{%
D}_{n}^{(1)},$ $\mathcal{A}_{2n}^{(2)},$ $\mathcal{A}_{2n-1}^{(2)},$ and $%
\mathcal{D}_{n+1}^{(2)}$. Furthermore, $\bar{\imath}$ have the form%
\begin{equation}
\bar{\imath}=\left\{ 
\begin{array}{c}
i-1/2\text{ \ \ \ \ \ \ \ \ \ }(1\leq i\leq n) \\ 
i+1/2\text{ \ \ }(n+1\leq i\leq 2n)%
\end{array}%
\right.  \label{C(n)indices}
\end{equation}%
for $\mathcal{C}_{n}^{(1)}$,%
\begin{equation}
\bar{\imath}=\left\{ 
\begin{array}{c}
i+1\text{ \ \ \ \ \ \ \ \ \ \ \ \ \ \ \ }(i<n+1) \\ 
n+3/2\text{ \ \ \ }(i=n+1,n+2) \\ 
i-1\text{ \ \ \ \ \ \ \ \ \ \ \ \ \ \ \ }(i>n+2)%
\end{array}%
\right.  \label{D(n+1)indices}
\end{equation}%
for $\mathcal{D}_{n+1}^{(2)}$, and%
\begin{equation}
\bar{\imath}=\left\{ 
\begin{array}{c}
i+1/2\text{ \ \ \ }(1\leq i<\frac{\mathbb{N}+1}{2}) \\ 
i\text{ \ \ \ \ \ \ \ \ \ \ \ \ \ \ \ \ \ }(i=\frac{\mathbb{N}+1}{2}) \\ 
i-1/2\text{ \ \ \ }(\frac{\mathbb{N}+1}{2}<i\leq \mathbb{N})%
\end{array}%
\right.  \label{A(n),B(n),D(n)indices}
\end{equation}%
in the remaining cases.

Regular solutions of the reflection equation (\ref{BYBE}) mean that the $K$%
-matrix in the form%
\begin{equation}
K^{-}(u)=\sum_{i,j=1}^{\mathbb{N}}k_{i,j}(u)E_{ij}  \label{K-matrix}
\end{equation}%
satisfies the condition%
\begin{equation}
k_{i,j}(0)=\delta _{i,j},\text{ \ \ \ }i,j=1,2,...,\mathbb{N}.
\label{K-matrixReg}
\end{equation}%
Substituting (\ref{K-matrix}) and the $\mathcal{R}$-matrices (\ref%
{A(n)R-matrices}), (\ref{B(n),C(n),D(n)R-matrices}) and (\ref%
{D(n+1)R-matrices}) into (\ref{BYBE}), we get $\mathbb{N}^{4}$ functional
equations for the matrix elements $k_{i,j}(u)$. Although we have many
equations, a few of them are actually independent. In order to solve them we
will proceed as follows. First we consider the component $(i,j)$ of the
matrix equation (\ref{BYBE}). By differentiating it with respect to $v$ and
by taking $v=0$, we obtain algebraic equations involving the single variable 
$u$ and $\mathbb{N}^{2}$ parameters%
\begin{equation}
\beta _{i,j}=\frac{dk_{i,j}(v)}{dv}\mid _{v=0},\text{ \ \ \ }i,j=1,2,...,%
\mathbb{N}.  \label{Parameter}
\end{equation}%
Next we denote these equations by $E[i,j]=0$ and collect them into blocks $%
B[i,j]$, $i=1,...,\mathbb{I}$ and $j=i,i+1,...,\mathbb{J}-i$ with $\mathbb{I}%
=\mathbb{J}=n\mathbb{N}$ for $\mathcal{A}_{n-1}^{(1)}$, $\mathcal{C}%
_{n}^{(1)}$, $\mathcal{D}_{n}^{(1)}$, $\mathcal{A}_{2n-1}^{(2)}$, $\mathbb{I}%
=\mathbb{J}=2n(n+1)+1$ for $\mathcal{B}_{n}^{(1)}$, $\mathcal{A}_{2n}^{(2)}$%
, $\mathbb{I}=2(n+1)^{2}$ and $\mathbb{J}=\mathbb{N}^{2}$ for $\mathcal{D}%
_{n+1}^{(2)}$. Such blocks $B[i,j]$ are defined by%
\begin{equation}
B[i,j]=\left\{ 
\begin{array}{c}
E[i,j]=0,\text{ \ \ \ \ \ \ \ \ \ \ \ \ \ \ \ \ \ \ \ \ \ \ \ } \\ 
E[j,i]=0,\text{ \ \ \ \ \ \ \ \ \ \ \ \ \ \ \ \ \ \ \ \ \ \ \ } \\ 
E[\mathbb{N}^{2}+1-i,\mathbb{N}^{2}+1-j]=0, \\ 
E[\mathbb{N}^{2}+1-j,\mathbb{N}^{2}+1-i]=0.%
\end{array}%
\right.  \label{Blocks}
\end{equation}%
For a given block $B[i,j]$, the equation $E[\mathbb{N}^{2}+1-i,\mathbb{N}%
^{2}+1-j]=0$ can be obtained from the equation $E[i,j]=0$ by exchanging%
\begin{equation}
k_{i,j}\leftrightarrow k_{n+1-i,n+1-j},\text{ \ \ \ }\beta
_{i,j}\leftrightarrow \beta _{n+1-i,n+1-j^{\prime }},\text{ \ \ \ }%
a_{3}\leftrightarrow a_{4}  \label{A(n)interchange}
\end{equation}%
for $\mathcal{A}_{n-1}^{(1)}$ models $(n\geq 2)$,%
\begin{eqnarray}
k_{i,j} &\leftrightarrow &k_{i^{\prime },j^{\prime }},\text{ \ \ \ }\beta
_{i,j}\leftrightarrow \beta _{i^{\prime },j^{\prime }},\text{ \ \ \ }%
b_{i}^{\pm }\leftrightarrow b_{i^{\prime }}^{\pm },\text{ \ \ \ }%
a_{i,j}\leftrightarrow a_{i^{\prime },j^{\prime }},  \notag \\
a_{3} &\leftrightarrow &a_{4},\text{ \ \ \ \ \ \ \ \ }a_{6}\leftrightarrow
a_{6},\text{ \ \ \ \ \ \ \ }a_{7}\leftrightarrow a_{8}
\label{D(n+1)interchange}
\end{eqnarray}%
for $\mathcal{D}_{n+1}^{(2)}$ models, 
\begin{equation}
k_{i,j}\leftrightarrow k_{i^{\prime },j^{\prime }},\text{ \ \ \ }\beta
_{i,j}\leftrightarrow \beta _{i^{\prime },j^{\prime }},\text{ \ \ \ }%
a_{3}\leftrightarrow a_{4},\text{ \ \ \ }a_{i,j}\leftrightarrow a_{i^{\prime
},j^{\prime }}  \label{B(n),C(n),D(n)interchange}
\end{equation}%
in the remaining cases, and the equation $E[j,i]=0$ is obtained from the
equation $E[i,j]=0$ by exchanging%
\begin{equation}
k_{i,j}\leftrightarrow k_{j,i},\text{ \ \ \ }\beta _{i,j}\leftrightarrow
\beta _{j,i}  \label{A(n)interchange II}
\end{equation}%
for $\mathcal{A}_{n-1}^{(1)}$ models, and%
\begin{equation}
k_{i,j}\leftrightarrow k_{j,i},\text{ \ \ \ }\beta _{i,j}\leftrightarrow
\beta _{j,i},\text{ \ \ \ }a_{i,j}\leftrightarrow a_{j^{\prime },i^{\prime }}
\label{B(n),C(n),D(n)interchange II}
\end{equation}%
in the remaining cases.

According to Jimbo \cite{Jim1} and Bazhanov \cite{Baz2,Baz1}, the $\mathcal{R%
}$-matrices for the vertex models associated with non-exceptional affine Lie
algebras in the fundamental representation have $PT$-symmetry%
\begin{equation}
\mathcal{P}_{12}\mathcal{R}_{12}(u)\mathcal{P}_{12}\equiv \mathcal{R}%
_{21}(u)=\mathcal{R}_{12}^{t_{1}t_{2}}(u),  \label{PT-symmetry}
\end{equation}%
where $\mathcal{P}_{12}$ is the permutation matrix, and satisfy the
unitarity condition%
\begin{equation}
\mathcal{R}_{12}(u)\mathcal{R}_{21}(-u)=\mathrm{\zeta }(u),
\label{Unitarity}
\end{equation}%
where $\mathrm{\zeta }(u)$ is some even scalar function of $u$. We will also
require the regularity property, i.e. $\mathcal{R}_{12}(0)\sim \mathcal{P}%
_{12}$.

In addition, for all cases except $\mathcal{A}_{n-1}^{(1)}$ $(n>2)$, the $%
\mathcal{R}$-matrices have crossing symmetry. Thus, for these models the
relation%
\begin{equation}
\mathcal{R}_{12}(u)=(U\otimes \mathbf{1})\mathcal{R}_{12}^{t_{2}}(-u-\rho
)(U\otimes \mathbf{1})^{-1}  \label{Crossing-unitarity}
\end{equation}%
holds with the crossing matrix $U$ given by%
\begin{equation}
U_{i,j}=\delta _{i^{\prime },j}\text{ }q^{(\bar{\imath}-\bar{j})/2},\text{ \
\ \ for }\mathcal{B}_{n}^{(1)},\mathcal{C}_{n}^{(1)},\mathcal{D}_{n}^{(1)},%
\mathcal{A}_{2n}^{(2)},\mathcal{A}_{2n-1}^{(2)},\mathcal{D}_{n+1}^{(2)}
\label{U-matrix}
\end{equation}%
and the crossing parameter $\rho $ reads as follows%
\begin{equation}
\rho =\left\{ 
\begin{array}{c}
-\frac{1}{2}\ln \xi ,\text{ \ \ \ \ \ \ \ \ \ \ \ \ \ \ \ \ \ \ \ \ \ \ for
\ }\mathcal{B}_{n}^{(1)}, \\ 
\ln \xi ,\text{ \ \ \ \ \ \ \ \ \ \ \ \ \ \ \ \ \ \ \ \ \ \ \ \ \ \ for \ }%
\mathcal{A}_{2n}^{(2)}, \\ 
-\ln \xi ,\text{ for }\mathcal{C}_{n}^{(1)},\mathcal{D}_{n}^{(1)},\mathcal{A}%
_{2n-1}^{(2)},\mathcal{D}_{n+1}^{(2)},%
\end{array}%
\right.  \label{Crossing}
\end{equation}%
where we have normalized the Boltzmann weights by a factor $\sqrt{\xi }%
\mathrm{e}^{u}$ for $\mathcal{C}_{n}^{(1)},$ $\mathcal{D}_{n}^{(1)},$ and $%
\mathcal{A}_{2n-1}^{(2)}$ models, and by $q^{n+1}\mathrm{e}^{2u}$ for $%
\mathcal{D}_{n+1}^{(2)}$ models.

The corresponding matrix $K^{+}(u)$ at the opposite boundary is obtained
from the matrix $K^{-}(u)$ by using the isomorphism (\ref{Isomorphism}),
where $M$ is a diagonal matrix related to the crossing matrix $U$ by $%
M=U^{t}U$, given by%
\begin{equation}
M_{i,j}=\left\{ 
\begin{array}{c}
\delta _{i,j}q^{2n+2-2\bar{\imath}},\text{ \ \ \ }i,j=1,2,...,2n+1,\text{ \
\ \ \ \ \ \ for }\mathcal{B}_{n}^{(1)},\mathcal{A}_{2n}^{(2)}, \\ 
\delta _{i,j}q^{2n+1-2\bar{\imath}},\text{ \ \ \ }i,j=1,2,...,2n,\text{ \ \
\ for }\mathcal{C}_{n}^{(1)},\mathcal{D}_{n}^{(1)},\mathcal{A}_{2n-1}^{(2)},
\\ 
\delta _{i,j}q^{2n+3-2\bar{\imath}},\text{ \ \ \ }i,j=1,2,...,2n+2,\text{ \
\ \ \ \ \ \ \ \ \ \ \ for }\mathcal{D}_{n+1}^{(2)}.%
\end{array}%
\right.  \label{M-matrices}
\end{equation}

We remark that the cases $\mathcal{B}_{n}^{(1)},$ $\mathcal{A}_{2n}^{(2)}$
for $n=1$ are well known: $\mathcal{B}_{1}^{(1)}$ is the
Zamolodchikov-Fateev model \cite{Fa1} (or the spin-1 representation of $%
\mathcal{A}_{1}^{(1)}$) which has $M=1$ and $\rho =\eta $, while $\mathcal{A}%
_{2}^{(2)}$ is the Izergin-Korepin model \cite{IK} with $M=$diag$\left( 
\mathrm{e}^{2\eta },1,\mathrm{e}^{-2\eta }\right) $ and $\rho =-6\eta -i\pi $%
.

Nevertheless, Nepomechie \cite{Ne3} shows that the $\mathcal{A}_{n-1}^{(1)}$
models $(n>2)$, which do not have crossing symmetry, can be treated in
almost the same way as the other cases. Indeed, no assumption of crossing
symmetry is necessary. According to Reshetikhin and Semenov-Tian-Shansky 
\cite{ReSh}, there exists a matrix $M$ such that%
\begin{equation}
\left\{ \left\{ \left\{ \mathcal{R}_{12}^{t_{2}}(u)\right\} ^{-1}\right\}
^{t_{2}}\right\} ^{-1}=\frac{\mathrm{\zeta }(u+\rho )}{\mathrm{\zeta }%
(u+2\rho )}(\mathbf{1}\otimes M)\mathcal{R}_{12}(u+2\rho )(\mathbf{1}\otimes
M)^{-1},  \label{Relation}
\end{equation}%
where $M$ is a symmetry of the $\mathcal{R}$-matrix,%
\begin{equation}
M^{t}=M,\text{ \ \ \ }[\mathcal{R}(u),M\otimes M]=0.  \label{M-matrix}
\end{equation}

Next we introduce matrices $K^{-}(u)$ and $K^{+}(u)$ which satisfy the
reflection equation (\ref{BYBE}) and the dual reflection equation (\ref%
{dualBYBE}), respectively. With the help of equations (\ref{Relation}) and (%
\ref{M-matrix}), one can verify that%
\begin{equation}
K^{+}(u)=\left( K^{-}\right) ^{t}(-u-\rho )M  \label{Automorphism}
\end{equation}%
is an automorphism. Therefore, we do not assume that the $\mathcal{A}%
_{n-1}^{(1)}$ $\mathcal{R}$-matrices have crossing symmetry for $n>2$, but
just replace the crossing relation by the weaker relation (\ref{Relation}),
and the integrability is consequently preserved for these models. Here we
have%
\begin{equation}
M_{i,j}=\delta _{i,j}q^{n+1-2i},\text{ \ \ \ }\rho =n\ln q.
\label{M-matrices II}
\end{equation}

\section{General Solutions}

Now the challenge consists in the calculation of the most general entire set
of solutions of the boundary Yang-Baxter equations (\ref{BYBE}) and (\ref%
{dualBYBE}) for the quantum $\mathcal{R}$-matrices associated with the
non-exceptional affine Lie algebras. In order to achieve this goal, we will
start our search by first looking for $K$-matrices containing only non-null
matrix elements which will be referred to as general solutions.

\subsection{The $\mathcal{A}_{n-1}^{(1)}$ Models}

Analyzing the reflection equation (\ref{BYBE}) for the $\mathcal{A}%
_{n-1}^{(1)}$ models $(n\geq 2)$ one can realize that there exists a very
special structure. There are several functional equations involving only the
non-diagonal matrix elements $k_{i,j}(u)$ $(i\neq j)$ which are the simplest
ones. Let us solve them first.

By direct inspection we verify that the diagonal blocks $B[i,i]$ are
uniquely solved by the relations 
\begin{equation}
\beta _{i,j}k_{j,i}(u)=\beta _{j,i}k_{i,j}(u),\text{ \ \ \ }\forall \text{ }%
i\neq j.  \label{A(n)gen1}
\end{equation}%
Thereby, we only need to find $\frac{n(n-1)}{2}$ elements $k_{i,j}(u)$ $%
(i<j) $. Now, we choose a particular $k_{i,j}(u)$ $(i<j)$ to be different
from zero, with $\beta _{i,j}\neq 0$, and try to express all the remaining
elements in terms of this particular element. We have verified that this is
possible provided that 
\begin{equation}
k_{p,q}(u)=\left\{ 
\begin{array}{c}
\frac{a_{4}(u)}{a_{3}(u)}\frac{\beta _{p,q}}{\beta _{i,j}}k_{i,j}(u),\text{
\ \ \ if }p>i\text{ and }q>j, \\ 
\frac{\beta _{p,q}}{\beta _{i,j}}k_{i,j}(u),\text{ \ \ \ \ \ \ \ \ \ \ if }%
p>i\text{ and }q<j,%
\end{array}%
\right.  \label{A(n)gen2}
\end{equation}%
for $p\neq q$. Combining (\ref{A(n)gen1}) with (\ref{A(n)gen2}) we will
obtain a very strong relation among the non-diagonal elements:%
\begin{equation}
k_{i,j}(u)\neq 0\Rightarrow \left\{ 
\begin{array}{c}
k_{p,j}(u)=0,\text{ \ \ \ for }p\neq i, \\ 
k_{i,q}(u)=0,\text{ \ \ \ for }q\neq j.%
\end{array}%
\right.  \label{A(n)gen3}
\end{equation}%
It means that, for a given $k_{i,j}(u)$, the only elements different from
zero in the $i$th row and in the $j$th column of $K^{-}(u)$ are $k_{i,i}(u)$%
, $k_{i,j}(u)$, $k_{j,j}(u)$ and $k_{j,i}(u)$.

After analyzing more carefully these equations with the conditions (\ref%
{A(n)gen1}) and (\ref{A(n)gen3}), we found from the $\frac{n(n-1)}{2}$
matrix elements $k_{i,j}(u)$ $(i<j)$ that there are two possibilities to
choose a particular $k_{i,j}(u)\neq 0$:

$\bullet $ Only one non-diagonal element and its symmetric one are allowed
to be different from zero. Thus we have $\frac{n(n-1)}{2}$ reflection $K$%
-matrices with $n+2$ non-null elements. Here we denote by $\mathcal{K}%
_{i,j}^{I}$ $(i<j)$ the $K$-matrix whose non-diagonal matrix element $%
k_{i,j}(u)$ was assigned as the non-null matrix element. These $K$-matrices
will be referred to as solutions of type $I$.

$\bullet $ For each $k_{i,j}(u)\neq 0$, additional non-diagonal elements and
their symmetric ones are allowed to be different from zero provided that
they satisfy the equations 
\begin{equation}
k_{i,j}(u)k_{j,i}(u)=k_{r,s}(u)k_{s,r}(u),\text{ \ \ \ with }i+j=r+s\text{
mod }n.  \label{A(n)gen4}
\end{equation}%
It follows that we will get $n$ reflection $K$-matrices whose number of
non-null elements depends on the parity of $n$. Next we choose $n$ possible
particular elements, namely $k_{1,j}(u)$, $j=2,...,n$, and $k_{2,n}(u)$. We
will denote the corresponding $K$-matrices by $\mathcal{K}_{1,j}^{II}$, $%
j=2,...,n$, and $\mathcal{K}_{2,n}^{II}$, respectively, and will be
referring to them as solutions of type $II$.

For example, the $\mathcal{A}_{2}^{(1)}$ model has the following solutions
of type $I$ which turn out to be equal to the corresponding solutions of
type $II$: 
\begin{eqnarray}
\mathcal{K}_{12}^{I} &=&\left( 
\begin{array}{ccc}
k_{11} & \mathbf{k}_{12} & 0 \\ 
k_{21} & k_{22} & 0 \\ 
0 & 0 & k_{33}%
\end{array}%
\right) ,\text{ \ \ \ }\mathcal{K}_{13}^{I}=\left( 
\begin{array}{ccc}
k_{11} & 0 & \mathbf{k}_{13} \\ 
0 & k_{22} & 0 \\ 
k_{31} & 0 & k_{33}%
\end{array}%
\right) ,  \notag \\
\mathcal{K}_{23}^{I} &=&\left( 
\begin{array}{ccc}
k_{11} & 0 & 0 \\ 
0 & k_{22} & \mathbf{k}_{23} \\ 
0 & k_{32} & k_{33}%
\end{array}%
\right) .  \label{A(n)gen5}
\end{eqnarray}%
These $K$-matrices are expected to be the three possibilities to write the
same solution for the $\mathcal{A}_{2}^{(1)}$ model.

For the $\mathcal{A}_{3}^{(1)}$ model we have six solutions of type $I$ \{$%
\mathcal{K}_{12}^{I}$, $\mathcal{K}_{13}^{I}$, $\mathcal{K}_{14}^{I}$, $%
\mathcal{K}_{23}^{I}$, $\mathcal{K}_{24}^{I}$, $\mathcal{K}_{34}^{I}$\} with
six non-null elements. In this case we also have two solutions of type $II$
\{$\mathcal{K}_{12}^{II}$, $\mathcal{K}_{14}^{II}$\} as follows 
\begin{eqnarray}
\mathcal{K}_{12}^{II} &=&\left( 
\begin{array}{cccc}
k_{11} & \mathbf{k}_{12} & 0 & 0 \\ 
k_{21} & k_{22} & 0 & 0 \\ 
0 & 0 & k_{33} & k_{34} \\ 
0 & 0 & k_{43} & k_{44}%
\end{array}%
\right) ,\text{ \ \ \ }k_{12}k_{21}=k_{34}k_{43},  \notag \\
\mathcal{K}_{14}^{II} &=&\left( 
\begin{array}{cccc}
k_{11} & 0 & 0 & \mathbf{k}_{14} \\ 
0 & k_{22} & k_{23} & 0 \\ 
0 & k_{32} & k_{33} & 0 \\ 
k_{41} & 0 & 0 & k_{44}%
\end{array}%
\right) ,\text{ \ \ \ }k_{14}k_{41}=k_{23}k_{32}.  \label{A(n)gen6}
\end{eqnarray}%
For $n\geq 5$, in addition to $\frac{n(n-1)}{2}$ solutions of type $I$ with $%
n+2$ non-null matrix elements, we also find $n$ solutions of type $II$ which
have the following property: if $n$ is odd the $K$-matrices have $2n-1$
non-null elements, but if $n$ is even, half of these $K$-matrices has $2n$
non-null elements while the remaining ones have $2n-2$ non-null elements.

Although we are able to count the $K$-matrices for the $\mathcal{A}%
_{n-1}^{(1)}$ models, we still have to identify which matrices equal one
another. Indeed, we can see a $\mathbb{Z}_{n}$ similarity transformation
that yields the matrix elements positions: 
\begin{equation}
K^{(\alpha )}=h_{\alpha }K^{(0)}h_{n-\alpha },\text{ \ \ \ }\alpha
=0,1,2,...,n-1,  \label{A(n)gen7}
\end{equation}%
where $h_{\alpha }$ are the $\mathbb{Z}_{n}$-matrices%
\begin{equation}
(h_{\alpha })_{i,j}=\delta _{i,j+\alpha }\text{ mod }n.  \label{A(n)gen8}
\end{equation}%
We can define $K^{(0)}$ as $\mathcal{K}_{12}^{II}$ and the above similarity
transformation (\ref{A(n)gen7}) will give the $K^{(\alpha )}$-matrices
possessing matrix elements lying in the same positions as found for the
solutions of type $II$ ($\mathcal{K}_{1,j}^{II}$ and $\mathcal{K}_{2,n}^{II}$%
). However, due to the fact that the relations (\ref{A(n)gen2}) involve the
ratio $\frac{a_{4}(u)}{a_{3}(u)}=\mathrm{e}^{u}$ as well as the additional
constraints (\ref{A(n)gen4}), we could not encounter a similarity
transformation among these matrices, even after performing a gauge
transformation. In fact, the similarity account is not simple due to the
presence of three types of scalar functions and the constraint equations for
the parameters $\beta _{i,j}$ related to the solutions of type $I$, for
instance. Nevertheless, as we have found a way to write all the solutions,
we can leave the similarity account to the reader.

At this point, we proceed in order to discover $n$ diagonal elements $%
k_{i,i}(u)$ in terms of the non-diagonal elements $k_{i,j}(u)$ for each $%
\mathcal{K}_{i,j}$-matrix. Such a procedure is now standard \cite{Li1}. For
instance, if we are looking at $\mathcal{K}_{12}^{II}$, the non-diagonal
elements $k_{i,j}(u)$ $(i+j=3$ mod $n)$ written in terms of $k_{12}(u)$ are
given by 
\begin{equation}
k_{i,j}(u)=\left\{ 
\begin{array}{c}
\frac{\beta _{i,j}}{\beta _{12}}k_{12}(u),\text{ \ \ \ \ \ \ \ \ \ \ \ \ \ \
\ \ for }i+j=3, \\ 
\frac{\beta _{i,j}}{\beta _{12}}\mathrm{e}^{u}k_{12}(u),\text{\ \ \ \ for }%
i+j=3\text{ mod }n, \\ 
0,\text{ \ \ \ \ \ \ \ \ \ \ \ \ \ \ \ \ \ \ \ \ \ \ \ \ \ \ \ \ \ \ \
otherwise},%
\end{array}%
\right.  \label{A(n)gen9}
\end{equation}%
for $i,j=1,2,...,n$ $(i\neq j)$.

By substituting (\ref{A(n)gen9}) into the functional equations, we can
easily find the elements $k_{i,i}(u)$ up to an arbitrary function, here
identified as $k_{12}(u)$. Moreover, their consistency relations will yield
some constraint equations for the parameters $\beta _{i,j}$.

Having found all diagonal elements in terms of $k_{i,j}(u)$, we can, without
loss of generality, choose the arbitrary function as 
\begin{equation}
k_{i,j}(u)=\frac{1}{2}\beta _{i,j}(\mathrm{e}^{2u}-1),\text{ \ \ \ }i<j.
\label{A(n)gen10}
\end{equation}%
This choice allows us to work out the solutions in terms of the functions $%
f_{i,i}(u)$ and $h_{i,j}(u)$, defined as%
\begin{equation}
f_{i,i}(u)=\beta _{i,i}(\mathrm{e}^{u}-1)+1\text{ \ \ \ and \ \ \ }%
h_{i,j}(u)=\frac{1}{2}\beta _{i,j}(\mathrm{e}^{2u}-1),  \label{A(n)gen11}
\end{equation}%
for $i,j=1,2,...,n$.

Next we present the $\mathcal{A}_{n-1}^{(1)}$ general $K$-matrices by
considering each type of solution separately. We remark that the general
solutions for the first values of $n$ are special and will be written
explicitly in Section 6.

\subsubsection{The $K$-matrices of type $I$}

Here we have $\frac{n(n-1)}{2}$ reflection $K$-matrices with $n+2$ non-null
elements. For $1<i<j\leq n$ we get $\frac{(n-2)(n-1)}{2}$ solutions%
\begin{eqnarray}
\mathcal{K}_{i,j}^{I} &=&f_{i,i}(u)E_{ii}+\mathrm{e}%
^{2u}f_{i,i}(-u)E_{jj}+h_{i,j}(u)E_{ij}+h_{j,i}(u)E_{ji}  \notag \\
&&+\mathcal{Z}_{i}(u)\sum_{l=1}^{i-1}E_{ll}+\mathcal{Y}_{i+1}^{(i)}(u)%
\sum_{l=i+1}^{j-1}E_{ll}+\mathrm{e}^{2u}\mathcal{Z}_{i}(u)%
\sum_{l=j+1}^{n}E_{ll},  \label{A(n)gen12}
\end{eqnarray}%
where $\mathcal{Z}_{i}(u)$ and $\mathcal{Y}_{i+1}^{(i)}(u)$ are scalar
functions defined as%
\begin{equation}
\mathcal{Z}_{i}(u)=f_{i,i}(-u)+\frac{1}{2}(\beta _{i,i}+\beta _{11})\mathrm{e%
}^{-u}(\mathrm{e}^{2u}-1)  \label{A(n)gen13}
\end{equation}%
and%
\begin{equation}
\mathcal{Y}_{l}^{(i)}(u)=f_{i,i}(u)+\frac{1}{2}(\beta _{l,l}-\beta _{i,i})(%
\mathrm{e}^{2u}-1).  \label{A(n)gen14}
\end{equation}%
For $i=1$ and $1<j\leq n$ we get the $n-1$ remaining solutions%
\begin{eqnarray}
\mathcal{K}_{1,j}^{I} &=&f_{11}(u)E_{11}+\mathrm{e}%
^{2u}f_{11}(-u)E_{jj}+h_{1,j}(u)E_{1j}+h_{j,1}(u)E_{j1}  \notag \\
&&+\mathcal{Y}_{2}^{(1)}(u)\sum_{l=2}^{j-1}E_{ll}+\mathcal{X}%
_{j+1}(u)\sum_{l=j+1}^{n}E_{ll},  \label{A(n)gen15}
\end{eqnarray}%
where a new scalar function $\mathcal{X}_{j+1}(u)$ appears, given by%
\begin{equation}
\mathcal{X}_{j+1}(u)=\mathrm{e}^{2u}f_{11}(-u)+\frac{1}{2}(\beta
_{j+1,j+1}+\beta _{11}-2)\mathrm{e}^{u}(\mathrm{e}^{2u}-1).
\label{A(n)gen16}
\end{equation}%
The number of free parameters is fixed by the constraint equations which
depend on the presence of the following scalar functions: when $\mathcal{Y}%
_{l}^{(i)}(u)$ is present in $\mathcal{K}_{i,j}^{I}$ we have constraint
equations of the type%
\begin{equation}
\beta _{i,j}\beta _{j,i}=(\beta _{l,l}+\beta _{i,i}-2)(\beta _{l,l}-\beta
_{i,i}),  \label{A(n)gen17}
\end{equation}%
but when $\mathcal{Z}_{i}(u)$ is present the corresponding constraint
equations are of the type 
\begin{equation}
\beta _{i,j}\beta _{j,i}=(\beta _{11}+\beta _{i,i})(\beta _{11}-\beta
_{i,i}).  \label{A(n)gen18}
\end{equation}%
The presence of $\mathcal{X}_{j+1}(u)$ yields a third type of constraint
equations as follows 
\begin{equation}
\beta _{i,j}\beta _{j,i}=(\beta _{j+1,j+1}+\beta _{11}-2)(\beta
_{j+1,j+1}-\beta _{11}-2).  \label{A(n)gen19}
\end{equation}%
From (\ref{A(n)gen12}) and (\ref{A(n)gen15}) we can see that each matrix $%
\mathcal{K}_{i,j}^{I}$ has two scalar functions. It means that all these $%
\mathcal{K}_{i,j}^{I}$-matrices are $3$-parameter general solutions of the
reflection equation (\ref{BYBE}) for the $\mathcal{A}_{n-1}^{(1)}$ models.

Finally, we observe that the solution with $i=1$ and $j=n$, i.e. 
\begin{eqnarray}
\mathcal{K}_{1,n}^{I} &=&f_{11}(u)E_{11}+\mathrm{e}%
^{2u}f_{11}(-u)E_{nn}+h_{1,n}(u)E_{1n}  \notag \\
&&+h_{n,1}(u)E_{n1}+\mathcal{Y}_{2}^{(1)}(u)\sum_{l=2}^{n-1}E_{ll}
\label{A(n)gen20}
\end{eqnarray}%
has the constraint%
\begin{equation}
\beta _{1,n}\beta _{n,1}=(\beta _{22}+\beta _{11}-2)(\beta _{22}-\beta _{11})
\label{A(n)gen21}
\end{equation}%
and it is just the solution derived by Abad and Rios \cite{Ab1}.

\subsubsection{The $K$-matrices of type $II$}

Due to the property (\ref{A(n)gen4}) we have found three general solutions
of type $II$ for each $\mathcal{A}_{n-1}^{(1)}$ model:%
\begin{eqnarray}
\text{Type }II\text{{\small a}} &=&\{\mathcal{K}_{1,2p}^{II}\},\text{ \ \ \
Type }II\text{{\small b}}=\{\mathcal{K}_{1,2p+1}^{II}\},\text{ \ \ \ Type }II%
\text{{\small c}}=\{\mathcal{K}_{2,n}^{II}\},  \notag \\
\text{with \ \ }p &=&1,2,...,\left[ \frac{n}{2}\right]  \label{A(n)gen22}
\end{eqnarray}%
where $\left[ \frac{n}{2}\right] $ is the integer part of $\frac{n}{2}$.

It turns out that all these general $K$-matrices of type $II$ depend on the
parity of $n$, leading us to two classes of solutions as follows.

\paragraph{Odd $n$}

For odd $n$, the solutions of type $II${\small a} are given by 
\begin{eqnarray}
\mathcal{K}_{1,2p}^{II} &=&f_{11}(u)\sum_{j=1}^{p}E_{jj}+\mathrm{e}%
^{2u}f_{11}(-u)\sum_{j=p+1}^{\left[ \frac{n}{2}\right] +p}E_{jj}+\mathrm{e}%
^{2u}f_{11}(u)\sum_{j=\left[ \frac{n}{2}\right] +p+2}^{n}E_{jj}  \notag \\
&&+\mathcal{X}_{\left[ \frac{n}{2}\right] +p+1}(u)E_{\left[ \frac{n}{2}%
\right] +p+1,\left[ \frac{n}{2}\right] +p+1}  \notag \\
&&+\left( \sum_{\substack{ i+j=1+2p  \\ i\neq j}}+\sum_{\substack{ i+j=1+2p%
\text{ mod }n  \\ i\neq j}}\mathrm{e}^{u}\right) h_{i,j}(u)E_{ij},
\label{A(n)gen23}
\end{eqnarray}%
with the constraint equations%
\begin{eqnarray}
\beta _{r,s}\beta _{s,r} &=&(\beta _{\left[ \frac{n}{2}\right] +p+1,\left[ 
\frac{n}{2}\right] +p+1}+\beta _{11}-2)(\beta _{\left[ \frac{n}{2}\right]
+p+1,\left[ \frac{n}{2}\right] +p+1}-\beta _{11}-2),  \notag \\
r+s &=&1+2p\text{ mod }n.  \label{A(n)gen24}
\end{eqnarray}%
The solutions of type $II${\small b} take the form%
\begin{eqnarray}
\mathcal{K}_{1,2p+1}^{II} &=&f_{11}(u)\sum_{j=1}^{p}E_{jj}+\mathrm{e}%
^{2u}f_{11}(-u)\sum_{j=p+2}^{\left[ \frac{n}{2}\right] +p+1}E_{jj}+\mathrm{e}%
^{2u}f_{11}(u)\sum_{j=\left[ \frac{n}{2}\right] +p+2}^{n}E_{jj}  \notag \\
&&+\mathcal{Y}_{p+1}^{(1)}(u)E_{p+1,p+1}+\left( \sum_{\substack{ i+j=2+2p 
\\ i\neq j}}+\sum_{\substack{ i+j=2+2p\text{ mod }n  \\ i\neq j}}\mathrm{e}%
^{u}\right) h_{i,j}(u)E_{ij},  \notag \\
&&  \label{A(n)gen25}
\end{eqnarray}%
with the following constraint equations%
\begin{eqnarray}
\beta _{r,s}\beta _{s,r} &=&(\beta _{p+1,p+1}+\beta _{11}-2)(\beta
_{p+1,p+1}-\beta _{11}),  \notag \\
r+s &=&2+2p\text{ mod }n.  \label{A(n)gen26}
\end{eqnarray}%
Finally, the $K$-matrices of type $II${\small c} are given by%
\begin{eqnarray}
\mathcal{K}_{2,n}^{II} &=&\mathcal{Z}_{2}(u)E_{11}+f_{22}(u)\sum_{j=2}^{%
\left[ \frac{n}{2}\right] +1}E_{jj}+\mathrm{e}^{2u}f_{22}(-u)\sum_{j=\left[ 
\frac{n}{2}\right] +2}^{n}E_{jj}  \notag \\
&&+\sum_{\substack{ i+j=2\text{ mod }n  \\ i\neq j}}h_{i,j}(u)E_{ij},
\label{A(n)gen27}
\end{eqnarray}%
where the constraint equations are%
\begin{eqnarray}
\beta _{r,s}\beta _{s,r} &=&(\beta _{11}+\beta _{22})(\beta _{11}-\beta
_{22}),  \notag \\
r+s &=&2\text{ mod }n.  \label{A(n)gen28}
\end{eqnarray}%
The function $\mathcal{Z}_{2}(u)$ is given by (\ref{A(n)gen13}) and the
functions $\mathcal{Y}_{p+1}^{(1)}(u)$ and $\mathcal{X}_{\left[ \frac{n}{2}%
\right] +p+1}(u)$ are given by (\ref{A(n)gen14}) and (\ref{A(n)gen16}),
respectively, while the functions $f_{11}(u)$, $f_{22}(u)$ and $h_{i,j}(u)$
are given by (\ref{A(n)gen11}). Therefore, for odd $n$ we have found $n$
reflection $K$-matrices, consisting of $\left( 2+\left[ \frac{n}{2}\right]
\right) $-parameter general solutions with $2n-1$ non-null matrix elements.

\paragraph{Even $n$}

Whether $n$ is even, the solutions of type $II${\small a} take the following
form%
\begin{eqnarray}
\mathcal{K}_{1,2p}^{II} &=&f_{11}(u)\sum_{j=1}^{p}E_{jj}+\mathrm{e}%
^{2u}f_{11}(-u)\sum_{j=p+1}^{\frac{n}{2}+p}E_{jj}+\mathrm{e}%
^{2u}f_{11}(u)\sum_{j=\frac{n}{2}+p+1}^{n}E_{jj}  \notag \\
&&+\left( \sum_{\substack{ i+j=1+2p  \\ i\neq j}}+\sum_{\substack{ i+j=1+2p%
\text{ mod }n  \\ i\neq j}}\mathrm{e}^{u}\right) h_{i,j}(u)E_{ij},
\label{A(n)gen29}
\end{eqnarray}%
with the constraint equations%
\begin{equation}
\beta _{1,2p}\beta _{2p,1}=\beta _{r,s}\beta _{s,r},\text{ \ \ \ \ \ \ \ }%
r+s=1+2p\text{ mod }n.  \label{A(n)gen30}
\end{equation}%
The solutions of type $II${\small b} are given by%
\begin{eqnarray}
\mathcal{K}_{1,2p+1}^{II} &=&f_{11}(u)\sum_{j=1}^{p}E_{jj}+\mathcal{Y}%
_{p+1}^{(1)}(u)E_{p+1,p+1}+\mathrm{e}^{2u}f_{11}(-u)\sum_{j=p+2}^{\frac{n}{2}%
+p}E_{jj}  \notag \\
&&+\mathcal{X}_{\frac{n}{2}+p+1}(u)E_{\frac{n}{2}+p+1,\frac{n}{2}+p+1}+%
\mathrm{e}^{2u}f_{11}(u)\sum_{j=\frac{n}{2}+p+2}^{n}E_{jj}  \notag \\
&&+\left( \sum_{\substack{ i+j=2+2p  \\ i\neq j}}+\sum_{\substack{ i+j=2+2p%
\text{ mod }n  \\ i\neq j}}\mathrm{e}^{u}\right) h_{i,j}(u)E_{ij},
\label{A(n)gen31}
\end{eqnarray}%
where the constraint equations are%
\begin{eqnarray}
\beta _{r,s}\beta _{s,r} &=&(\beta _{p+1,p+1}+\beta _{11}-2)(\beta
_{p+1,p+1}-\beta _{11})  \notag \\
&=&(\beta _{\frac{n}{2}+p+1,\frac{n}{2}+p+1}+\beta _{11}-2)(\beta _{\frac{n}{%
2}+p+1,\frac{n}{2}+p+1}-\beta _{11}-2),  \notag \\
r+s &=&2+2p\text{ mod }n.  \label{A(n)gen32}
\end{eqnarray}%
Again, the $K$-matrices of type $II${\small c} take the form%
\begin{eqnarray}
\mathcal{K}_{2,n}^{II} &=&\mathcal{Z}_{2}(u)E_{11}+f_{22}(u)\sum_{j=2}^{%
\frac{n}{2}}E_{jj}+\mathcal{Y}_{\frac{n}{2}+1}^{(2)}(u)E_{\frac{n}{2}+1,%
\frac{n}{2}+1}  \notag \\
&&+\mathrm{e}^{2u}f_{22}(-u)\sum_{j=\frac{n}{2}+2}^{n}E_{jj}+\sum_{\substack{
i+j=2\text{ mod }n  \\ i\neq j}}h_{i,j}(u)E_{ij},  \label{A(n)gen33}
\end{eqnarray}%
with the following constraint equations%
\begin{eqnarray}
\beta _{r,s}\beta _{s,r} &=&(\beta _{11}+\beta _{22})(\beta _{11}-\beta
_{22})  \notag \\
&=&(\beta _{\frac{n}{2}+1,\frac{n}{2}+1}+\beta _{22}-2)(\beta _{\frac{n}{2}%
+1,\frac{n}{2}+1}-\beta _{22}),  \notag \\
r+s &=&2\text{ mod }n,  \label{A(n)gen34}
\end{eqnarray}%
where the scalar functions $\mathcal{Z}_{2}(u)$, $\mathcal{Y}_{l+1}^{(i)}(u)$
and $\mathcal{X}_{\frac{n}{2}+p+1}(u)$ are given by (\ref{A(n)gen13}), (\ref%
{A(n)gen14}) and (\ref{A(n)gen16}), respectively. Thereby, for even $n$ we
have found $\frac{n}{2}$ $(2+\frac{n}{2})$-parameter general solutions of
type $II${\small a} with $2n$ non-null matrix elements, and $\frac{n}{2}$ $%
(1+\frac{n}{2})$-parameter general $K$-matrices of types $II${\small b} and $%
II${\small c}, both solutions with $2(n-1)$ non-null matrix elements.

\subsection{The $\mathcal{B}_{n}^{(1)},$ $\mathcal{C}_{n}^{(1)},$ $\mathcal{D%
}_{n}^{(1)},$ $\mathcal{A}_{2n}^{(2)},$ and $\mathcal{A}_{2n-1}^{(2)}$ Models%
}

\subsubsection{Non-diagonal matrix elements}

Looking into the reflection equation (\ref{BYBE}) for the models of types $%
\mathcal{B}_{n}^{(1)},$ $\mathcal{C}_{n}^{(1)},$ $\mathcal{D}_{n}^{(1)},$ $%
\mathcal{A}_{2n}^{(2)},$ and $\mathcal{A}_{2n-1}^{(2)}$, we note that the
simplest functional equations turn out to be those involving only two matrix
elements $k_{i,i^{\prime }}(u)$ on the secondary diagonal of $K^{-}(u)$,
belonging to the blocks $B[1,2n+3]$, $B[1,4n+5]$, $B[1,6n+7]$, $...$, and we
choose to express their solutions in terms of the element $k_{1,\mathbb{N}%
}(u)$ with $\beta _{1,\mathbb{N}}\neq 0$: 
\begin{equation}
k_{i,i^{\prime }}(u)=\left( \frac{\beta _{i,i^{\prime }}}{\beta _{1,\mathbb{N%
}}}\right) k_{1,\mathbb{N}}(u).  \label{B(n),A(2n)gen1}
\end{equation}%
Next we look at the last blocks of the collection $\{B[1,j]\}$. Here we can
write the matrix elements in the first row $k_{1,j}(u)$ $(j\neq 1,\mathbb{N}%
) $ in terms of the element $k_{1,\mathbb{N}}(u)$ and their transpose in
terms of the element $k_{\mathbb{N},1}(u)$. From the last blocks of the
collection $\{B[2n+3,j]\}$, the matrix elements in the second row $%
k_{2,j}(u) $ $(j\neq 2,\mathbb{N}-1)$ are expressed in terms of $k_{2,%
\mathbb{N}-1}(u)$ and their transpose in terms of $k_{\mathbb{N}-1,2}(u)$.
Applying this procedure to the collections $\{B[4n+5,j]\}$, $\{B[6n+7,j]\}$, 
$...$, we will succeed in writing all non-diagonal matrix elements as%
\begin{equation}
k_{i,j}(u)=\left( \frac{a_{1}a_{11}-a_{2}^{2}}{%
a_{3}a_{4}a_{11}^{2}-a_{2}^{2}a_{12}a_{21}}\right) \left( \beta
_{i,j}a_{3}a_{11}-\beta _{j^{\prime },i^{\prime }}a_{2}a_{i,j^{\prime
}}\right) \frac{k_{1,\mathbb{N}}(u)}{\beta _{1,\mathbb{N}}}\text{ \ \ }%
(j<i^{\prime })  \label{B(n),A(2n)gen2}
\end{equation}%
and%
\begin{equation}
k_{i,j}(u)=\left( \frac{a_{1}a_{11}-a_{2}^{2}}{%
a_{3}a_{4}a_{11}^{2}-a_{2}^{2}a_{12}a_{21}}\right) \left( \beta
_{i,j}a_{4}a_{11}-\beta _{j^{\prime },i^{\prime }}a_{2}a_{i,j^{\prime
}}\right) \frac{k_{1,\mathbb{N}}(u)}{\beta _{1,\mathbb{N}}}\text{ \ \ }%
(j>i^{\prime }),  \label{B(n),A(2n)gen3}
\end{equation}%
where we have used (\ref{B(n),A(2n)gen1}) and the identities%
\begin{equation}
a_{i,j}=a_{j^{\prime },i^{\prime }}\text{ \ \ \ and \ \ \ }%
a_{1j}a_{j1}=a_{12}a_{21}\text{ \ \ \ }(j\neq 1).  \label{B(n),A(2n)gen4}
\end{equation}

Taking into account the Boltzmann weights of each model (\ref%
{B(n),C(n),D(n)weights}), we substitute these expressions into the remaining
functional equations and turn our attention to those without diagonal
entries $k_{i,i}(u)$ aiming to fix some parameters $\beta _{i,j}$ $(i\neq j)$%
. For example, from the diagonal blocks $B[i,i]$ one can see that the
equations are solved by the following relations 
\begin{equation}
\beta _{i,j}k_{j,i}(u)=\beta _{j,i}k_{i,j}(u)  \label{B(n),A(2n)gen5}
\end{equation}%
provided that%
\begin{equation}
\beta _{i,j}\beta _{j^{\prime },i^{\prime }}=\beta _{j,i}\beta _{i^{\prime
},j^{\prime }}.  \label{B(n),A(2n)gen6}
\end{equation}

This procedure is carried out to a large number of equations involving
non-diagonal terms. After performing some algebraic manipulations, we found
two possibilities to express the parameters for the matrix elements lying
below the secondary diagonal $(\beta _{i,j}$ with $j>i^{\prime })$ in terms
of those lying above the secondary diagonal: 
\begin{equation}
\beta _{i,j}=\left\{ 
\begin{array}{c}
\pm \theta _{i}\frac{1}{\sqrt{\xi }}q^{\frac{1}{2}(\bar{\imath}-\bar{\imath}%
^{\prime })+j-n-1}\beta _{j^{\prime },i^{\prime }}\text{ \ \ \ for }j>%
\mathbb{L}, \\ 
\pm \theta _{j}\frac{1}{\sqrt{\xi }}q^{\frac{1}{2}(\bar{j}-\bar{j}^{\prime
})+i-n-1}\beta _{j^{\prime },i^{\prime }}\text{ \ \ for }j\leq \mathbb{L},%
\end{array}%
\right.  \label{B(n),A(2n)gen7}
\end{equation}%
where $\theta _{i}=q\varepsilon _{i}$ for $\mathcal{C}_{n}^{(1)}$, $\theta
_{i}=\frac{1}{\sqrt{q}}$ for $\mathcal{B}_{n}^{(1)},$ $\mathcal{A}%
_{2n}^{(2)} $, and $\theta _{i}=1$ for $\mathcal{D}_{n}^{(1)},$ $\mathcal{A}%
_{2n-1}^{(2)} $, with $\mathbb{L}=n+1$ for $\mathcal{B}_{n}^{(1)},$ $%
\mathcal{A}_{2n}^{(2)} $, $=n$ for $\mathcal{C}_{n}^{(1)},$ $\mathcal{D}%
_{n}^{(1)},$ $\mathcal{A}_{2n-1}^{(2)}$.

These relations simplify significantly the expressions for the non-diagonal
matrix elements (\ref{B(n),A(2n)gen2}) and (\ref{B(n),A(2n)gen3})%
\begin{equation}
k_{i,j}(u)=\left\{ 
\begin{array}{c}
\beta _{i,j}\mathcal{G}^{(\pm )}(u),\text{ \ \ \ \ \ \ \ \ \ \ \ \ \ \ \ \ \
\ }(j<i^{\prime }), \\ 
\beta _{i,i^{\prime }}\left( \frac{q\mathrm{e}^{u}\pm \sqrt{\xi }}{q\pm 
\sqrt{\xi }}\right) \mathcal{G}^{(\pm )}(u),\text{ \ \ \ }(j=i^{\prime }),
\\ 
\beta _{i,j}\mathrm{e}^{u}\mathcal{G}^{(\pm )}(u),\text{ \ \ \ \ \ \ \ \ \ \
\ \ \ \ \ }(j>i^{\prime }),%
\end{array}%
\right.  \label{B(n),A(2n)gen8}
\end{equation}%
where $\mathcal{G}^{(\pm )}(u)$ is defined by assigning a suitable
normalization to $k_{1,\mathbb{N}}(u)$:%
\begin{equation}
\mathcal{G}^{(\pm )}(u)=\frac{1}{\beta _{1,\mathbb{N}}}\left( \frac{q\pm 
\sqrt{\xi }}{q\mathrm{e}^{u}\pm \sqrt{\xi }}\right) k_{1,\mathbb{N}}(u).
\label{B(n),A(2n)gen9}
\end{equation}

We substitute these expressions into the remaining equations and search for
equations of the type 
\begin{equation}
F(u)\mathcal{G}^{(\pm )}(u)=0,  \label{B(n),A(2n)gen10}
\end{equation}%
where $F(u)=\sum_{k}f_{k}(\{\beta _{i,j}\})\mathrm{e}^{ku}$. The constraint
equations $f_{k}(\{\beta _{i,j}\})\equiv 0$, $\forall k$, can be solved in
terms of $\mathbb{N}$ parameters. Of course, the expressions for $k_{i,j}(u)$
will depend on our choice of these parameters. After some attempts, we
concluded that the choice $\beta _{12}$, $\beta _{13}$, ..., $\beta _{1,%
\mathbb{N}}$ and $\beta _{21}$ is the most appropriate for our purpose.
Taking into account all fixed parameters in terms of these $\mathbb{N}$
parameters, we are able to rewrite the matrix elements $k_{i,j}(u)$ $(i\neq
j)$ for $n>\tau $, where $\tau =1$ for $\mathcal{B}_{n}^{(1)},$ $\mathcal{A}%
_{2n}^{(2)}$, $=2$ for $\mathcal{C}_{n}^{(1)},$ $\mathcal{D}_{n}^{(1)},$ $%
\mathcal{A}_{2n-1}^{(2)}$:

The secondary diagonal has the matrix elements%
\begin{equation}
k_{i,i^{\prime }}(u)=\varepsilon _{i}\frac{\beta _{1,i^{\prime }}^{2}}{\beta
_{1,\mathbb{N}}}\frac{q^{\bar{\imath}-\bar{1}}}{\xi }\frac{(q\pm \sqrt{\xi })%
}{(q+1)^{2}}(q\mathrm{e}^{u}\pm \sqrt{\xi })\mathcal{G}^{(\pm )}(u)\text{ \
\ \ }(i\neq 1,\mathbb{N})  \label{B(n),A(2n)gen11}
\end{equation}%
and%
\begin{equation}
k_{\mathbb{N},1}(u)=\varepsilon _{\mathbb{N}}\beta _{1,\mathbb{N}}\frac{%
\beta _{21}^{2}}{\beta _{1,\mathbb{N}-1}^{2}}q^{(\bar{2}n-\bar{1})-\tau
}\left( \frac{q\mathrm{e}^{u}\pm \sqrt{\xi }}{q\pm \sqrt{\xi }}\right) 
\mathcal{G}^{(\pm )}(u).  \label{B(n),A(2n)gen12}
\end{equation}%
In the first row and in the first column of $K^{-}(u)$ the entries are given
by%
\begin{eqnarray}
k_{1,j}(u) &=&\beta _{1,j}\mathcal{G}^{(\pm )}(u)\text{ \ \ \ }(j\neq 1,%
\mathbb{N}),  \label{B(n),A(2n)gen13} \\
k_{i,1}(u) &=&\varepsilon _{i}\beta _{21}\frac{\beta _{1,i^{\prime }}}{\beta
_{1,\mathbb{N}-1}}q^{\bar{\imath}-\bar{2}}\mathcal{G}^{(\pm )}(u)\text{ \ \
\ }(i\neq 1,\mathbb{N}),  \label{B(n),A(2n)gen14}
\end{eqnarray}%
while in the last column and in the last row we have%
\begin{eqnarray}
k_{i,\mathbb{N}}(u) &=&\pm \varepsilon _{i}\beta _{1,i^{\prime }}\frac{q^{%
\bar{\imath}-\bar{1}}}{\sqrt{\xi }}\mathrm{e}^{u}\mathcal{G}^{(\pm )}(u)%
\text{ \ \ \ }(i\neq 1,\mathbb{N}),  \label{B(n),A(2n)gen15} \\
k_{\mathbb{N},j}(u) &=&\pm \beta _{21}\frac{\beta _{1,j}}{\beta _{1,\mathbb{N%
}-1}}\frac{q^{\bar{2}n-\bar{\tau}}}{\sqrt{\xi }}\mathrm{e}^{u}\mathcal{G}%
^{(\pm )}(u)\text{ \ \ \ }(j\neq 1,\mathbb{N}).  \label{B(n),A(2n)gen16}
\end{eqnarray}%
The remaining non-diagonal matrix elements are given by%
\begin{equation}
k_{i,j}(u)=\left\{ 
\begin{array}{c}
\pm \varepsilon _{i}\beta _{1,j}\frac{\beta _{1,i^{\prime }}}{\beta _{1,%
\mathbb{N}}}\frac{q^{\bar{\imath}-\bar{1}}}{\sqrt{\xi }}\left( \frac{q\pm 
\sqrt{\xi }}{q+1}\right) \mathcal{G}^{(\pm )}(u)\text{ \ \ \ \ \ \ \ \ }%
(j<i^{\prime }), \\ 
\varepsilon _{i}\beta _{1,j}\frac{\beta _{1,i^{\prime }}}{\beta _{1,\mathbb{N%
}}}\frac{q^{\bar{\imath}-\bar{1}}}{\xi }\left( \frac{q\pm \sqrt{\xi }}{q+1}%
\right) \mathrm{e}^{u}\mathcal{G}^{(\pm )}(u)\text{ \ \ \ \ \ \ \ }%
(j>i^{\prime }).%
\end{array}%
\right.  \label{B(n),A(2n)gen17}
\end{equation}%
These relations solve all functional equations without diagonal entries $%
k_{i,i}(u)$, $i=1,2,...,\mathbb{N}$. Our next task is to consider the
remaining functional equations which involve the diagonal matrix elements $%
k_{i,i}(u)$, the function $\mathcal{G}^{(\pm )}(u)$ and $2\mathbb{N}$
parameters. By virtue of distinct features of these models, in the following
two subsections, we will direct our attention separately to the $\mathcal{B}%
_{n}^{(1)},$ $\mathcal{A}_{2n}^{(2)}$ series and to the $\mathcal{C}%
_{n}^{(1)},$ $\mathcal{D}_{n}^{(1)},$ $\mathcal{A}_{2n-1}^{(2)}$ series.

\subsubsection{The $\mathcal{B}_{n}^{(1)}$ and $\mathcal{A}_{2n}^{(2)}$
diagonal matrix elements}

It should be first pointed out that almost all equations have only two
diagonal matrix elements. By working out those containing consecutive
elements we can get the following relations%
\begin{equation}
k_{i+1,i+1}(u)=\left\{ 
\begin{array}{c}
k_{i,i}(u)+(\beta _{i+1,i+1}-\beta _{i,i})\mathcal{G}^{(\pm )}(u)\text{ \ \
\ \ \ \ \ \ \ \ \ \ \ \ \ \ \ }(1\leq i\leq n), \\ 
k_{i,i}(u)+(\beta _{i+1,i+1}-\beta _{i,i})\mathrm{e}^{u}\mathcal{G}^{(\pm
)}(u)\text{ \ \ }(n+2<i\leq 2n+1), \\ 
\end{array}%
\right.  \label{B(n),A(2n)gen18}
\end{equation}%
and two special relations%
\begin{eqnarray}
k_{n+1,n+1}(u) &=&k_{n,n}(u)+(\beta _{n+1,n+1}-\beta _{n,n})\mathcal{G}%
^{(\pm )}(u)-\mathcal{J}^{(\pm )}(u),  \label{B(n),A(2n)gen19} \\
k_{n+2,n+2}(u) &=&k_{n+1,n+1}(u)+(\beta _{n+2,n+2}-\beta _{n+1,n+1})\mathrm{e%
}^{u}\mathcal{G}^{(\pm )}(u)  \notag \\
&&-\mathcal{F}^{(\pm )}(u),  \label{B(n),A(2n)gen20}
\end{eqnarray}%
where $\mathcal{J}^{(\pm )}(u)$ and $\mathcal{F}^{(\pm )}(u)$ are scalar
functions defined as%
\begin{equation}
\mathcal{J}^{(\pm )}(u)=\frac{\beta _{1,n}\beta _{1,n+2}}{\beta _{1,2n+1}}%
\frac{q^{n}}{(q+1)^{2}}\left( \frac{q\pm \sqrt{\xi }}{\xi }\right) (\mathrm{e%
}^{u}-1)\mathcal{G}^{(\pm )}(u)  \label{B(n),A(2n)gen20a}
\end{equation}%
and%
\begin{equation}
\mathcal{F}^{(\pm )}(u)=\pm \frac{\beta _{1,n}\beta _{1,n+2}}{\beta _{1,2n+1}%
}\frac{q^{n}}{(q+1)^{2}}\left( \frac{q\pm \sqrt{\xi }}{\sqrt{\xi }}\right) (%
\mathrm{e}^{u}-1)\mathcal{G}^{(\pm )}(u).  \label{B(n),A(2n)gen21}
\end{equation}%
It means that we can express all diagonal matrix elements in terms of $%
\mathcal{G}^{(\pm )}(u)$ and $k_{11}(u)$:%
\begin{eqnarray}
k_{i,i}(u) &=&k_{11}(u)+(\beta _{i,i}-\beta _{11})\mathcal{G}^{(\pm )}(u)%
\text{ \ \ \ for }1\leq i\leq n,  \label{B(n),A(2n)gen22} \\
k_{n+1,n+1}(u) &=&k_{11}(u)+(\beta _{n+1,n+1}-\beta _{11})\mathcal{G}^{(\pm
)}(u)-\mathcal{J}^{(\pm )}(u),  \label{B(n),A(2n)gen23} \\
k_{i,i}(u) &=&k_{11}(u)+(\beta _{n+1,n+1}-\beta _{11})\mathcal{G}^{(\pm
)}(u)+(\beta _{i,i}-\beta _{n,n})\mathrm{e}^{u}\mathcal{G}^{(\pm )}(u) 
\notag \\
&&\text{ }-\mathcal{J}^{(\pm )}(u)-\mathcal{F}^{(\pm )}(u)\text{ \ \ \ for }%
n+2\leq i\leq 2n+1.  \label{B(n),A(2n)gen24}
\end{eqnarray}

The most important fact is that these recurrent relations are closed by the
solution of the block $B[2n+2,4n+2]$. From this block we can get another
expression for $k_{2n+1,2n+1}(u)$: 
\begin{equation}
k_{2n+1,2n+1}(u)=\mathrm{e}^{2u}k_{11}(u)+(\beta _{2n+1,2n+1}-\beta _{11}-2)%
\mathrm{e}^{u}\left( \frac{q\mathrm{e}^{u}\pm \sqrt{\xi }}{q\pm \sqrt{\xi }}%
\right) \mathcal{G}^{(\pm )}(u).  \label{B(n),A(2n)gen25}
\end{equation}%
Taking $i=2n+1$ into (\ref{B(n),A(2n)gen24}) and comparing with (\ref%
{B(n),A(2n)gen25}) we can find $k_{11}(u)$ without solving any additional
equation:%
\begin{eqnarray}
k_{11}(u) &=&\left( \frac{2\mathrm{e}^{u}-(\beta _{n+1,n+1}-\beta _{11})(%
\mathrm{e}^{u}-1)}{\mathrm{e}^{2u}-1}\right) \mathcal{G}^{(\pm )}(u)-\frac{%
\mathcal{J}^{(\pm )}(u)+\mathcal{F}^{(\pm )}(u)}{\mathrm{e}^{2u}-1}  \notag
\\
&&-q\left( \frac{\beta _{2n+1,2n+1}-\beta _{11}-2}{q\pm \sqrt{\xi }}\right) 
\frac{\mathrm{e}^{u}}{\mathrm{e}^{u}+1}\mathcal{G}^{(\pm )}(u).
\label{B(n),A(2n)gen26}
\end{eqnarray}

The above relations were derived for $n>1$. It turns out that the cases $%
\mathcal{B}_{1}^{(1)}$ and $\mathcal{A}_{2}^{(2)}$ are ruled out and their
general solutions will be presented in Section 6. Indeed, these relations
held for the $\mathcal{A}_{2}^{(2)}$ model after involving some
modifications.

Before substituting these expressions into the functional equations, we
first have to fix some parameters. We can, for instance, look at the
combination $\mathrm{e}^{u}k_{11}(u)+k_{22}(u)$ that many equations display.
Consistency conditions of the results will give us all the constraint
equations to find $4n+2$ remaining parameters. Following this procedure, we
can fix $2n-1$ diagonal parameters $\beta _{i,i}$ $(i\neq 1,2n+1)$: 
\begin{equation}
\beta _{i,i}=\left\{ 
\begin{array}{c}
\beta _{11}\pm (-1)^{n}\left( \frac{q\pm \sqrt{\xi }}{\sqrt{\xi }}\right)
\left( \dsum\limits_{j=0}^{i-2}(-q)^{j}\right) \frac{\beta _{1,n}\beta
_{1,n+2}}{\beta _{1,2n+1}}\text{ \ \ \ \ \ \ \ \ \ }(1<i\leq n), \\ 
\beta _{n+2,n+2}-q^{n}\left( \frac{q\pm \sqrt{\xi }}{\xi }\right) \left(
\dsum\limits_{j=0}^{i-n-3}(-q)^{j}\right) \frac{\beta _{1,n}\beta _{1,n+2}}{%
\beta _{1,2n+1}}\text{ \ \ \ \ \ \ \ \ \ \ \ \ \ \ \ \ \ \ \ } \\ 
\text{ \ \ \ }(n+2<i<2n+1),\text{ \ \ \ \ \ \ \ \ \ \ \ \ \ \ \ \ \ \ \ \ \
\ \ \ \ \ \ \ \ \ \ \ \ \ \ \ \ \ \ \ \ \ \ \ \ \ \ \ \ \ \ \ \ \ \ \ \ \ }%
\end{array}%
\right.  \label{B(n),A(2n)gen27}
\end{equation}%
where%
\begin{eqnarray}
\beta _{n+1,n+1} &=&\beta _{11}\pm \frac{q^{n-1/2}}{q+1}\left( \frac{q\pm 
\sqrt{\xi }}{\sqrt{\xi }}\right)  \notag \\
&&\times \left\{ \frac{\beta _{1,n+1}^{2}}{\beta _{1,2n+1}}+\left( \frac{%
q^{n}+(-1)^{n}(q+1)\mp \frac{q^{n}}{\sqrt{\xi }}}{q^{n-1/2}(q+1)}\right) 
\frac{\beta _{1,n}\beta _{1,n+2}}{\beta _{1,2n+1}}\right\} ,  \notag \\
&&  \label{B(n),A(2n)gen28} \\
\beta _{n+2,n+2} &=&\beta _{11}-\frac{q^{n-3/2}(q\pm \sqrt{\xi })(q\mp q%
\sqrt{\xi })}{\xi (q+1)}\times  \notag \\
&&\left\{ \frac{\beta _{1,n+1}^{2}}{\beta _{1,2n+1}}+\left( \frac{%
2q^{n}+(-1)^{n}(q+1)\pm \frac{q^{n}}{\sqrt{\xi }}(q-1)}{q^{n-3/2}\left( 
\frac{q\sqrt{\xi }\mp q}{\sqrt{\xi }}\right) (q+1)}\right) \frac{\beta
_{1,n}\beta _{1,n+2}}{\beta _{1,2n+1}}\right\}  \notag \\
&&  \label{B(n),A(2n)gen29}
\end{eqnarray}%
and $n-1$ non-diagonal parameters%
\begin{eqnarray}
\beta _{21} &=&-\frac{q}{\xi }\left( \frac{q\pm \sqrt{\xi }}{q+1}\right) ^{2}%
\frac{\beta _{12}\beta _{1,2n}^{2}}{\beta _{1,2n+1}^{2}},
\label{B(n),A(2n)gen30} \\
\beta _{1,j} &=&(-1)^{n+j}\frac{\beta _{1,n}\beta _{1,n+2}}{\beta _{1,2n+2-j}%
},\text{ \ \ \ }j=2,3,...,n-1.  \label{B(n),A(2n)gen31}
\end{eqnarray}%
Next we substitute these expressions into the block $B[2n+1,2n+2]$ to fix
the last parameters%
\begin{equation}
\beta _{2n+1,2n+1}=\beta _{11}+2\pm (-1)^{n}\frac{q^{\bar{2}n+\bar{1}}}{\xi
^{\bar{2}}}(q^{2n-1}-\xi )\left( \frac{q\pm \sqrt{\xi }}{q+1}\right) ^{2}%
\frac{\beta _{1,n}\beta _{1,n+2}}{\beta _{1,2n+1}}  \label{B(n),A(2n)gen32}
\end{equation}%
and%
\begin{equation}
\beta _{1,n}=\pm (-1)^{n}\frac{\frac{q^{n}}{\sqrt{\xi q}}(q+1)}{\left[ \pm 
\frac{q^{n}}{\sqrt{\xi }}-(-1)^{n}\right] ^{2}}\left[ \frac{\beta
_{1,n+1}^{2}}{\beta _{1,n+2}}+2\xi \frac{(q+1)}{q^{n-\bar{1}}(q\pm \sqrt{\xi 
})(q\mp q\sqrt{\xi })}\frac{\beta _{1,2n+1}}{\beta _{1,n+2}}\right] .
\label{B(n),A(2n)gen33}
\end{equation}%
Hence, we have derived two $(n+2)$-parameter general solutions for the $%
\mathcal{B}_{n}^{(1)}$ case and two $(n+2)$-parameter complex conjugate
general solutions for the $\mathcal{A}_{2n}^{(2)}$ case, whose $n+2$ free
parameters are $\beta _{1,n+1}$, $\beta _{1,n+2}$, $...$, $\beta _{1,2n+1}$
and $\beta _{11}$. Nevertheless, the number of free parameters turns out to
be $n+1$ because we still need to make use of the regularity property (\ref%
{K-matrixReg}). For example, we choose the arbitrary function as%
\begin{equation}
k_{1,2n+1}(u)=\frac{1}{2}\beta _{1,2n+1}(\mathrm{e}^{2u}-1)
\label{B(n),A(2n)gen34}
\end{equation}%
and fix the parameter $\beta _{11}$ by the regular condition.

Let us summarize our results: Firstly, we had from (\ref{B(n),A(2n)gen11})
to (\ref{B(n),A(2n)gen17}) all non-diagonal matrix elements. Secondly, the
diagonal matrix elements were obtained by using (\ref{B(n),A(2n)gen22}), (%
\ref{B(n),A(2n)gen23}) and (\ref{B(n),A(2n)gen24}) with $k_{11}(u)$ given by
(\ref{B(n),A(2n)gen26}). Finally, we substituted into these matrix elements
all fixed parameters given by (\ref{B(n),A(2n)gen27})-(\ref{B(n),A(2n)gen33}%
).

\subsubsection{The $\mathcal{C}_{n}^{(1)},$ $\mathcal{D}_{n}^{(1)},$ and $%
\mathcal{A}_{2n-1}^{(2)}$ diagonal matrix elements}

From the equations $E[1,2]$ and $E[1,2n+1]$ we can find $k_{11}(u)$ and $%
k_{22}(u)$, and from the equations $E[4n^{2},4n^{2}-1]$ and $%
E[4n^{2},4n^{2}+2n]$ we also find $k_{2n,2n}(u)$ and $k_{2n-1,2n-1}(u)$.
Next we turn our attention to the equations $E[2,j]$, $j=3,4,...,2n-2$, in
order to get the matrix elements $k_{j,j}(u)$.

We notice that the expressions thus obtained for the diagonal elements are
too large. However, after finding the following $n-1$ non-diagonal parameters%
\begin{eqnarray}
\beta _{21} &=&-\frac{q}{\xi }\left( \frac{q\pm \sqrt{\xi }}{q+1}\right) ^{2}%
\frac{\beta _{12}\beta _{1,2n-1}^{2}}{\beta _{1,2n}^{2}},
\label{C(n),D(n),A(2n-1)gen16a} \\
\beta _{1,j} &=&(-1)^{n+j}\frac{\beta _{1,n}\beta _{1,n+1}}{\beta _{1,2n+1-j}%
},\text{ \ \ \ }j=2,3,...,n-1,  \label{C(n),D(n),A(2n-1)gen16}
\end{eqnarray}%
we noted that they are related to $k_{11}(u)$ in a very simple way:%
\begin{eqnarray}
&&k_{i,i}(u)=k_{11}(u)+(\beta _{i,i}-\beta _{11})\mathcal{G}^{(\pm )}(u)%
\text{ \ \ \ }(2\leq i\leq n),  \label{C(n),D(n),A(2n-1)gen17} \\
&&k_{n+1,n+1}(u)=k_{n,n}(u)+(\beta _{n+1,n+1}-\beta _{n,n})\mathrm{e}^{u}%
\mathcal{G}^{(\pm )}(u)+\mathcal{H}^{(\pm )}(u),
\label{C(n),D(n),A(2n-1)gen18} \\
&&k_{i,i}(u)=k_{n+1,n+1}(u)+(\beta _{i,i}-\beta _{n+1,n+1})\mathrm{e}^{u}%
\mathcal{G}^{(\pm )}(u)\text{ \ }(n+2\leq i\leq 2n),  \notag \\
&&  \label{C(n),D(n),A(2n-1)gen19}
\end{eqnarray}%
where%
\begin{equation}
\mathcal{H}^{(\pm )}(u)=-\Delta _{n}^{(\pm )}(-q)^{n-1}\frac{(\theta
_{n+1}^{2}-\varepsilon _{n+1})}{(q+1)^{2}}(\mathrm{e}^{u}-1)\mathcal{G}%
^{(\pm )}(u)  \label{C(n),D(n),A(2n-1)gen20}
\end{equation}%
with%
\begin{equation}
\Delta _{n}^{(\pm )}=\pm (-1)^{n}\frac{\beta _{1,n}\beta _{1,n+1}}{\beta
_{1,2n}}\left( \frac{q\pm \sqrt{\xi }}{\sqrt{\xi }}\right) .
\label{C(n),D(n),A(2n-1)gen21}
\end{equation}%
Here we point out that $\mathcal{H}^{(\pm )}(u)=0$ for the $\mathcal{D}%
_{n}^{(1)}$ and $\mathcal{A}_{2n-1}^{(2)}$ models.

An important simplification occurs when we consider the equation $E[2n+1,4n]$
separately. This equation yields an additional relation between $%
k_{2n,2n}(u) $ and $k_{11}(u)$:%
\begin{equation}
k_{2n,2n}(u)=\mathrm{e}^{2u}k_{11}(u)+(\beta _{2n,2n}-\beta _{11}-2)\mathrm{e%
}^{u}\left( \frac{q\mathrm{e}^{u}\pm \sqrt{\xi }}{q\pm \sqrt{\xi }}\right) 
\mathcal{G}^{(\pm )}(u).  \label{C(n),D(n),A(2n-1)gen22}
\end{equation}

Taking $i=2n$ into (\ref{C(n),D(n),A(2n-1)gen19}) and comparing with (\ref%
{C(n),D(n),A(2n-1)gen22}) we can find the following expression for $%
k_{11}(u) $:%
\begin{eqnarray}
k_{11}(u) &=&\frac{\mathcal{H}^{(\pm )}(u)}{\mathrm{e}^{2u}-1}+\left\{ \beta
_{n,n}-\beta _{11}+(\beta _{2n,2n}-\beta _{n,n})\mathrm{e}^{u}\right.  \notag
\\
&&\left. -(\beta _{2n,2n}-\beta _{11}-2)\mathrm{e}^{u}\left( \frac{q\mathrm{e%
}^{u}\pm \sqrt{\xi }}{q\pm \sqrt{\xi }}\right) \right\} \frac{\mathcal{G}%
^{(\pm )}(u)}{\mathrm{e}^{2u}-1}.  \label{C(n),D(n),A(2n-1)gen23}
\end{eqnarray}

Substituting these expressions into the functional equations we get
constraint equations which will allow us to fix some of the $3n+1$ remaining
parameters. We recall the equations $E[2,2n+j]$ to find $\beta _{j,j}$, $%
j=3,4,...,2n-2$, in terms of the diagonal parameter $\beta _{22}$ given by
the equation $E[2,2n+1]$. After performing this calculation, we used the
equation $E[2,2n-1]$ to identify $\beta _{2n-1,2n-1}$ and $\beta _{2n,2n}$.
These parameters can be written in terms of $\beta _{11}$, $\beta _{1,n}$, $%
\beta _{1,n+1}$ and $\beta _{1,2n}$ in the following way: 
\begin{eqnarray}
\beta _{i,i} &=&\beta _{11}+\Delta _{n}^{(\pm )}\sum_{j=0}^{i-2}(-q)^{j}%
\text{ \ \ \ }(1<i\leq n),  \label{C(n),D(n),A(2n-1)gen24} \\
\beta _{n+1,n+1} &=&\beta _{11}+\Delta _{n}^{(\pm )}\left[ \frac{1-(-q)^{n-1}%
}{q+1}\pm (-q)^{n-1}\frac{(\theta _{n+1}^{2}-\varepsilon _{n+1})(q\pm \sqrt{%
\xi })}{\sqrt{\xi }(q+1)^{2}}\right] ,  \notag \\
&&  \label{C(n),D(n),A(2n-1)gen25} \\
\beta _{i,i} &=&\beta _{n+1,n+1}+\Delta _{n}^{(\pm )}\left[ \pm \frac{\theta
_{n+1}^{2}\varepsilon _{n+1}}{\sqrt{\xi }}\sum_{j=n-1}^{i-3}(-q)^{j}\right] 
\notag \\
&&\text{ }(n+2\leq i\leq 2n-1),  \label{C(n),D(n),A(2n-1)gen26} \\
\beta _{2n,2n} &=&\beta _{11}+2+\Delta _{n}^{(\pm )}\frac{(q\pm \sqrt{\xi })%
}{\xi }\frac{(\xi -\varepsilon _{2n}q^{\bar{2}n-\bar{1}})}{(q+1)^{2}}.
\label{C(n),D(n),A(2n-1)gen27}
\end{eqnarray}%
Note that $\beta _{n+1,n+1}=\beta _{n,n}$ or that $k_{n+1,n+1}(u)=k_{n,n}(u)$
for $\mathcal{D}_{n}^{(1)}$ and $\mathcal{A}_{2n-1}^{(2)}$ models.

Finally, we can, for example, use the equation $E[2,4n]$ to fix $\beta
_{1,n} $: 
\begin{equation}
\beta _{1,n}=\pm (-1)^{n}\frac{2\xi \sqrt{\xi }(q+1)^{2}}{(1\mp \sqrt{\xi }%
)[\theta _{1}q^{n-1}\mp (-1)^{n}\sqrt{\xi }][\theta _{n+1}q^{n}\mp (-1)^{n}%
\sqrt{\xi }](q\pm \sqrt{\xi })}\frac{\beta _{1,2n}}{\beta _{1,n+1}}.
\label{C(n),D(n),A(2n-1)gen28}
\end{equation}

Although it has been possible to treat these solutions simultaneously in the
above calculations, it is now necessary to separate them in order to take
into account the existence of the amplitude $k_{1,n}(u)$ for each model:

$\bullet $ For $\mathcal{A}_{2n-1}^{(2)}$ models we have $\xi =-q^{2n}$ and $%
\theta _{k}=\varepsilon _{k}=1$, $\forall k$, and there is no restriction in
(\ref{C(n),D(n),A(2n-1)gen28}). It follows that the solution with $\mathcal{G%
}^{(+)}(u)$ (upper sign) is related to the solution with $\mathcal{G}%
^{(-)}(u)$ (lower sign) by complex conjugation.

$\bullet $ For $\mathcal{D}_{n}^{(1)}$ models we have $\xi =q^{2n-2}$ and $%
\theta _{k}=\varepsilon _{k}=1$, $\forall k$. It means that the factors $%
[q^{n-1}\mp (-1)^{n}\sqrt{\xi }]$ are different from zero for the solution
with $\mathcal{G}^{(+)}(u)$ only if $n$ is odd and for the solution with $%
\mathcal{G}^{(-)}(u)$ if $n$ is even.

$\bullet $ For $\mathcal{C}_{n}^{(1)}$ models we have $\xi =q^{2n+2}$ and $%
\theta _{1}=-\theta _{n+1}=q$. In this case, the factors $[-q^{n+1}\mp
(-1)^{n}\sqrt{\xi }]$ are different from zero for the solution with $%
\mathcal{G}^{(+)}(u)$ if $n$ is even and for the solution with $\mathcal{G}%
^{(-)}(u)$ if $n$ is odd.

Therefore, we have found two general solutions for the $\mathcal{A}%
_{2n-1}^{(2)}$ models, and one general solution for the $\mathcal{C}%
_{n}^{(1)}$ and $\mathcal{D}_{n}^{(1)}$ models.

Substituting all fixed parameters into (\ref{C(n),D(n),A(2n-1)gen23}) we
obtain the following expressions for the amplitude $k_{11}(u)$: 
\begin{eqnarray}
k_{11}(u) &=&\frac{2\mathrm{e}^{u}\mathcal{G}^{(\pm )}(u)}{\mathrm{e}^{2u}-1}
\notag \\
&&-\frac{2\mathcal{G}^{(\pm )}(u)}{\mathrm{e}^{u}+1}\left\{ \frac{\xi
\lbrack 1+q-(-q)^{n-1}+(-q)^{n}]+q\mathrm{e}^{u}(\xi -q^{2n-2})}{(1\mp \sqrt{%
\xi })[q^{n-1}\mp (-1)^{n}\sqrt{\xi }][q^{n}\mp (-1)^{n}\sqrt{\xi }]}\right\}
\notag \\
&&  \label{C(n),D(n),A(2n-1)gen29}
\end{eqnarray}%
for $\mathcal{D}_{n}^{(1)}$ and $\mathcal{A}_{2n-1}^{(2)}$ models, and 
\begin{eqnarray}
k_{11}(u) &=&\frac{2\mathrm{e}^{u}\mathcal{G}^{(\pm )}(u)}{\mathrm{e}^{2u}-1}
\notag \\
&&+\frac{2\mathcal{G}^{(\pm )}(u)}{\mathrm{e}^{u}+1}\left\{ \frac{\xi
\lbrack 1+q+(-q)^{n}+(-q)^{n+1}]+q\mathrm{e}^{u}(\xi +q^{2n})}{(1\mp \sqrt{%
\xi })[q^{n}\mp (-1)^{n}\sqrt{\xi }][q^{n+1}\pm (-1)^{n}\sqrt{\xi }]}\right\}
\notag \\
&&  \label{C(n),D(n),A(2n-1)gen30}
\end{eqnarray}%
for $\mathcal{C}_{n}^{(1)}$ models.

From (\ref{C(n),D(n),A(2n-1)gen29}) we realize that $k_{11}(u)$ is quite
simple for the $\mathcal{D}_{n}^{(1)}$ models: 
\begin{equation}
k_{11}(u)=\frac{\mathcal{G}^{(+)}(u)}{\mathrm{e}^{u}-1}\text{ \ \ \ }(\text{%
odd }n),\text{ \ \ \ \ \ \ \ }k_{11}(u)=\frac{\mathcal{G}^{(-)}(u)}{\mathrm{e%
}^{u}-1}\text{ \ \ \ }(\text{even }n).  \label{C(n),D(n),A(2n-1)gen31}
\end{equation}%
Moreover, by substituting (\ref{C(n),D(n),A(2n-1)gen30}) into (\ref%
{C(n),D(n),A(2n-1)gen17})-(\ref{C(n),D(n),A(2n-1)gen19}) we find a simple
relation between the diagonal matrix elements for the $\mathcal{C}_{n}^{(1)}$
models:%
\begin{equation}
k_{n+i,n+i}(u)=\mathrm{e}^{u}k_{i,i}(u)\Rightarrow \beta _{n+i,n+i}=\beta
_{i,i}+1,\text{ \ \ \ }1\leq i\leq n.  \label{C(n),D(n),A(2n-1)gen32}
\end{equation}

Now, let us summarize these results: Firstly, we had from (\ref%
{B(n),A(2n)gen11}) to (\ref{B(n),A(2n)gen17}) all non-diagonal matrix
elements after substituting $n$ fixed non-diagonal parameters $\beta _{21}$
and $\beta _{1,j}$ $(j=2,...,n)$ given by (\ref{C(n),D(n),A(2n-1)gen16a}), (%
\ref{C(n),D(n),A(2n-1)gen16}) and (\ref{C(n),D(n),A(2n-1)gen28}),
respectively. Secondly, the diagonal matrix elements were obtained by using (%
\ref{C(n),D(n),A(2n-1)gen17}), (\ref{C(n),D(n),A(2n-1)gen18}) and (\ref%
{C(n),D(n),A(2n-1)gen19}) with $k_{11}(u)$ given by (\ref%
{C(n),D(n),A(2n-1)gen29}) for $\mathcal{D}_{n}^{(1)}$ and $\mathcal{A}%
_{2n-1}^{(2)}$ models and by (\ref{C(n),D(n),A(2n-1)gen30}) for $\mathcal{C}%
_{n}^{(1)}$ models, and by substituting the diagonal parameters given by (%
\ref{C(n),D(n),A(2n-1)gen24}), (\ref{C(n),D(n),A(2n-1)gen25}), (\ref%
{C(n),D(n),A(2n-1)gen26}) and (\ref{C(n),D(n),A(2n-1)gen27}).

These calculations lead to two complex conjugate general solutions with $n+1$
free parameters, $\beta _{1,n+1}$, $\beta _{1,n+2}$, $...$, $\beta _{1,2n}$
and $\beta _{11}$, for the $\mathcal{A}_{2n-1}^{(2)}$ models, one $(n+1)$%
-parameter general solution for the $\mathcal{C}_{n}^{(1)}$ models, and one $%
(n+1)$-parameter general solution for the $\mathcal{D}_{n}^{(1)}$ models.
However, the number of free parameters turns out to be $n$ since we still
have to apply the regular condition (\ref{K-matrixReg}), which will fix the
parameter $\beta _{11}$.

Here we remark that the above classification holds for $n>2$. Thus, the
cases $\mathcal{A}_{1}^{(2)},$ $\mathcal{A}_{3}^{(2)},$ $\mathcal{C}%
_{1}^{(1)},$ $\mathcal{C}_{2}^{(1)},$ $\mathcal{D}_{1}^{(1)},$ and $\mathcal{%
D}_{2}^{(1)}$ are special and will be treated in Section 6.

\subsection{The $\mathcal{D}_{n+1}^{(2)}$ Models}

\subsubsection{Non-diagonal matrix elements}

The reflection equation (\ref{BYBE}) for the $\mathcal{D}_{n+1}^{(2)}$
models exhibits a special feature. There are lots of functional equations
involving only the elements lying out of a block diagonal structure which
consists of the diagonal elements $k_{i,i}(u)$ plus the central elements on
the secondary diagonal, namely $k_{n+1,n+2}(u)$ and $k_{n+2,n+1}(u)$. The
simplest functional equations possess only the elements on the secondary
diagonal, and we choose to express their solutions in terms of the element $%
k_{1,2n+2}(u)$: 
\begin{equation}
k_{i,i^{\prime }}(u)=\left( \frac{\beta _{i,i^{\prime }}}{\beta _{1,2n+2}}%
\right) k_{1,2n+2}(u),\text{ \ \ \ }i\neq n+1,n+2.  \label{D(n+1)gen1}
\end{equation}%
From the collections $\{B[i,j]\}$, $i=1,2,...,n-1$, one can note that the
equations from the last blocks of each collection are simple and can be
easily solved by expressing the elements $k_{i,j}(u)$ with $j\neq i^{\prime
} $ in terms of $k_{1,2n+2}(u)$: 
\begin{equation}
k_{i,j}(u)=\left\{ 
\begin{array}{c}
\mathbb{F}_{i,j}\left( \beta _{i,j}a_{3}a_{i,i}-\beta _{j^{\prime
},i^{\prime }}a_{2}a_{i,j^{\prime }}\right) k_{1,2n+2}(u)\text{ \ \ \ }%
(i<j^{\prime }), \\ 
\mathbb{F}_{i,j}\left( \beta _{i,j}a_{4}a_{i,i}-\beta _{j^{\prime
},i^{\prime }}a_{2}a_{i,j^{\prime }}\right) k_{1,2n+2}(u)\text{ \ \ \ }%
(i>j^{\prime }),%
\end{array}%
\right.  \label{D(n+1)gen2}
\end{equation}%
with%
\begin{equation}
\mathbb{F}_{i,j}=\frac{a_{1}a_{i,i}-a_{2}^{2}}{\beta
_{1,2n+2}(a_{3}a_{4}a_{i,i}^{2}-a_{2}^{2}a_{i,j^{\prime }}a_{j^{\prime },i})}%
.  \label{D(n+1)gen3}
\end{equation}%
Moreover, for $j=n+1,n+2$ with $i\neq j,j^{\prime }$ we have%
\begin{equation}
k_{i,j}=\left\{ 
\begin{array}{c}
\Delta _{i}[a_{i,i}(\beta _{i,j}a_{5}+\beta _{i,j^{\prime
}}a_{7})-a_{2}(\beta _{j^{\prime },i^{\prime }}b_{i}^{+}+\beta _{j,i^{\prime
}}b_{i}^{-})]k_{1,2n+2}\text{ \ \ \ }(i^{\prime }>j), \\ 
\Delta _{i}[a_{i,i}(\beta _{i,j}a_{6}+\beta _{i,j^{\prime
}}a_{8})-a_{2}(\beta _{j^{\prime },i^{\prime }}b_{i}^{+}+\beta _{j,i^{\prime
}}b_{i}^{-})]k_{1,2n+2}\text{ \ \ \ }(i^{\prime }<j),%
\end{array}%
\right.  \label{D(n+1)gen4}
\end{equation}%
and for $i=n+1,n+2$ with $j\neq i,i^{\prime }$ we get 
\begin{equation}
k_{i,j}=\left\{ 
\begin{array}{c}
\Delta _{j}[a_{j,j}(\beta _{i,j}a_{5}+\beta _{i^{\prime
},j}a_{7})-a_{2}(\beta _{j^{\prime },i^{\prime }}b_{j}^{+}+\beta _{j^{\prime
},i}b_{j}^{-})]k_{1,2n+2}\text{ \ \ \ }(i^{\prime }>j), \\ 
\Delta _{j}[a_{j,j}(\beta _{i,j}a_{6}+\beta _{i^{\prime
},j}a_{8})-a_{2}(\beta _{j^{\prime },i^{\prime }}b_{j}^{+}+\beta _{j^{\prime
},i}b_{j}^{-})]k_{1,2n+2}\text{ \ \ \ }(i^{\prime }<j),%
\end{array}%
\right.  \label{D(n+1)gen5}
\end{equation}%
where%
\begin{equation}
\Delta _{l}=\frac{a_{1}a_{l,l}-a_{2}^{2}}{\beta
_{1,2n+2}[a_{l,l}^{2}(a_{6}+a_{8})(a_{5}+a_{7})-a_{2}^{2}(b_{l}^{+}+b_{l}^{-})(b_{l^{\prime }}^{+}+b_{l^{\prime }}^{-})]%
}.  \label{D(n+1)gen6}
\end{equation}%
Here we observe that $\mathbb{F}_{i,j}=0$ and $\Delta _{l}=\frac{0}{0}$ for
the $\mathcal{D}_{2}^{(2)}$ model. However, by choosing $\Delta _{l}$
suitably the case $n=1$ can be included in our discussion. We will carry out
this calculation in Section 6.

Next we substitute the above expressions back to the remaining functional
equations. In fact, it is enough to consider the equations from the
collections $\{B[1,j]\}$ and $\{B[2,j]\}$. Let us look for equations of the
type%
\begin{equation}
G(u)k_{1,2n+2}(u)=0,  \label{D(n+1)gen7}
\end{equation}%
where $G(u)=\sum_{k}f_{k}(\{\beta _{i,j}\})\mathrm{e}^{ku}$. The constraint
equations $f_{k}(\{\beta _{i,j}\})\equiv 0$, $\forall k$, can be solved in
terms of $2n+2$ parameters which allow us to find all elements $k_{i,j}(u)$
lying\ out of the block diagonal structure in terms of $k_{1,2n+2}(u)$.

Following the same patterns outlined in Section 3.2, the expressions for $%
k_{i,j}(u)$ will depend on our choice of these parameters. As before, the
choice $\beta _{12}$, $\beta _{13}$, ..., $\beta _{1,2n+2}$ and $\beta _{21}$
turns out to be the most suitable for our purpose. Taking into account the
fixed parameters and the Boltzmann weights of the $\mathcal{D}_{n+1}^{(2)}$
models (\ref{D(n+1)weights1}) and (\ref{D(n+1)weights2}), we can rewrite
these matrix elements $k_{i,j}(u)$ for $n>1$ in the following way:

The elements on the secondary diagonal of $K^{-}(u)$ are given by 
\begin{equation}
k_{i,i^{\prime }}(u)=q^{\bar{\imath}-2n}\left( \frac{q^{n-1}+1}{q+1}\right)
^{2}\left( \frac{\beta _{1,i^{\prime }}}{\beta _{1,2n+2}}\right)
^{2}k_{1,2n+2}(u)\text{ \ \ \ }(i\neq 1,2n+2)  \label{D(n+1)gen8}
\end{equation}%
and%
\begin{equation}
k_{2n+2,1}(u)=q^{2n-3}\left( \frac{\beta _{21}}{\beta _{1,2n+1}}\right)
^{2}k_{1,2n+2}(u).  \label{D(n+1)gen9}
\end{equation}%
The matrix elements in the first row and in the first column are,
respectively,%
\begin{eqnarray}
k_{1,j}(u) &=&\frac{q^{n-1}+1}{\mathrm{e}^{2u}+q^{n-1}}\frac{\Gamma _{1,j}(u)%
}{\beta _{1,2n+2}}k_{1,2n+2}(u)\text{ \ \ \ }(j\neq 2n+2),
\label{D(n+1)gen10} \\
k_{i,1}(u) &=&q^{\bar{\imath}-3}\frac{q^{n-1}+1}{\mathrm{e}^{2u}+q^{n-1}}%
\frac{\Gamma _{1,i^{\prime }}(u)}{\beta _{1,2n+2}}\frac{\beta _{21}}{\beta
_{1,2n+1}}k_{1,2n+2}(u)\text{ \ \ \ }(i\neq 2n+2),  \notag \\
&&  \label{D(n+1)gen11}
\end{eqnarray}%
while the elements in the last column and in the last row are%
\begin{eqnarray}
k_{i,2n+2}(u) &=&q^{\bar{\imath}-n-2}\frac{q^{n-1}+1}{\mathrm{e}^{2u}+q^{n-1}%
}\frac{\Pi _{1,i^{\prime }}(u)}{\beta _{1,2n+2}}\mathrm{e}^{2u}k_{1,2n+2}(u)%
\text{ \ \ \ }(i\neq 1),  \label{D(n+1)gen12} \\
k_{2n+2,j}(u) &=&q^{n-2}\frac{q^{n-1}+1}{\mathrm{e}^{2u}+q^{n-1}}\frac{\beta
_{21}}{\beta _{1,2n+1}}\frac{\Pi _{1,j}(u)}{\beta _{1,2n+2}}\mathrm{e}%
^{2u}k_{1,2n+2}(u)\text{ \ \ \ }(j\neq 1).  \notag \\
&&  \label{D(n+1)gen13}
\end{eqnarray}%
The remaining matrix elements are given by%
\begin{equation}
k_{i,j}(u)=q^{\bar{\imath}-n-1}\frac{q^{n-1}+1}{q+1}\frac{q^{n-1}+1}{\mathrm{%
e}^{2u}+q^{n-1}}\frac{\Gamma _{1,i^{\prime }}(u)}{\beta _{1,2n+2}}\frac{%
\Gamma _{1,j}(u)}{\beta _{1,2n+2}}k_{1,2n+2}(u)  \label{D(n+1)gen14}
\end{equation}%
for $i^{\prime }>j$, and%
\begin{equation}
k_{i,j}(u)=q^{\bar{\imath}-2n-1}\frac{q^{n-1}+1}{q+1}\frac{q^{n-1}+1}{%
\mathrm{e}^{2u}+q^{n-1}}\frac{\Pi _{1,i^{\prime }}(u)}{\beta _{1,2n+2}}\frac{%
\Pi _{1,j}(u)}{\beta _{1,2n+2}}\mathrm{e}^{2u}k_{1,2n+2}(u)
\label{D(n+1)gen15}
\end{equation}%
for $i^{\prime }<j$.

In these expressions we are making use of a compact notation defined as%
\begin{equation}
\Gamma _{1,a}(u)=\left\{ 
\begin{array}{c}
\beta _{1,a}\text{ \ \ \ \ \ \ \ \ \ \ }(a\neq n+1,n+2), \\ 
\frac{1}{2}(\mathrm{e}^{u}\beta _{-}+\beta _{+})\text{ \ \ \ \ \ }(a=n+1),
\\ 
\frac{1}{2}(-\mathrm{e}^{u}\beta _{-}+\beta _{+})\text{ \ \ \ }(a=n+2),%
\end{array}%
\right.  \label{D(n+1)gen16}
\end{equation}%
and%
\begin{equation}
\Pi _{1,a}(u)=\left\{ 
\begin{array}{c}
\beta _{1,a}\text{ \ \ \ \ \ \ \ \ \ \ \ \ \ \ \ }(a\neq n+1,n+2), \\ 
\frac{1}{2}(q^{n}\mathrm{e}^{-u}\beta _{-}+\beta _{+})\text{ \ \ \ \ \ }%
(a=n+1), \\ 
\frac{1}{2}(-q^{n}\mathrm{e}^{-u}\beta _{-}+\beta _{+})\text{ \ \ \ }(a=n+2),%
\end{array}%
\right.  \label{D(n+1)gen17}
\end{equation}%
where $\beta _{\pm }=\beta _{1,n+1}\pm \beta _{1,n+2}$.

\subsubsection{Block diagonal matrix elements}

We have reached a point in which we have $2n(2n+3)$ matrix elements in terms
of $2n+2$ parameters. Nevertheless, we still need to find $2n+4$ matrix
elements that belong to the block diagonal structure.

Such a block diagonal structure has the form%
\begin{equation}
\text{Diag}(k_{11},k_{22},...,k_{n,n},\mathbb{B}%
,k_{n+3,n+3},...,k_{2n+2,2n+2}),  \label{D(n+1)gen18}
\end{equation}%
where $\mathbb{B}$ contains the central elements%
\begin{equation}
\mathbb{B}=\left( 
\begin{array}{cc}
k_{n+1,n+1} & k_{n+1,n+2} \\ 
k_{n+2,n+1} & k_{n+2,n+2}%
\end{array}%
\right) .  \label{D(n+1)gen19}
\end{equation}%
Here the situation is a bit different. Although it is very cumbersome to
write these matrix elements in terms of the Boltzmann weights, after
performing some algebraic manipulations we succeeded in finding out that the
diagonal elements satisfy two distinct recurrent relations:%
\begin{equation}
k_{i,i}(u)=\left\{ 
\begin{array}{c}
k_{11}(u)-\frac{q^{n-1}+1}{\mathrm{e}^{2u}+q^{n-1}}\frac{\beta _{11}-\beta
_{i,i}}{\beta _{1,2n+2}}k_{1,2n+2}(u)\text{ \ \ \ \ \ \ \ \ \ \ \ \ \ \ \ \
\ \ \ \ }(i<n+1), \\ 
k_{n+3,n+3}(u)-\frac{q^{n-1}+1}{\mathrm{e}^{2u}+q^{n-1}}\frac{\beta
_{n+3,n+3}-\beta _{i,i}}{\beta _{1,2n+2}}\mathrm{e}^{2u}k_{1,2n+2}(u)\text{
\ \ \ }(i>n+2).%
\end{array}%
\right.  \label{D(n+1)gen20}
\end{equation}

Substituting (\ref{D(n+1)gen20}) into the functional equations we get $%
k_{n+3,n+3}(u)$ and $k_{11}(u)$, and consequently all elements $%
k_{i,i}(u)\notin \mathbb{B}$ will be known after finding $2n$ parameters $%
\beta _{i,i}$.

The solution of this problem depends on whether $n$ is even or odd. Besides,
at this stage, all remaining parameters $\beta _{i,j}$, including those
associated with the central elements, are fixed in terms of $n+3$
parameters. In order to solve this problem, we turn our approach toward
treating separately two classes of $\mathcal{D}_{n+1}^{(2)}$ general $K$%
-matrices, according to the parity of $n$ manifested.

\subsubsection{The $K$-matrices for odd $n$}

Whether $n$ is odd we have the expressions for $k_{11}(u)$ and $%
k_{n+3,n+3}(u)$ given by%
\begin{eqnarray}
&&k_{11}(u)=  \notag \\
&&\left\{ \left( \frac{q^{n-1}+1}{q+1}\right) \frac{(q+1)(\mathrm{e}%
^{2u}+1)(q^{n}\beta _{-}^{2}-\beta _{+}^{2})+2(\mathrm{e}^{2u}-q^{n})(q^{n}%
\beta _{-}^{2}+\beta _{+}^{2})}{8\beta _{1,2n+2}^{2}q^{n-1/2}(\mathrm{e}%
^{2u}+1)}\right.  \notag \\
&&\left. +\frac{2q(q^{n-1}-1)+(q-1)(\mathrm{e}^{2u}+1)}{\beta
_{1,2n+2}(q^{n}-1)(\mathrm{e}^{2u}-1)}\right\} \left( \frac{q^{n-1}+1}{%
\mathrm{e}^{2u}+q^{n-1}}\right) k_{1,2n+2}(u),  \notag \\
&&  \label{D(n+1)gen21} \\
&&k_{n+3,n+3}(u)=  \notag \\
&&\left\{ \left( \frac{q^{n-1}+1}{q+1}\right) \frac{(q+1)(\mathrm{e}%
^{2u}+1)(q^{n}\beta _{-}^{2}-\beta _{+}^{2})-2(\mathrm{e}^{2u}-q^{n})(q^{n}%
\beta _{-}^{2}+\beta _{+}^{2})}{8\beta _{1,2n+2}^{2}q^{n-1/2}(\mathrm{e}%
^{2u}+1)}\right.  \notag \\
&&\left. +\frac{2q(q^{n-1}-1)+(q-1)(\mathrm{e}^{2u}+1)}{\beta
_{1,2n+2}(q^{n}-1)(\mathrm{e}^{2u}-1)}\right\} \left( \frac{q^{n-1}+1}{%
\mathrm{e}^{2u}+q^{n-1}}\right) \mathrm{e}^{2u}k_{1,2n+2}(u).  \notag \\
&&  \label{D(n+1)gen22}
\end{eqnarray}

The parameters $\beta _{i,i}$, $i\neq n+1,n+2$, are fixed by the following
recurrent relations:%
\begin{equation}
\beta _{i+1,i+1}=\left\{ 
\begin{array}{c}
\beta _{i,i}+(-q)^{i-1}\Theta _{odd}\text{ \ \ \ \ \ \ \ \ \ \ \ \ }(i<n),
\\ 
\beta _{i,i}+(-q)^{i-n-3}\Theta _{odd}\text{ \ \ \ }(i>n+2),%
\end{array}%
\right.  \label{D(n+1)gen23}
\end{equation}%
with%
\begin{equation}
\beta _{n+3,n+3}=\beta _{11}+2+\left( \frac{q^{n-1}+1}{q+1}\right) \frac{%
(q^{n}-1)(q^{n}\beta _{-}^{2}+\beta _{+}^{2})}{4q^{n-1/2}\beta _{1,2n+2}}
\label{D(n+1)gen24}
\end{equation}%
and%
\begin{equation}
\Theta _{odd}=-\frac{(q+1)^{2}}{q^{n}-1}-\frac{(q+1)(q^{n-1}+1)(q^{n}\beta
_{-}^{2}-\beta _{+}^{2})}{8q^{n-1/2}\beta _{1,2n+2}}.  \label{D(n+1)gen25}
\end{equation}

Finally, we can solve the last functional equations in order to find the
central elements. The solution is as follows%
\begin{eqnarray}
k_{n+1,n+1}(u) &=&-\left\{ \left( \frac{q^{n-1}+1}{q+1}\right) \frac{%
q^{n}\beta _{-}^{2}-\beta _{+}^{2}}{8q^{n-1/2}}(\mathrm{e}^{2u}+q^{n})\right.
\notag \\
&&\left. +\beta _{1,2n+2}\frac{(\mathrm{e}^{2u}+1)(\mathrm{e}^{2u}-q^{n})}{(%
\mathrm{e}^{2u}-1)(q^{n}-1)}\right\} \left( \frac{q^{n-1}+1}{\mathrm{e}%
^{2u}+q^{n-1}}\right) \frac{k_{1,2n+2}(u)}{\beta _{1,2n+2}^{2}},  \notag \\
k_{n+2,n+2}(u) &=&k_{n+1,n+1}(u),  \label{D(n+1)gen26} \\
k_{n+1,n+2}(u) &=&\frac{(q^{n-1}+1)^{2}[(q^{n}+1)(q^{n}\beta _{-}^{2}+\beta
_{+}^{2})\mathrm{e}^{u}-2q^{n}\beta _{-}\beta _{+}(\mathrm{e}^{2u}+1)]}{%
4q^{n-1/2}(q+1)(\mathrm{e}^{2u}+1)(\mathrm{e}^{2u}+q^{n-1})}  \notag \\
&&\times \frac{\mathrm{e}^{u}k_{1,2n+2}(u)}{\beta _{1,2n+2}^{2}},
\label{D(n+1)gen27} \\
k_{n+2,n+1}(u) &=&\frac{(q^{n-1}+1)^{2}[(q^{n}+1)(q^{n}\beta _{-}^{2}+\beta
_{+}^{2})\mathrm{e}^{u}+2q^{n}\beta _{-}\beta _{+}(\mathrm{e}^{2u}+1)]}{%
4q^{n-1/2}(q+1)(\mathrm{e}^{2u}+1)(\mathrm{e}^{2u}+q^{n-1})}  \notag \\
&&\times \frac{\mathrm{e}^{u}k_{1,2n+2}(u)}{\beta _{1,2n+2}^{2}}.
\label{D(n+1)gen28}
\end{eqnarray}%
Moreover, there are $n$ fixed parameters given by 
\begin{eqnarray}
\beta _{21} &=&\frac{1}{q^{2n-3}}\left( \frac{q^{n-1}+1}{q+1}\right) ^{2}%
\frac{\beta _{1,n}\beta _{1,n+3}\beta _{1,2n+1}}{\beta _{1,2n+2}^{2}}\text{
\ \ \ }(n\neq 1),  \label{D(n+1)gen29} \\
\beta _{1,n} &=&\frac{[(q^{n}-1)(q^{n-1}+1)(q^{n}\beta _{-}^{2}-\beta
_{+}^{2})+8q^{n-1/2}(q+1)\beta _{1,2n+2}](q+1)}{8\sqrt{q}%
(q^{n}-1)(q^{n-1}+1)\beta _{1,n+3}}  \notag \\
&&\text{ }(n\neq 1),  \label{D(n+1)gen30} \\
\beta _{1,j} &=&(-1)^{j-1}\frac{\beta _{1,n}\beta _{1,n+3}}{\beta _{1,2n+3-j}%
},\text{ \ \ \ }j=2,3,...,n-1.  \label{D(n+1)gen31}
\end{eqnarray}%
We thus get one general solution with $n+3$ free parameters, $\beta _{11}$, $%
\beta _{1,n+1}$, $\beta _{1,n+2}$, $...$, $\beta _{1,2n+2}$. By choosing%
\begin{equation}
k_{1,2n+2}(u)=\frac{1}{2}\beta _{1,2n+2}(\mathrm{e}^{2u}-1),
\label{D(n+1)gen32}
\end{equation}%
one can, for instance, fix the parameter $\beta _{11}$ by using the regular
condition (\ref{K-matrixReg}). Therefore, we have found one $(n+2)$%
-parameter general solution in the $\mathcal{D}_{n+1}^{(2)}$ case for odd $n$%
.

\subsubsection{The $K$-matrices for even $n$}

Whether $n$ is even we obtain the following expressions for $k_{11}(u)$ and $%
k_{n+3,n+3}(u)$, respectively, 
\begin{eqnarray}
k_{11}(u) &=&\frac{k_{1,2n+2}(u)}{2q^{n-1/2}\beta _{1,2n+2}^{2}(\mathrm{e}%
^{4u}-1)(\mathrm{e}^{2u}+q^{n-1})}  \notag \\
&&\times \{(q+1)(\mathrm{e}^{2u}-q^{n})(q^{n}\beta _{-}^{2}\mathrm{e}%
^{2u}-\beta _{+}^{2})  \notag \\
&&+2\sqrt{q}\beta _{1,n}\beta _{1,n+3}(\mathrm{e}^{2u}-1)[2(\mathrm{e}%
^{2u}-q^{n})-(q+1)(\mathrm{e}^{2u}+1)]\},  \notag \\
&&  \label{D(n+1)gen33} \\
k_{n+3,n+3}(u) &=&\frac{\mathrm{e}^{2u}k_{1,2n+2}(u)}{2q^{n-1/2}\beta
_{1,2n+2}^{2}(\mathrm{e}^{4u}-1)(\mathrm{e}^{2u}+q^{n-1})}  \notag \\
&&\times \{(q+1)(\mathrm{e}^{2u}-q^{n})(q^{n}\beta _{-}^{2}-\beta _{+}^{2}%
\mathrm{e}^{2u})  \notag \\
&&-2\sqrt{q}\beta _{1,n}\beta _{1,n+3}(\mathrm{e}^{2u}-1)[2(\mathrm{e}%
^{2u}-q^{n})-(q+1)(\mathrm{e}^{2u}+1)]\}.  \notag \\
&&  \label{D(n+1)gen34}
\end{eqnarray}%
The central elements are given by%
\begin{eqnarray}
k_{n+1,n+1}(u) &=&k_{n+2,n+2}(u)=2\frac{\mathrm{e}^{2u}}{\mathrm{e}^{2u}-1}%
\frac{q^{n-1}+1}{\mathrm{e}^{2u}+q^{n-1}}\frac{k_{1,2n+2}(u)}{\beta _{1,2n+2}%
},  \label{D(n+1)gen35} \\
k_{n+1,n+2}(u) &=&\frac{k_{1,2n+2}(u)}{4q^{n-1/2}\beta _{1,2n+2}^{2}(\mathrm{%
e}^{2u}+q^{n-1})(\mathrm{e}^{2u}+1)}\left( \frac{q^{n-1}+1}{q+1}\right) ^{2}
\notag \\
&&\times \left\{ (q^{n}\beta _{-}^{2}+\beta _{+}^{2})^{2}(q+1)(q^{n}+1)%
\mathrm{e}^{2u}\right.  \notag \\
&&-2\beta _{-}\beta _{+}(q+1)q^{n}\mathrm{e}^{u}(\mathrm{e}^{2u}+1)  \notag
\\
&&\left. -4\sqrt{q}\beta _{1,n}\beta _{1,n+3}(\mathrm{e}^{2u}-q^{n})(\mathrm{%
e}^{2u}-1)\right\} ,  \label{D(n+1)gen36}
\end{eqnarray}%
\begin{eqnarray}
k_{n+2,n+1}(u) &=&\frac{k_{1,2n+2}(u)}{4q^{n-1/2}\beta _{1,2n+2}^{2}(\mathrm{%
e}^{2u}+q^{n-1})(\mathrm{e}^{2u}+1)}\left( \frac{q^{n-1}+1}{q+1}\right) ^{2}
\notag \\
&&\times \left\{ (q^{n}\beta _{-}^{2}+\beta _{+}^{2})^{2}(q+1)(q^{n}+1)%
\mathrm{e}^{2u}\right.  \notag \\
&&+2\beta _{-}\beta _{+}(q+1)q^{n}\mathrm{e}^{u}(\mathrm{e}^{2u}+1)  \notag
\\
&&\left. -4\sqrt{q}\beta _{1,n}\beta _{1,n+3}(\mathrm{e}^{2u}-q^{n})(\mathrm{%
e}^{2u}-1)\right\} .  \label{D(n+1)gen37}
\end{eqnarray}%
The parameters $\beta _{i,i}$ are fixed by the following recurrent relations:%
\begin{equation}
\beta _{i+1,i+1}=\left\{ 
\begin{array}{c}
\beta _{i,i}+(-q)^{i-1}\Theta _{even}\text{ \ \ \ \ \ \ \ \ \ \ \ \ }(i<n),
\\ 
\beta _{i,i}-(-q)^{i-n-3}\Theta _{even}\text{ \ \ \ }(i>n+2),%
\end{array}%
\right.  \label{D(n+1)gen38}
\end{equation}%
with%
\begin{equation}
\beta _{n+3,n+3}=\beta _{11}+2-\frac{2(q^{n}-1)(q+1)(q^{n}\beta
_{-}^{2}+\beta _{+}^{2})+8\beta _{1,n}\beta _{1,n+3}q^{3/2}(q^{n-1}+1)}{%
(q+1)(q^{n}-1)(q^{n}\beta _{-}^{2}-\beta _{+}^{2})}  \label{D(n+1)gen39}
\end{equation}%
and%
\begin{equation}
\Theta _{even}=-8\frac{\sqrt{q}(q+1)}{q^{n}-1}\frac{\beta _{1,n}\beta
_{1,n+3}}{q^{n}\beta _{-}^{2}-\beta _{+}^{2}}.  \label{D(n+1)gen40}
\end{equation}%
Now we have $n$ fixed parameters given by%
\begin{eqnarray}
\beta _{21} &=&-q^{3-2n}\frac{q^{n-1}+1}{q+1}\beta _{1,2n+1}\frac{\beta
_{1,n}\beta _{1,n+3}}{\beta _{1,2n+2}^{2}},  \label{D(n+1)gen41} \\
\beta _{1,2n+2} &=&-\frac{1}{8}\frac{q^{n}-1}{q^{n-1/2}}\frac{q^{n-1}+1}{q+1}%
(q^{n}\beta _{-}^{2}-\beta _{+}^{2}),  \label{D(n+1)gen42} \\
\beta _{1,j} &=&(-1)^{j-1}\frac{\beta _{1,n}\beta _{1,n+3}}{\beta _{1,2n+3-j}%
},\text{ \ \ \ }j=2,3,...,n-1,  \label{D(n+1)gen43}
\end{eqnarray}%
and $n+3$ free parameters, $\beta _{11}$, $\beta _{1,n}$, $...$, $\beta
_{1,2n+1}$. Again, we can fix $\beta _{11}$ by making use of the regular
condition (\ref{K-matrixReg}). Hence, we have also obtained one $(n+2)$%
-parameter general solution in the $\mathcal{D}_{n+1}^{(2)}$ case for even $%
n $.

\section{Reduced Solutions}

In this section we move on to discuss $K$-matrices generated by employing a
reduction procedure described below for each class of our general solutions.
Here we do not take account of the $\mathcal{A}_{n-1}^{(1)}$ models since
they exhibit an unique structure and have very different properties as well.

\subsection{The $\mathcal{B}_{n}^{(1)}$ and $\mathcal{A}_{2n}^{(2)}$ Models}

We concentrate on the possible reduced $K$-matrices of types $\mathcal{B}%
_{n}^{(1)}$ and $\mathcal{A}_{2n}^{(2)}$ generated by considering all
parameters $\beta _{i,j}=0$ $(j\neq i,i^{\prime })$. Thereby, the only
non-null matrix elements are given by $k_{i,i}(u)$ on the main diagonal and $%
k_{i,i^{\prime }}(u)$ on the secondary diagonal of the $K$-matrix.

Taking the limit $\beta _{i,j}\rightarrow 0$ $(j\neq i,i^{\prime })$ into
the $\mathcal{B}_{n}^{(1)}$ general solutions $($for $n\geq 1)$ we find one
reduced solution whose normalized matrix elements are given by%
\begin{align}
& k_{11}(u)=k_{22}(u)=...=k_{n,n}(u)=1,  \notag \\
& k_{n+1,n+1}(u)=\frac{\mathrm{e}^{2u}-q}{1-q},  \notag \\
& k_{n+2,n+2}(u)=k_{n+3,n+3}(u)=...=k_{2n+1,2n+1}(u)=\mathrm{e}^{2u}
\label{B(n)red1}
\end{align}%
and%
\begin{equation}
k_{i,i^{\prime }}(u)=\left\{ 
\begin{array}{c}
\frac{1}{2}\beta _{i,i^{\prime }}(\mathrm{e}^{2u}-1),\text{ \ \ \ \ \ \ \ \
\ }i<n+1, \\ 
\frac{2q}{\beta _{i^{\prime },i}(q-1)^{2}}(\mathrm{e}^{2u}-1),\text{ \ \ \ }%
i>n+1.%
\end{array}%
\right.  \label{B(n)red2}
\end{equation}%
Comparing the $\mathcal{B}_{n}^{(1)}$ general $K$-matrices with the above
solution (\ref{B(n)red1}), (\ref{B(n)red2}), we observe that the number of
free parameters reduces to $n$ in this limit procedure.

Let us consider the $\mathcal{A}_{2n}^{(2)}$ $K$-matrices $(n\geq 1)$. We
have the following reduced solution given by%
\begin{eqnarray}
&&k_{11}(u)=k_{22}(u)=...=k_{n,n}(u)=1+\beta _{11}(\mathrm{e}^{u}-1),  \notag
\\
&&k_{n+1,n+1}(u)=\beta _{11}\mathrm{e}^{u}-\frac{\mathrm{e}^{2u}-q}{1-q}%
(\beta _{11}-1),  \notag \\
&&k_{n+2,n+2}(u)=k_{n+3,n+3}(u)=...=k_{2n+1,2n+1}(u)=\mathrm{e}^{2u}[1+\beta
_{11}(\mathrm{e}^{-u}-1)]  \notag \\
&&  \label{A(2n)red1}
\end{eqnarray}%
and%
\begin{equation}
k_{i,i^{\prime }}(u)=\left\{ 
\begin{array}{c}
\frac{1}{2}\beta _{i,i^{\prime }}(\mathrm{e}^{2u}-1),\text{\ \ \ \ \ \ \ \ \
\ \ \ \ \ \ }i<n+1, \\ 
\left( \frac{\beta _{11}-1}{q-1}\right) ^{2}\frac{2q}{\beta _{i^{\prime },i}}%
(\mathrm{e}^{2u}-1),\text{ \ \ \ }i>n+1.%
\end{array}%
\right.  \label{A(2n)red2}
\end{equation}%
Here we note that the number of free parameters remains the same as obtained
for the $\mathcal{A}_{2n}^{(2)}$ general solutions. We argue that this
feature is responsible for the fact that both solutions can be regarded as
general $K$-matrices. Therefore, there exists another type of $\mathcal{A}%
_{2n}^{(2)}$ general solution with $n+1$ free parameters, revealed by taking
the limit $\beta _{i,j}\rightarrow 0$ $(j\neq i,i^{\prime })$ into the $%
\mathcal{A}_{2n}^{(2)}$ general $K$-matrices $(n\geq 1)$ previously
presented in Section 3.2. In particular, for the Izergin-Korepin model, it
has the form%
\begin{equation}
K^{-}=\left( 
\begin{array}{ccc}
1+\beta _{11}(\mathrm{e}^{u}-1) & 0 & \frac{1}{2}\beta _{13}(\mathrm{e}%
^{2u}-1) \\ 
0 & \beta _{11}\mathrm{e}^{u}-\frac{\mathrm{e}^{2u}-q}{1-q}(\beta _{11}-1) & 
0 \\ 
\frac{2q}{\beta _{13}}\frac{(\beta _{11}-1)^{2}}{(q-1)^{2}}(\mathrm{e}%
^{2u}-1) & 0 & \mathrm{e}^{2u}[1+\beta _{11}(\mathrm{e}^{-u}-1)]%
\end{array}%
\right) .  \label{A(2n)red3}
\end{equation}%
We remark that this $2$-parameter $K$-matrix for the $\mathcal{A}_{2}^{(2)}$
model has been derived by Kim in \cite{Kim1}.

The $\mathcal{B}_{n}^{(1)}$ and $\mathcal{A}_{2n}^{(2)}$ general $K$%
-matrices bring $n+1$ free parameters, and there are in fact several reduced
solutions which can be found by setting one or more free parameters equal to
zero. For instance, by analyzing the functional equations for $n>1$ we
observe that the vanishing of any element $k_{i,i^{\prime }}(u)$, by setting 
$\beta _{i,i^{\prime }}=0$, implies that the only non-null entries are those
on the main diagonal of $K^{-}(u)$, i.e.%
\begin{equation}
\beta _{i,i^{\prime }}=0\Rightarrow \{k_{i,l}(u)=0\text{ }(l\neq i)\text{
and }k_{l,i^{\prime }}(u)=0\text{ }(l\neq i^{\prime })\}.  \label{A(2n)red4}
\end{equation}%
Similar consideration holds for the transpose of\ the matrix. In addition,
the parameters $\beta _{i,j}$ and $\beta _{j,i}$ $(i\neq j)$ are linked by
the relation (\ref{B(n),A(2n)gen5}) and the constraint equations (\ref%
{B(n),A(2n)gen6}) and (\ref{B(n),A(2n)gen7}). As a consequence, we find that%
\begin{equation}
\text{if }\beta _{i,j}=0,\text{ then }\left\{ 
\begin{array}{c}
k_{i,j}(u)=0,\text{ \ \ \ for }i\neq j, \\ 
k_{i,j}(u)=1,\text{ \ \ \ for }i=j,%
\end{array}%
\right.  \label{A(2n)red5}
\end{equation}%
where we have considered the normalization condition (\ref{K-matrixReg}).

Hence, by applying this limit procedure and by choosing suitably the free
parameters it is possible to obtain other reduced $K$-matrices from the
general solution.

\subsection{The $\mathcal{C}_{n}^{(1)},$ $\mathcal{D}_{n}^{(1)},$ and $%
\mathcal{A}_{2n-1}^{(2)}$ Models}

Now we start off to elucidate the possible reduced $K$-matrices of types $%
\mathcal{C}_{n}^{(1)},$ $\mathcal{D}_{n}^{(1)},$ and $\mathcal{A}%
_{2n-1}^{(2)}$. In Section 3.2, we have firstly considered general solutions
which contain only non-null matrix elements. In particular, the $\mathcal{C}%
_{n}^{(1)}$ and $\mathcal{D}_{n}^{(1)}$ $K$-matrices depend on the parity of 
$n$: one general solution with $\mathcal{G}^{(+)}(u)$ for odd $n$ and one
general solution with $\mathcal{G}^{(-)}(u)$ for even $n$ in the $\mathcal{D}%
_{n}^{(1)}$ case, and the opposite occurring for the $\mathcal{C}_{n}^{(1)}$
models. However, we have found that there exist $\mathcal{D}_{n}^{(1)}$ $K$%
-matrices with $\mathcal{G}^{(+)}(u)$ for even $n$ and $\mathcal{G}^{(-)}(u)$
for odd $n$, as well as $\mathcal{C}_{n}^{(1)}$ $K$-matrices with $\mathcal{G%
}^{(+)}(u)$ for odd $n$ and $\mathcal{G}^{(-)}(u)$ for even $n$, provided
that some matrix elements are set equal to zero. Let us recall (\ref%
{B(n),A(2n)gen1})-(\ref{B(n),A(2n)gen3}) to see that the vanishing of the
element $k_{n+1,n}(u)$ on the secondary diagonal of $K^{-}(u)$ implies that $%
k_{i,n}(u)=0$ $(i\neq n)$ and $k_{n+1,j}(u)=0$ $(j\neq n+1)$.

Therefore, we can consider the case $k_{n+1,n}(u)=0$ which will imply that $%
k_{n,n+1}(u)=k_{i,n+1}(u)=k_{n,j}(u)=0$ for these models. It means that we
are dealing with $K$-matrices that contain $2(4n-3)$ null entries, and
particularly $k_{1,n}(u)=0$. The non-diagonal matrix elements are directly
obtained from (\ref{B(n),A(2n)gen11})-(\ref{B(n),A(2n)gen17}) by taking the
limits $\beta _{1,n}\rightarrow 0$ and $\beta _{1,n+1}\rightarrow 0$.

On the secondary diagonal we have 
\begin{eqnarray}
k_{i,i^{\prime }}(u) &=&\varepsilon _{i}\frac{\beta _{1,i^{\prime }}^{2}}{%
\beta _{1,2n}}\frac{q^{\bar{\imath}-\bar{1}}}{\xi }\frac{(q\pm \sqrt{\xi })}{%
(q+1)^{2}}(q\mathrm{e}^{u}\pm \sqrt{\xi })\mathcal{G}^{(\pm )}(u)  \notag \\
&&\text{ }(i\neq 1,n,n+1,2n),  \label{C(n),D(n),A(2n-1)red1} \\
k_{n,n+1}(u) &=&k_{n+1,n}(u)=0,  \label{C(n),D(n),A(2n-1)red2} \\
k_{2n,1}(u) &=&\varepsilon _{2n}\beta _{1,2n}\frac{\beta _{21}^{2}}{\beta
_{1,2n-1}^{2}}q^{(\bar{2}n-\bar{1})-2}\left( \frac{q\mathrm{e}^{u}\pm \sqrt{%
\xi }}{q\pm \sqrt{\xi }}\right) \mathcal{G}^{(\pm )}(u).
\label{C(n),D(n),A(2n-1)red3}
\end{eqnarray}%
The boundary rows and columns are%
\begin{eqnarray}
k_{1,j}(u) &=&\beta _{1,j}\mathcal{G}^{(\pm )}(u),\text{\ \ }%
k_{i,1}(u)=\varepsilon _{i}\beta _{21}\frac{\beta _{1,i^{\prime }}}{\beta
_{1,2n-1}}q^{\bar{\imath}-\bar{2}}\mathcal{G}^{(\pm )}(u),
\label{C(n),D(n),A(2n-1)red4} \\
k_{1,n}(u) &=&k_{1,n+1}(u)=k_{n,1}(u)=k_{n+1,1}(u)=0,
\label{C(n),D(n),A(2n-1)red5} \\
k_{i,2n}(u) &=&\pm \varepsilon _{i}\beta _{1,i^{\prime }}\frac{q^{\bar{\imath%
}-\bar{1}}}{\sqrt{\xi }}\mathrm{e}^{u}\mathcal{G}^{(\pm )}(u),\text{\ }%
k_{2n,j}(u)=\pm \beta _{21}\frac{\beta _{1,j}}{\beta _{1,2n-1}}\frac{q^{\bar{%
2}n-\bar{2}}}{\sqrt{\xi }}\mathrm{e}^{u}\mathcal{G}^{(\pm )}(u)  \notag \\
&&  \label{C(n),D(n),A(2n-1)red6} \\
k_{n,2n}(u) &=&k_{n+1,2n}(u)=k_{2n,n}(u)=k_{2n,n+1}(u)=0,
\label{C(n),D(n),A(2n-1)red7}
\end{eqnarray}%
and the remaining non-diagonal matrix elements are given by%
\begin{eqnarray}
k_{i,j}(u) &=&\pm \varepsilon _{i}\beta _{1,j}\frac{\beta _{1,i^{\prime }}}{%
\beta _{1,2n}}\frac{q^{\bar{\imath}-\bar{1}}}{\sqrt{\xi }}\left( \frac{q\pm 
\sqrt{\xi }}{q+1}\right) \mathcal{G}^{(\pm )}(u)\text{ \ \ \ }(j<i^{\prime
}),  \label{C(n),D(n),A(2n-1)red8} \\
k_{i,j}(u) &=&\varepsilon _{i}\beta _{1,j}\frac{\beta _{1,i^{\prime }}}{%
\beta _{1,2n}}\frac{q^{\bar{\imath}-\bar{1}}}{\xi }\left( \frac{q\pm \sqrt{%
\xi }}{q+1}\right) \mathrm{e}^{u}\mathcal{G}^{(\pm )}(u)\text{ \ \ \ }%
(j>i^{\prime }),  \label{C(n),D(n),A(2n-1)red9} \\
k_{i,n}(u) &=&k_{i,n+1}(u)=k_{n,j}(u)=k_{n+1,j}(u)=0.
\label{C(n),D(n),A(2n-1)red10}
\end{eqnarray}%
In order to find the corresponding diagonal elements we will repeat the same
steps followed in Section 3.2, but now using the equations $E[n,2n(n-1)+2]$
and $E[n+1,2n^{2}+2]$ to get $k_{n,n}(u)$ and $k_{n+1,n+1}(u)$,
respectively. Next we identify $n-2$ non-diagonal parameters 
\begin{eqnarray}
\beta _{21} &=&-\frac{q}{\xi }\left( \frac{q\pm \sqrt{\xi }}{q+1}\right) ^{2}%
\frac{\beta _{12}\beta _{1,2n-1}^{2}}{\beta _{1,2n}^{2}},
\label{C(n),D(n),A(2n-1)red11a} \\
\beta _{1,j} &=&(-1)^{n-1+j}\frac{\beta _{1,n-1}\beta _{1,n+2}}{\beta
_{1,2n+1-j}},\text{ \ \ \ }j=2,3,...,n-2  \label{C(n),D(n),A(2n-1)red11}
\end{eqnarray}%
for $n>3$, which are also related to $k_{11}(u)$ in a very simple way:%
\begin{eqnarray}
k_{i,i}(u) &=&k_{11}(u)+(\beta _{i,i}-\beta _{11})\mathcal{G}^{(\pm )}(u)%
\text{ \ \ \ }(1<i\leq n-1),  \label{C(n),D(n),A(2n-1)red12} \\
k_{n,n}(u) &=&k_{n+1,n+1}(u)=k_{n-1,n-1}(u)+(\beta _{n,n}-\beta _{n-1,n-1})%
\mathrm{e}^{u}\mathcal{G}^{(\pm )}(u)  \notag \\
&&+\mathcal{F}_{n-1}^{(\pm )}(u),  \label{C(n),D(n),A(2n-1)red13} \\
k_{i,i}(u) &=&k_{n,n}(u)+(\beta _{i,i}-\beta _{n,n})\mathrm{e}^{u}\mathcal{G}%
^{(\pm )}(u)-\varepsilon _{i}\theta _{i}^{2}q^{2}\mathcal{F}_{n-1}^{(\pm
)}(u)  \notag \\
&&\text{ }(n+2\leq i\leq 2n),\text{ }  \label{C(n),D(n),A(2n-1)red14}
\end{eqnarray}%
where%
\begin{equation}
\mathcal{F}_{n-1}^{(\pm )}(u)=-\Delta _{n-1}^{(\pm )}(-q)^{n-2}\frac{(%
\mathrm{e}^{u}-1)}{(q+1)^{2}}\mathcal{G}^{(\pm )}(u)
\label{C(n),D(n),A(2n-1)red15}
\end{equation}%
with%
\begin{equation}
\Delta _{n-1}^{(\pm )}=\pm (-1)^{n-1}\frac{\beta _{1,n-1}\beta _{1,n+2}}{%
\beta _{1,2n}}\left( \frac{q\pm \sqrt{\xi }}{\sqrt{\xi }}\right) .
\label{C(n),D(n),A(2n-1)red16}
\end{equation}%
Here we point out that $\Delta _{n-1}^{(\pm )}$ can be understood as a limit
of $\Delta _{n}^{(\pm )}$ given by (\ref{C(n),D(n),A(2n-1)gen21}), i.e. $%
\beta _{1,n}\rightarrow -\beta _{1,n-1};\beta _{1,n+1}\rightarrow \beta
_{1,n+2}$.

Again, the equation $E[2n+1,4n]$ gives another relation between $%
k_{2n,2n}(u) $ and $k_{11}(u)$: 
\begin{equation}
k_{2n,2n}(u)=\mathrm{e}^{2u}k_{11}(u)+(\beta _{2n,2n}-\beta _{11}-2)\mathrm{e%
}^{u}\left( \frac{q\mathrm{e}^{u}\pm \sqrt{\xi }}{q\pm \sqrt{\xi }}\right) 
\mathcal{G}^{(\pm )}(u),  \label{C(n),D(n),A(2n-1)red17}
\end{equation}%
which allows to write $k_{11}(u)$ as follows%
\begin{eqnarray}
k_{11}(u) &=&\frac{(1-\varepsilon _{2n}\theta _{2n}^{2}q^{2})\mathcal{F}%
_{n-1}^{(\pm )}(u)}{\mathrm{e}^{2u}-1}+\left\{ \beta _{n-1,n-1}-\beta
_{11}+(\beta _{2n,2n}-\beta _{n-1,n-1})\mathrm{e}^{u}\right.  \notag \\
&&\left. -(\beta _{2n,2n}-\beta _{11}-2)\mathrm{e}^{u}\left( \frac{q\mathrm{e%
}^{u}\pm \sqrt{\xi }}{q\pm \sqrt{\xi }}\right) \right\} \frac{\mathcal{G}%
^{(\pm )}(u)}{\mathrm{e}^{2u}-1}.  \label{C(n),D(n),A(2n-1)red18}
\end{eqnarray}

Substituting these expressions back to the functional equations we get
constraint equations which will enable us to fix some of the $3n-1$
remaining parameters. We can make use of the equations $E[2,2n+j]$ $(j\neq
n,n+1)$ to find $\beta _{j,j}$, $j=3,4,...,2n-2,$ in terms of $\beta _{22}$
given by the equation $E[2,2n+1]$. The parameters $\beta _{n,n}$ and $\beta
_{n+1,n+1}$ are fixed in terms of $\beta _{22}$ by requiring the equations $%
E[n,2n^{2}-n+2]$ and $E[n+1,2n^{2}+n+2]$, respectively. After carrying out
this calculation, we used the equation $E[2,2n-1]$ to obtain $\beta
_{2n-1,2n-1}$ and $\beta _{2n,2n}$. These parameters can be written in terms
of $\beta _{11}$, $\beta _{1,n-1}$, $\beta _{1,n+2}$ and $\beta _{1,2n}$ as
follows:%
\begin{eqnarray}
\beta _{i,i} &=&\beta _{11}+\Delta _{n-1}^{(\pm )}\sum_{j=0}^{i-2}(-q)^{j}%
\text{ \ \ \ }(1<i\leq n-1),  \label{C(n),D(n),A(2n-1)red19} \\
\beta _{n,n} &=&\beta _{n+1,n+1}=\beta _{11}+\Delta _{n-1}^{(\pm )}\left[ 
\frac{1-(-q)^{n-2}}{q+1}\right] -q^{n-2}\Sigma _{n-1}^{(\pm )},
\label{C(n),D(n),A(2n-1)red20} \\
\beta _{n+2,n+2} &=&\beta _{11}+\Delta _{n-1}^{(\pm )}\left[ \frac{%
1-(-q)^{n-2}}{q+1}\right] -(q^{n-2}-\varepsilon _{n+2}\theta
_{n+2}^{2}q^{n})\Sigma _{n-1}^{(\pm )},  \notag \\
&&  \label{C(n),D(n),A(2n-1)red21} \\
\beta _{i,i} &=&\beta _{n+2,n+2}+\Delta _{n-1}^{(\pm )}\left[ \pm \frac{%
\varepsilon _{n+2}\theta _{n+2}^{2}}{\sqrt{\xi }}\sum_{j=n-1}^{i-3}(-q)^{j}%
\right]  \notag \\
\text{ } &&\text{\ }(n+3\leq i\leq 2n-1),  \label{C(n),D(n),A(2n-1)red22}
\end{eqnarray}%
and%
\begin{equation}
\beta _{2n,2n}=\beta _{11}+2+\Delta _{n-1}^{(\pm )}\frac{(q\pm \sqrt{\xi })}{%
\xi }\frac{(\xi -\varepsilon _{2n}q^{\bar{2}n-\bar{1}})}{(q+1)^{2}},
\label{C(n),D(n),A(2n-1)red23}
\end{equation}%
where%
\begin{equation}
\Sigma _{n-1}^{(\pm )}=\frac{\beta _{1,n-1}\beta _{1,n+2}}{\beta _{1,2n}}%
\frac{1}{\xi }\left( \frac{q\pm \sqrt{\xi }}{q+1}\right) ^{2}.
\label{C(n),D(n),A(2n-1)red24}
\end{equation}%
Next we can, for instance, make use of the equation $E[2,4n]$ to fix $\beta
_{1,n-1}$:%
\begin{eqnarray}
&&\beta _{1,n-1}=\pm (-1)^{n-1}  \notag \\
&&\times \frac{2\xi \sqrt{\xi }(q+1)^{2}}{(1\mp \sqrt{\xi })[q^{n-1}\pm
(-1)^{n}\sqrt{\xi }][\varepsilon _{n+2}\theta _{n+2}^{2}q^{n}\pm (-1)^{n}%
\sqrt{\xi }](q\pm \sqrt{\xi })}\frac{\beta _{1,2n}}{\beta _{1,n+2}}.  \notag
\\
&&  \label{C(n),D(n),A(2n-1)red25}
\end{eqnarray}%
At this point we must treat these solutions separately in order to take
account of the existence of the amplitude $k_{1,n-1}(u)$ for each model:

$\bullet $ For $\mathcal{A}_{2n-1}^{(2)}$ models we have $\xi =-q^{2n}$ and $%
\theta _{k}=\varepsilon _{k}=1$, $\forall k$, and there is no restriction in
(\ref{C(n),D(n),A(2n-1)red25}). It follows that the solution with $\mathcal{G%
}^{(+)}(u)$ (upper sign) is related to the solution with $\mathcal{G}%
^{(-)}(u)$ (lower sign) by complex conjugation.

$\bullet $ For $\mathcal{D}_{n}^{(1)}$ models we have $\xi =q^{2n-2}$ and $%
\theta _{k}=\varepsilon _{k}=1$, $\forall k$. It means that the factors $%
[q^{n-1}\pm (-1)^{n}\sqrt{\xi }]$ are different from zero for the solution
with $\mathcal{G}^{(+)}(u)$ only if $n$ is even and for the solution with $%
\mathcal{G}^{(-)}(u)$ if $n$ is odd.

$\bullet $ For $\mathcal{C}_{n}^{(1)}$ models we have $\xi =q^{2n+2}$, $%
\varepsilon _{n+2}=-1$ and $\theta _{n+2}^{2}=q^{2}$. In this case there is
also no restriction because both factors $[q^{n-1}\pm (-1)^{n}\sqrt{\xi }]$
and $[-q^{n+2}\pm (-1)^{n}\sqrt{\xi }]$ are different from zero. It means
that we have two independent solutions, one with $\mathcal{G}^{(+)}(u)$ and
another with $\mathcal{G}^{(-)}(u)$, for all $n>3$.

On comparing these results with those presented in Section 3.2, one could
conclude that we have simply made a reduction of the general solution by an
appropriate choice of the free parameters. Nevertheless, new solutions are
appearing for the $\mathcal{C}_{n}^{(1)}$ and $\mathcal{D}_{n}^{(1)}$ models.

Substituting all fixed parameters into (\ref{C(n),D(n),A(2n-1)red18}) we
find the following expressions for the amplitude $k_{11}(u)$:%
\begin{eqnarray}
k_{11}(u) &=&\frac{2\mathrm{e}^{u}\mathcal{G}^{(\pm )}(u)}{\mathrm{e}^{2u}-1}
\notag \\
&&-\frac{2\mathcal{G}^{(\pm )}(u)}{\mathrm{e}^{u}+1}\left\{ \frac{\xi
\lbrack 1+q-(-q)^{n}+(-q)^{n-1}]+q\mathrm{e}^{u}(\xi -q^{2n-2})}{(1\mp \sqrt{%
\xi })[q^{n-1}\pm (-1)^{n}\sqrt{\xi }][q^{n}\pm (-1)^{n}\sqrt{\xi }]}\right\}
\notag \\
&&  \label{C(n),D(n),A(2n-1)red26}
\end{eqnarray}%
for $\mathcal{A}_{2n-1}^{(2)}$ and $\mathcal{D}_{n}^{(1)}$ models, and%
\begin{eqnarray}
k_{11}(u) &=&\frac{2\mathrm{e}^{u}\mathcal{G}^{(\pm )}(u)}{\mathrm{e}^{2u}-1}
\notag \\
&&+\frac{2\mathcal{G}^{(\pm )}(u)}{\mathrm{e}^{u}+1}\left\{ \frac{\xi
\lbrack 1+q+(-q)^{n-1}+(-q)^{n+2}]+q\mathrm{e}^{u}(\xi +q^{2n})}{(1\mp \sqrt{%
\xi })[q^{n-1}\pm (-1)^{n}\sqrt{\xi }][q^{n+2}\mp (-1)^{n}\sqrt{\xi }]}%
\right\}  \notag \\
&&  \label{C(n),D(n),A(2n-1)red27}
\end{eqnarray}
for $\mathcal{C}_{n}^{(1)}$ models.

In the $\mathcal{D}_{n}^{(1)}$ case we still have the simplified expression
for $k_{11}(u)$, but now the parity of the solutions with $\mathcal{G}^{(\pm
)}(u)$ is exchanged, 
\begin{equation}
k_{11}(u)=\frac{\mathcal{G}^{(+)}(u)}{\mathrm{e}^{u}-1}\text{ \ \ \ }(\text{%
even }n),\text{ \ \ \ \ \ \ \ }k_{11}(u)=\frac{\mathcal{G}^{(-)}(u)}{\mathrm{%
e}^{u}-1}\text{ \ \ \ }(\text{odd }n).  \label{C(n),D(n),A(2n-1)red28}
\end{equation}%
For $\mathcal{C}_{n}^{(1)}$ models we lost the relations given by (\ref%
{C(n),D(n),A(2n-1)gen32}) between the diagonal entries, but (\ref%
{C(n),D(n),A(2n-1)red27}) defines a new solution with $\mathcal{G}^{(+)}(u)$
when $n$ is odd and another one with $\mathcal{G}^{(-)}(u)$ when $n$ is even.

Now we sum up our results: Firstly, we had from (\ref{C(n),D(n),A(2n-1)red1}%
) to (\ref{C(n),D(n),A(2n-1)red10}) all non-diagonal matrix elements after
substituting $n-1$ fixed non-diagonal parameters $\beta _{21}$ and $\beta
_{1,j}$ $(j=2,...,n-1)$ given by (\ref{C(n),D(n),A(2n-1)red11a}), (\ref%
{C(n),D(n),A(2n-1)red11}) and (\ref{C(n),D(n),A(2n-1)red25}). Secondly, the
diagonal matrix elements were obtained by using (\ref{C(n),D(n),A(2n-1)red12}%
), (\ref{C(n),D(n),A(2n-1)red13}) and (\ref{C(n),D(n),A(2n-1)red14}) with $%
k_{11}(u)$ given by (\ref{C(n),D(n),A(2n-1)red26}) for $\mathcal{A}%
_{2n-1}^{(2)}$ and $\mathcal{D}_{n}^{(1)}$ models and by (\ref%
{C(n),D(n),A(2n-1)red27}) for $\mathcal{C}_{n}^{(1)}$ models, and by
substituting the diagonal parameters given by (\ref{C(n),D(n),A(2n-1)red19}%
)-(\ref{C(n),D(n),A(2n-1)red23}).

These calculations lead to two reduced and new solutions with $n$
parameters, $\beta _{1,n+2}$, $\beta _{1,n+3}$, $...$, $\beta _{1,2n}$ and $%
\beta _{11}$, for these models. Once again, the number of free parameters is 
$n-1$ since we have to require the regular condition (\ref{K-matrixReg}),
which will fix the parameter $\beta _{11}$.

The general solutions which we have previously found in Section 3.2 have $n$
free parameters. Thus, solutions with $n-1$ free parameters can be
understood as reductions generated by employing a complicated limit
procedure we described above. This fact concerns the results for $\mathcal{A}%
_{2n-1}^{(2)}$ models, whereas for $\mathcal{D}_{n}^{(1)}$ models our
general solutions with $\mathcal{G}^{(+)}(u)$ have $n$ free parameters, but
are defined only for $n=3,5,7,...$, while our general solutions with $%
\mathcal{G}^{(-)}(u)$ are defined only for $n=4,6,8,...$. We emphasize that
our limit procedure has revealed new $(n-1)$-parameter solutions with $%
\mathcal{G}^{(-)}(u)$ for $n=5,7,9,...$, as well as new solutions with $%
\mathcal{G}^{(+)}(u)$ for $n=4,6,8,...$, in the $\mathcal{D}_{n}^{(1)}$
case. Similar considerations held for $\mathcal{C}_{n}^{(1)}$ models after
exchanging the parity of $n$. We remark that the cases $\mathcal{C}%
_{3}^{(1)} $ and $\mathcal{D}_{3}^{(1)}$ are special, each one featuring a $%
3 $-parameter solution, and will be treated in Section 6.

We should continue applying our reduction procedure in order to verify if
other solutions can be discovered. The next step is to consider $%
k_{1,n-1}(u)=0$ together with $k_{1,n}(u)=0$. On following this approach, we
find solutions with $16n-20$ null entries and with $n-2$ free parameters for 
$n>4$. However, all these solutions turn out to be reductions of those with $%
n$ and $n-1$ free parameters. Hence, there is no new solution. In
particular, for $n=4$ the solutions have $n-1$ free parameters. We observe
that the rational limit of some reduced solutions has been presented in \cite%
{AADF}.

After exhausting all the possible reductions, we attained to the last
reduction which was obtained after applying $n-1$ reduction steps. For $%
\mathcal{A}_{2n-1}^{(2)}$ and $\mathcal{D}_{n}^{(1)}$ the final reduction is
the non-diagonal $K$-matrix with the following non-null entries%
\begin{eqnarray}
&&k_{11}(u)=1,\text{\ \ \ \ \ \ \ \ }k_{2n,2n}(u)=\mathrm{e}^{2u},  \notag \\
&&k_{22}(u)=k_{33}(u)=...=k_{2n-1,2n-1}(u)=\frac{q^{2n-2}-\mathrm{e}^{2u}}{%
q^{2n-2}-1},  \notag \\
&&k_{1,2n}(u)=\frac{1}{2}\beta _{1,2n}(\mathrm{e}^{2u}-1),  \notag \\
&&k_{2n,1}(u)=\frac{2}{\beta _{1,2n}}\frac{q^{2n-2}}{(q^{2n-2}-1)^{2}}(%
\mathrm{e}^{2u}-1),  \label{C(n),D(n),A(2n-1)red29}
\end{eqnarray}%
while for $\mathcal{C}_{n}^{(1)}$ the corresponding reduction is%
\begin{align}
& k_{11}(u)=1,\text{ \ \ \ \ \ \ \ }k_{2n,2n}(u)=\mathrm{e}^{2u},  \notag \\
& k_{22}(u)=k_{33}(u)=...=k_{2n-1,2n-1}(u)=\frac{q^{2n}+\mathrm{e}^{2u}}{%
q^{2n}+1},  \notag \\
& k_{1,2n}(u)=\frac{1}{2}\beta _{1,2n}(\mathrm{e}^{2u}-1),  \notag \\
& k_{2n,1}(u)=-\frac{2}{\beta _{1,2n}}\frac{q^{2n}}{(q^{2n}+1)^{2}}(\mathrm{e%
}^{2u}-1).  \label{C(n),D(n),A(2n-1)red30}
\end{align}%
Here we note that these final reductions do not depend on both $\mathcal{G}%
^{(\pm )}(u)$ and $\xi $. Therefore, they can be regarded as new solutions
indeed.

We conclude this analysis by listing the following new $K$-matrices gotten
through our limit approach for $n>3$:

$\bullet $ For $\mathcal{A}_{2n-1}^{(2)}$ models we have found one $1$%
-parameter solution given by (\ref{C(n),D(n),A(2n-1)red29}).

$\bullet $ For $\mathcal{D}_{n}^{(1)}$ models we have found one $(n-1)$%
-parameter solution with $8n-6$ null entries depending on the parity of $n$,
and one $1$-parameter solution given by (\ref{C(n),D(n),A(2n-1)red29}).

$\bullet $ For $\mathcal{C}_{n}^{(1)}$ models we have found one $(n-1)$%
-parameter solution with $8n-6$ null entries depending on the parity of $n$,
and one $1$-parameter solution given by (\ref{C(n),D(n),A(2n-1)red30}).

These results yield all the independent solutions of the reflection equation
(\ref{BYBE}) with at least two non-diagonal entries.

\subsection{The $\mathcal{D}_{n+1}^{(2)}$ Models}

Here we are interested in looking for reduced solutions of the reflection
equation (\ref{BYBE}) for the $\mathcal{D}_{n+1}^{(2)}$ models. We managed
to identify the only possible $K$-matrices, which are encountered when the
recurrent relations (\ref{D(n+1)gen20}) degenerate into $k_{11}(u)$ and $%
k_{n+3,n+3}(u)$, respectively, 
\begin{equation}
k_{n,n}(u)=k_{n-1,n-1}(u)=...=k_{22}(u)=k_{11}(u),  \label{D(n+1)red1}
\end{equation}%
and%
\begin{equation}
k_{2n+2,2n+2}(u)=k_{2n+1,2n+1}(u)=...=k_{n+4,n+4}(u)=k_{n+3,n+3}(u).
\label{D(n+1)red2}
\end{equation}%
This reduced solution can be obtained by the same procedure developed
previously in Section 3.3, and for the sake of brevity we will only quote
the final results.

We have found two classes of solutions for any value of $n$, which are block
diagonal $K$-matrices with one free parameter, $\beta _{n+1,n+2}$. The first
class is given by%
\begin{eqnarray}
k_{11}(u) &=&\frac{1}{2}\frac{(\mathrm{e}^{2u}+q^{n})[(q^{n}-1)(\mathrm{e}%
^{2u}+1)-\beta _{n+1,n+2}(q^{n}+1)(\mathrm{e}^{2u}-1)]}{\mathrm{e}%
^{2u}(q^{2n}-1)},  \notag \\
&&  \label{D(n+1)red3} \\
k_{n+3,n+3}(u) &=&\frac{1}{2}\frac{(\mathrm{e}^{2u}+q^{n})[(q^{n}-1)(\mathrm{%
e}^{2u}+1)+\beta _{n+1,n+2}(q^{n}+1)(\mathrm{e}^{2u}-1)]}{(q^{2n}-1)}, 
\notag \\
&&  \label{D(n+1)red4}
\end{eqnarray}%
with central elements%
\begin{eqnarray}
k_{n+1,n+2}(u) &=&k_{n+2,n+1}(u)=\frac{1}{2}\beta _{n+1,n+2}(\mathrm{e}%
^{2u}-1),  \label{D(n+1)red5} \\
k_{n+1,n+1}(u) &=&\frac{1}{2}(\mathrm{e}^{2u}+1)\left\{ 1+\frac{(\mathrm{e}%
^{2u}-1)}{\mathrm{e}^{u}(q^{2n}-1)}\Gamma _{\pm }\right\} ,
\label{D(n+1)red6} \\
k_{n+2,n+2}(u) &=&\frac{1}{2}(\mathrm{e}^{2u}+1)\left\{ 1+\frac{(\mathrm{e}%
^{2u}-1)}{\mathrm{e}^{u}(q^{2n}-1)}\Gamma _{\mp }\right\} ,
\label{D(n+1)red7}
\end{eqnarray}%
where%
\begin{eqnarray}
\Gamma _{\pm } &=&\frac{1}{\Sigma _{\pm }}\left\{ 2q^{n}[(q^{n}+1)^{2}\beta
_{n+1,n+2}^{2}-(q^{n}-1)^{2}]\right.  \notag \\
&&\pm \lbrack (q^{n}+1)^{2}\beta _{n+1,n+2}+(q^{n}-1)^{2}]  \notag \\
&&\left. \times \sqrt{q^{n}[(q^{n}+1)^{2}\beta _{n+1,n+2}^{2}-(q^{n}-1)^{2}]}%
\right\}  \label{D(n+1)red8}
\end{eqnarray}%
and%
\begin{eqnarray}
\Sigma _{\pm } &=&[(q^{n}+1)^{2}\beta _{n+1,n+2}+(q^{n}-1)^{2}]  \notag \\
&&\pm 2\sqrt{q^{n}[(q^{n}+1)^{2}\beta _{n+1,n+2}^{2}-(q^{n}-1)^{2}]}.
\label{D(n+1)red9}
\end{eqnarray}%
The signs $(\pm )$ and $(\mp )$ represent the existence of two conjugate
solutions. Here we notice that these solutions degenerate into two complex
diagonal solutions by setting $\beta _{n+1,n+2}=0$.

The second family is given by 
\begin{eqnarray}
k_{11}(u) &=&\frac{1}{2}\frac{(\mathrm{e}^{2u}-q^{n})}{\mathrm{e}%
^{u}(q^{n}-1)^{2}}\left\{ (\mathrm{e}^{2u}-1)[(q^{n}+1)\beta
_{n+1,n+2}\right.  \notag \\
&&\left. \pm 2\sqrt{q^{n}[\beta _{n+1,n+2}^{2}-1]}]-(\mathrm{e}%
^{2u}+1)(q^{n}-1)\right\} ,  \label{D(n+1)red10} \\
k_{n+3,n+3}(u) &=&-\frac{1}{2}\frac{\mathrm{e}^{u}(\mathrm{e}^{2u}-q^{n})}{%
(q^{n}-1)^{2}}\left\{ (\mathrm{e}^{2u}-1)[(q^{n}+1)\beta _{n+1,n+2}\right. 
\notag \\
&&\left. \pm 2\sqrt{q^{n}[\beta _{n+1,n+2}^{2}-1]}]+(\mathrm{e}%
^{2u}+1)(q^{n}-1)\right\} ,  \label{D(n+1)red11}
\end{eqnarray}%
with the following central elements%
\begin{eqnarray}
k_{n+1,n+1}(u) &=&k_{n+2,n+2}(u)=\frac{1}{2}\mathrm{e}^{u}(\mathrm{e}%
^{2u}+1),  \label{D(n+1)red12} \\
k_{n+1,n+2}(u) &=&\frac{1}{2}\frac{(\mathrm{e}^{2u}-1)}{(q^{n}-1)^{2}}%
\left\{ \beta _{n+1,n+2}[\mathrm{e}^{u}(q^{n}+1)^{2}-2q^{n}(\mathrm{e}%
^{2u}+1)]\right.  \notag \\
&&\left. \mp (q^{n}+1)\sqrt{q^{n}[\beta _{n+1,n+2}^{2}-1]}(\mathrm{e}%
^{u}-1)^{2}\right\} ,  \label{D(n+1)red13} \\
k_{n+2,n+1}(u) &=&\frac{1}{2}\frac{(\mathrm{e}^{2u}-1)}{(q^{n}-1)^{2}}%
\left\{ \beta _{n+1,n+2}[\mathrm{e}^{u}(q^{n}+1)^{2}+2q^{n}(\mathrm{e}%
^{2u}+1)]\right.  \notag \\
&&\left. \pm (q^{n}+1)\sqrt{q^{n}[\beta _{n+1,n+2}^{2}-1]}(\mathrm{e}%
^{u}+1)^{2}\right\} .  \label{D(n+1)red14}
\end{eqnarray}%
In particular, one cannot derive a diagonal solution from the second family
of solutions showed above.

We remark that these $\mathcal{D}_{n+1}^{(2)}$ reduced $K$-matrices have
been discussed by Martins and Guan in \cite{Ma2}.

\section{Diagonal Solutions}

We start this section by presenting the list of diagonal $K$-matrices
related to the vertex models associated with each non-exceptional affine Lie
algebra. In order to reveal as well as avoid missing any solution, we have
solved the reflection equation (\ref{BYBE}) again.

\subsection{The $\mathcal{A}_{n-1}^{(1)}$ Diagonal $K$-Matrices}

We derive the set of $\mathcal{A}_{n-1}^{(1)}$ regular diagonal solutions
for $n\geq 5$ by considering separately each type of solution presented in
Section 3.1. We remark that the diagonal $K$-matrices for the first values
of $n$ are special and will be shown in Section 6. Aside from the following
classification, we have the trivial diagonal solution $K^{-}(u)=\mathbf{1}$
for these models.

\subsubsection{The diagonal $K$-matrices of type $I$}

Performing the following reductions for the scalar functions $\mathcal{Z}%
_{i}(u)$ (\ref{A(n)gen13}), $\mathcal{Y}_{i+1}^{(i)}(u)$ (\ref{A(n)gen14})
and $\mathcal{X}_{j+1}(u)$ (\ref{A(n)gen16}), namely%
\begin{eqnarray}
\lim_{\beta _{j+1,j+1}\rightarrow -\beta _{11}+2}\mathcal{X}_{j+1}(u) &=&%
\mathrm{e}^{2u}f_{11}(-u),  \notag \\
\lim_{\beta _{j+1,j+1}\rightarrow \beta _{11}+2}\mathcal{X}_{j+1}(u) &=&%
\mathrm{e}^{2u}f_{11}(u),  \label{A(n)diag1}
\end{eqnarray}%
and%
\begin{eqnarray}
\lim_{\beta _{i+1,i+1}\rightarrow -\beta _{i,i}+2}\mathcal{Y}_{i+1}^{(i)}(u)
&=&\mathrm{e}^{2u}f_{i,i}(-u),\text{ \ \ \ }\lim_{\beta
_{i+1,i+1}\rightarrow \beta _{i,i}}\mathcal{Y}_{i+1}^{(i)}(u)=f_{i,i}(u), 
\notag \\
\lim_{\beta _{22}\rightarrow -\beta _{11}+2}\mathcal{Y}_{2}^{(1)}(u) &=&%
\mathrm{e}^{2u}f_{11}(-u),\text{ \ \ \ \ \ \ }\lim_{\beta _{22}\rightarrow
\beta _{11}}\mathcal{Y}_{2}^{(1)}(u)=f_{11}(u),  \notag \\
\lim_{\beta _{11}\rightarrow -\beta _{i,i}}\mathcal{Z}_{i}(u) &=&f_{i,i}(-u),%
\text{ \ \ \ \ \ \ \ \ \ \ }\lim_{\beta _{11}\rightarrow \beta _{i,i}}%
\mathcal{Z}_{i}(u)=f_{i,i}(u),  \notag \\
&&  \label{A(n)diag2}
\end{eqnarray}%
we solve the constraint equations (\ref{A(n)gen17}), (\ref{A(n)gen18}), (\ref%
{A(n)gen19}) and obtain four $1$-parameter diagonal solutions $\mathbb{K}%
_{i,j}^{I}$ for $1<i<j\leq n$ given by%
\begin{eqnarray}
\mathbb{K}_{i,j}^{I} &=&f_{i,i}(u)E_{ii}+\mathrm{e}%
^{2u}f_{i,i}(-u)E_{jj}+f_{i,i}(-u)\sum_{l=1}^{i-1}E_{ll}  \notag \\
&&+\mathrm{e}^{2u}f_{i,i}(-u)\sum_{l=i+1}^{j-1}E_{ll}+\mathrm{e}%
^{2u}f_{i,i}(-u)\sum_{l=j+1}^{n}E_{ll},  \label{A(n)diag3}
\end{eqnarray}%
\begin{eqnarray}
\mathbb{K}_{i,j}^{I} &=&f_{i,i}(u)E_{ii}+\mathrm{e}%
^{2u}f_{i,i}(-u)E_{jj}+f_{i,i}(-u)\sum_{l=1}^{i-1}E_{ll}  \notag \\
&&+f_{i,i}(u)\sum_{l=i+1}^{j-1}E_{ll}+\mathrm{e}^{2u}f_{i,i}(-u)%
\sum_{l=j+1}^{n}E_{ll},  \label{A(n)diag4}
\end{eqnarray}%
\begin{eqnarray}
\mathbb{K}_{i,j}^{I} &=&f_{i,i}(u)E_{ii}+\mathrm{e}%
^{2u}f_{i,i}(-u)E_{jj}+f_{i,i}(u)\sum_{l=1}^{i-1}E_{ll}  \notag \\
&&+\mathrm{e}^{2u}f_{i,i}(-u)\sum_{l=i+1}^{j-1}E_{ll}+\mathrm{e}%
^{2u}f_{i,i}(u)\sum_{l=j+1}^{n}E_{ll},  \label{A(n)diag5}
\end{eqnarray}%
\begin{eqnarray}
\mathbb{K}_{i,j}^{I} &=&f_{i,i}(u)E_{ii}+\mathrm{e}%
^{2u}f_{i,i}(-u)E_{jj}+f_{i,i}(u)\sum_{l=1}^{i-1}E_{ll}  \notag \\
&&+f_{i,i}(u)\sum_{l=i+1}^{j-1}E_{ll}+\mathrm{e}^{2u}f_{i,i}(u)%
\sum_{l=j+1}^{n}E_{ll},  \label{A(n)diag6}
\end{eqnarray}%
where the functions $f_{i,i}(u)$ are given by (\ref{A(n)gen11}) and $\beta
_{i,i}$ is the free parameter.

Moreover, for $i=1$ and $1<j\leq n$ we get four $1$-parameter diagonal
matrices $\mathbb{K}_{1,j}^{I}$ as follows%
\begin{eqnarray}
\mathbb{K}_{1,j}^{I} &=&f_{11}(u)E_{11}+\mathrm{e}^{2u}f_{11}(-u)E_{jj}+%
\mathrm{e}^{2u}f_{11}(-u)\sum_{l=2}^{j-1}E_{ll}  \notag \\
&&+\mathrm{e}^{2u}f_{11}(-u)\sum_{l=j+1}^{n}E_{ll},  \label{A(n)diag7}
\end{eqnarray}%
\begin{eqnarray}
\mathbb{K}_{1,j}^{I} &=&f_{11}(u)E_{11}+\mathrm{e}^{2u}f_{11}(-u)E_{jj}+%
\mathrm{e}^{2u}f_{11}(-u)\sum_{l=2}^{j-1}E_{ll}  \notag \\
&&+\mathrm{e}^{2u}f_{11}(u)\sum_{l=j+1}^{n}E_{ll},  \label{A(n)diag8}
\end{eqnarray}%
\begin{eqnarray}
\mathbb{K}_{1,j}^{I} &=&f_{11}(u)E_{11}+\mathrm{e}%
^{2u}f_{11}(-u)E_{jj}+f_{11}(u)\sum_{l=2}^{j-1}E_{ll}  \notag \\
&&+\mathrm{e}^{2u}f_{11}(-u)\sum_{l=j+1}^{n}E_{ll},  \label{A(n)diag9}
\end{eqnarray}%
\begin{eqnarray}
\mathbb{K}_{1,j}^{I} &=&f_{11}(u)E_{11}+\mathrm{e}%
^{2u}f_{11}(-u)E_{jj}+f_{11}(u)\sum_{l=2}^{j-1}E_{ll}  \notag \\
&&+\mathrm{e}^{2u}f_{11}(u)\sum_{l=j+1}^{n}E_{ll},  \label{A(n)diag10}
\end{eqnarray}%
and $\beta _{11}$ is the free parameter.

In particular, for $i=1$ and $j=n$ we have two $1$-parameter diagonal
solutions $\mathbb{K}_{1,n}^{I}$ given by%
\begin{equation}
\mathbb{K}_{1,n}^{I}=f_{11}(u)E_{11}+\mathrm{e}^{2u}f_{11}(-u)E_{nn}+\mathrm{%
e}^{2u}f_{11}(-u)\sum_{l=2}^{n-1}E_{ll}  \label{A(n)diag11}
\end{equation}%
and%
\begin{equation}
\mathbb{K}_{1,n}^{I}=f_{11}(u)E_{11}+\mathrm{e}%
^{2u}f_{11}(-u)E_{nn}+f_{11}(u)\sum_{l=2}^{n-1}E_{ll},  \label{A(n)diag12}
\end{equation}%
where $\beta _{11}$ is the free parameter.

\subsubsection{The diagonal $K$-matrices of type $II$}

Now we use the reductions for the scalar functions (\ref{A(n)gen13}), (\ref%
{A(n)gen14}) and (\ref{A(n)gen16}) as follows%
\begin{eqnarray}
\lim_{\beta _{\left[ \frac{n}{2}\right] +p+1,\left[ \frac{n}{2}\right]
+p+1}\rightarrow -\beta _{11}+2}\mathcal{X}_{\left[ \frac{n}{2}\right]
+p+1}(u) &=&\mathrm{e}^{2u}f_{11}(-u),  \notag \\
\lim_{\beta _{\left[ \frac{n}{2}\right] +p+1,\left[ \frac{n}{2}\right]
+p+1}\rightarrow \beta _{11}+2}\mathcal{X}_{\left[ \frac{n}{2}\right]
+p+1}(u) &=&\mathrm{e}^{2u}f_{11}(u),  \label{A(n)diag13}
\end{eqnarray}%
and%
\begin{eqnarray}
\lim_{\beta _{p+1,p+1}\rightarrow -\beta _{11}+2}\mathcal{Y}_{p+1}^{(1)}(u)
&=&\mathrm{e}^{2u}f_{11}(-u),\text{ \ \ }\lim_{\beta _{p+1,p+1}\rightarrow
\beta _{11}}\mathcal{Y}_{p+1}^{(1)}(u)=f_{11}(u),  \notag \\
\lim_{\beta _{\frac{n}{2}+1,\frac{n}{2}+1}\rightarrow -\beta _{22}+2}%
\mathcal{Y}_{\frac{n}{2}+1}^{(2)}(u) &=&\mathrm{e}^{2u}f_{22}(-u),\text{\ \ }%
\lim_{\beta _{\frac{n}{2}+1,\frac{n}{2}+1}\rightarrow \beta _{22}}\mathcal{Y}%
_{\frac{n}{2}+1}^{(2)}(u)=f_{22}(u),  \notag \\
\lim_{\beta _{11}\rightarrow -\beta _{22}}\mathcal{Z}_{2}(u) &=&f_{22}(-u),%
\text{ \ \ \ \ \ \ \ \ \ }\lim_{\beta _{11}\rightarrow \beta _{22}}\mathcal{Z%
}_{2}(u)=f_{22}(u)  \notag \\
&&  \label{A(n)diag14}
\end{eqnarray}%
in order to solve the constraint equations yielded by three solutions of
type $II$ for each $\mathcal{A}_{n-1}^{(1)}$ model:%
\begin{eqnarray}
\text{Type }II\text{{\small a}} &=&\{\mathbb{K}_{1,2p}^{II}\},\text{ \ \ \
Type }II\text{{\small b}}=\{\mathbb{K}_{1,2p+1}^{II}\},\text{ \ \ \ Type }II%
\text{{\small c}}=\{\mathbb{K}_{2,n}^{II}\},  \notag \\
\text{with \ \ }p &=&1,2,...,\left[ \frac{n}{2}\right] ,  \label{A(n)diag15}
\end{eqnarray}%
where $\left[ \frac{n}{2}\right] $ is the integer part of $\frac{n}{2}$. The
diagonal $K$-matrices of type $II$ depend on the parity of $n$, leading us
to the following families of solutions:

\paragraph{Odd $n$}

We have the $1$-parameter diagonal solutions of type $II${\small a} given by%
\begin{eqnarray}
\mathbb{K}_{1,2p}^{II} &=&f_{11}(u)\sum_{j=1}^{p}E_{jj}+\mathrm{e}%
^{2u}f_{11}(-u)\sum_{j=p+1}^{\left[ \frac{n}{2}\right] +p}E_{jj}+\mathrm{e}%
^{2u}f_{11}(u)\sum_{j=\left[ \frac{n}{2}\right] +p+2}^{n}E_{jj}  \notag \\
&&+\mathrm{e}^{2u}f_{11}(-u)E_{\left[ \frac{n}{2}\right] +p+1,\left[ \frac{n%
}{2}\right] +p+1}  \label{A(n)diag16}
\end{eqnarray}%
and%
\begin{eqnarray}
\mathbb{K}_{1,2p}^{II} &=&f_{11}(u)\sum_{j=1}^{p}E_{jj}+\mathrm{e}%
^{2u}f_{11}(-u)\sum_{j=p+1}^{\left[ \frac{n}{2}\right] +p}E_{jj}+\mathrm{e}%
^{2u}f_{11}(u)\sum_{j=\left[ \frac{n}{2}\right] +p+2}^{n}E_{jj}  \notag \\
&&+\mathrm{e}^{2u}f_{11}(u)E_{\left[ \frac{n}{2}\right] +p+1,\left[ \frac{n}{%
2}\right] +p+1},  \label{A(n)diag17}
\end{eqnarray}%
where the scalar function $f_{11}(u)$ is given by (\ref{A(n)gen11}) and $%
\beta _{11}$ is the free parameter.

The $1$-parameter diagonal $K$-matrices of type $II${\small b} take the form%
\begin{eqnarray}
\mathbb{K}_{1,2p+1}^{II} &=&f_{11}(u)\sum_{j=1}^{p}E_{jj}+\mathrm{e}%
^{2u}f_{11}(-u)\sum_{j=p+2}^{\left[ \frac{n}{2}\right] +p+1}E_{jj}+\mathrm{e}%
^{2u}f_{11}(u)\sum_{j=\left[ \frac{n}{2}\right] +p+2}^{n}E_{jj}  \notag \\
&&+\mathrm{e}^{2u}f_{11}(-u)E_{p+1,p+1}  \label{A(n)diag18}
\end{eqnarray}%
and%
\begin{eqnarray}
\mathbb{K}_{1,2p+1}^{II} &=&f_{11}(u)\sum_{j=1}^{p}E_{jj}+\mathrm{e}%
^{2u}f_{11}(-u)\sum_{j=p+2}^{\left[ \frac{n}{2}\right] +p+1}E_{jj}+\mathrm{e}%
^{2u}f_{11}(u)\sum_{j=\left[ \frac{n}{2}\right] +p+2}^{n}E_{jj}  \notag \\
&&+f_{11}(u)E_{p+1,p+1}.  \label{A(n)diag19}
\end{eqnarray}%
Finally, the $1$-parameter diagonal solutions of type $II${\small c} are
given by%
\begin{equation}
\mathbb{K}_{2,n}^{II}=f_{22}(-u)E_{11}+f_{22}(u)\sum_{j=2}^{\left[ \frac{n}{2%
}\right] +1}E_{jj}+\mathrm{e}^{2u}f_{22}(-u)\sum_{j=\left[ \frac{n}{2}\right]
+2}^{n}E_{jj}  \label{A(n)diag20}
\end{equation}%
and%
\begin{equation}
\mathbb{K}_{2,n}^{II}=f_{22}(u)E_{11}+f_{22}(u)\sum_{j=2}^{\left[ \frac{n}{2}%
\right] +1}E_{jj}+\mathrm{e}^{2u}f_{22}(-u)\sum_{j=\left[ \frac{n}{2}\right]
+2}^{n}E_{jj},  \label{A(n)diag21}
\end{equation}%
where $\beta _{22}$ is the free parameter.

\paragraph{Even $n$}

Here we have the $1$-parameter diagonal solutions of type $II${\small a} as
follows%
\begin{equation}
\mathbb{K}_{1,2p}^{II}=f_{11}(u)\sum_{j=1}^{p}E_{jj}+\mathrm{e}%
^{2u}f_{11}(-u)\sum_{j=p+1}^{\frac{n}{2}+p}E_{jj}+\mathrm{e}%
^{2u}f_{11}(u)\sum_{j=\frac{n}{2}+p+1}^{n}E_{jj},  \label{A(n)diag22}
\end{equation}%
with the scalar function $f_{11}(u)$ given by (\ref{A(n)gen11}) and $\beta
_{11}$ is the free parameter.

The $1$-parameter diagonal $K$-matrices of type $II${\small b} take the
following form%
\begin{eqnarray}
\mathbb{K}_{1,2p+1}^{II} &=&f_{11}(u)\sum_{j=1}^{p}E_{jj}+\mathrm{e}%
^{2u}f_{11}(-u)E_{p+1,p+1}+\mathrm{e}^{2u}f_{11}(-u)\sum_{j=p+2}^{\frac{n}{2}%
+p}E_{jj}  \notag \\
&&+\mathrm{e}^{2u}f_{11}(-u)E_{\frac{n}{2}+p+1,\frac{n}{2}+p+1}+\mathrm{e}%
^{2u}f_{11}(u)\sum_{j=\frac{n}{2}+p+2}^{n}E_{jj},  \label{A(n)diag23}
\end{eqnarray}%
\begin{eqnarray}
\mathbb{K}_{1,2p+1}^{II} &=&f_{11}(u)\sum_{j=1}^{p}E_{jj}+\mathrm{e}%
^{2u}f_{11}(-u)E_{p+1,p+1}+\mathrm{e}^{2u}f_{11}(-u)\sum_{j=p+2}^{\frac{n}{2}%
+p}E_{jj}  \notag \\
&&+\mathrm{e}^{2u}f_{11}(u)E_{\frac{n}{2}+p+1,\frac{n}{2}+p+1}+\mathrm{e}%
^{2u}f_{11}(u)\sum_{j=\frac{n}{2}+p+2}^{n}E_{jj},  \label{A(n)diag24}
\end{eqnarray}%
\begin{eqnarray}
\mathbb{K}_{1,2p+1}^{II}
&=&f_{11}(u)\sum_{j=1}^{p}E_{jj}+f_{11}(u)E_{p+1,p+1}+\mathrm{e}%
^{2u}f_{11}(-u)\sum_{j=p+2}^{\frac{n}{2}+p}E_{jj}  \notag \\
&&+\mathrm{e}^{2u}f_{11}(-u)E_{\frac{n}{2}+p+1,\frac{n}{2}+p+1}+\mathrm{e}%
^{2u}f_{11}(u)\sum_{j=\frac{n}{2}+p+2}^{n}E_{jj},  \label{A(n)diag25}
\end{eqnarray}%
\begin{eqnarray}
\mathbb{K}_{1,2p+1}^{II}
&=&f_{11}(u)\sum_{j=1}^{p}E_{jj}+f_{11}(u)E_{p+1,p+1}+\mathrm{e}%
^{2u}f_{11}(-u)\sum_{j=p+2}^{\frac{n}{2}+p}E_{jj}  \notag \\
&&+\mathrm{e}^{2u}f_{11}(u)E_{\frac{n}{2}+p+1,\frac{n}{2}+p+1}+\mathrm{e}%
^{2u}f_{11}(u)\sum_{j=\frac{n}{2}+p+2}^{n}E_{jj}.  \label{A(n)diag26}
\end{eqnarray}%
We also have the $1$-parameter diagonal solutions of type $II${\small c}
given by%
\begin{eqnarray}
\mathbb{K}_{2,n}^{II} &=&f_{22}(-u)E_{11}+f_{22}(u)\sum_{j=2}^{\frac{n}{2}%
}E_{jj}+\mathrm{e}^{2u}f_{22}(-u)E_{\frac{n}{2}+1,\frac{n}{2}+1}  \notag \\
&&+\mathrm{e}^{2u}f_{22}(-u)\sum_{j=\frac{n}{2}+2}^{n}E_{jj},
\label{A(n)diag27}
\end{eqnarray}%
\begin{eqnarray}
\mathbb{K}_{2,n}^{II} &=&f_{22}(-u)E_{11}+f_{22}(u)\sum_{j=2}^{\frac{n}{2}%
}E_{jj}+f_{22}(u)E_{\frac{n}{2}+1,\frac{n}{2}+1}  \notag \\
&&+\mathrm{e}^{2u}f_{22}(-u)\sum_{j=\frac{n}{2}+2}^{n}E_{jj},
\label{A(n)diag28}
\end{eqnarray}%
\begin{eqnarray}
\mathbb{K}_{2,n}^{II} &=&f_{22}(u)E_{11}+f_{22}(u)\sum_{j=2}^{\frac{n}{2}%
}E_{jj}+\mathrm{e}^{2u}f_{22}(-u)E_{\frac{n}{2}+1,\frac{n}{2}+1}  \notag \\
&&+\mathrm{e}^{2u}f_{22}(-u)\sum_{j=\frac{n}{2}+2}^{n}E_{jj},
\label{A(n)diag29}
\end{eqnarray}%
\begin{eqnarray}
\mathbb{K}_{2,n}^{II} &=&f_{22}(u)E_{11}+f_{22}(u)\sum_{j=2}^{\frac{n}{2}%
}E_{jj}+f_{22}(u)E_{\frac{n}{2}+1,\frac{n}{2}+1}  \notag \\
&&+\mathrm{e}^{2u}f_{22}(-u)\sum_{j=\frac{n}{2}+2}^{n}E_{jj},
\label{A(n)diag30}
\end{eqnarray}%
and $\beta _{22}$ is the free parameter.

\subsection{The $\mathcal{B}_{n}^{(1)}$ Diagonal $K$-Matrices}

For $n\geq 1$ we have one $1$-parameter solution $\mathbb{K}_{\beta }$ given
by%
\begin{eqnarray}
&&k_{11}(u)=\left( \frac{\beta (\mathrm{e}^{-u}-1)+2}{\beta (\mathrm{e}%
^{u}-1)+2}\right) ,  \notag \\
&&k_{22}(u)=...=k_{n+1,n+1}(u)=...=k_{2n,2n}(u)=1,  \notag \\
&&k_{2n+1,2n+1}(u)=\left( \frac{\beta (q^{2n-3}\mathrm{e}^{u}-1)+2}{\beta
(q^{2n-3}\mathrm{e}^{-u}-1)+2}\right) ,  \label{B(n)diag1}
\end{eqnarray}%
where $\beta =\beta _{n+1,n+1}-\beta _{11}$ is the free parameter.

We also get $2n-2$ solutions $\mathbb{K}^{[p]}$, $p=2,3,...,n$, with no free
parameter given by%
\begin{eqnarray}
&&k_{11}(u)=k_{22}(u)=...=k_{p,p}(u)=\mathrm{e}^{-u},  \notag \\
&&k_{p+1,p+1}(u)=k_{p+2,p+2}(u)=...=k_{2n-p+1,2n-p+1}(u)=\frac{q^{2p-n-1/2}%
\mathrm{e}^{u}\pm 1}{q^{2p-n-1/2}\pm \mathrm{e}^{u}},  \notag \\
&&k_{2n-p+2,2n-p+2}(u)=k_{2n-p+3,2n-p+3}(u)=...=k_{2n+1,2n+1}(u)=\mathrm{e}%
^{u}.  \notag \\
&&  \label{B(n)diag2}
\end{eqnarray}%
Therefore, we have found $2n-1$ regular diagonal $K$-matrices for the $%
\mathcal{B}_{n}^{(1)}$ models. We remark that the solutions $\mathbb{K}%
^{[p=n]}$ have been computed by Batchelor et al. in \cite{Bat1}.

\subsection{The $\mathcal{C}_{n}^{(1)}$ Diagonal $K$-Matrices}

Here we have one $1$-parameter solution $\mathbb{K}_{\beta }$ as follows%
\begin{eqnarray}
&&k_{11}(u)=k_{22}(u)=...=k_{n,n}(u)=1,  \notag \\
&&k_{n+1,n+1}(u)=k_{n+2,n+2}(u)=...=k_{2n,2n}(u)=\frac{\beta (\mathrm{e}%
^{u}-1)-2}{\beta (\mathrm{e}^{-u}-1)-2},  \notag \\
&&  \label{C(n)diag1}
\end{eqnarray}%
where $\beta $ is the free parameter.

For $n>2$, in addition to the trivial diagonal solution $K^{-}(u)=\mathbf{1}$%
, we also get $n-1$ solutions $\mathbb{K}^{[p]}$, $p=2,3,...,n$, which have
no free parameter given by the following matrix elements%
\begin{eqnarray}
&&k_{11}(u)=k_{22}(u)=...=k_{p-1,p-1}(u)=1,  \notag \\
&&k_{p,p}(u)=k_{p+1,p+1}(u)=...=k_{2n-p+1,2n-p+1}(u)=\mathrm{e}^{2u}\frac{%
\mathrm{e}^{-u}+\epsilon _{p}q^{2p-n-2}}{\mathrm{e}^{u}+\epsilon
_{p}q^{2p-n-2}},  \notag \\
&&k_{2n-p+2,2n-p+2}(u)=k_{2n-p+3,2n-p+3}(u)=...=k_{2n,2n}(u)=\mathrm{e}^{2u},
\notag \\
&&  \label{C(n)diag2}
\end{eqnarray}%
where $\epsilon _{p}=\pm 1$ for $2p\neq n+2$ and $\epsilon _{p}=1$ for $%
2p=n+2$. We have thus found $2n$ regular diagonal $K$-matrices if $n$ is
odd, and $2n-1$ regular diagonal $K$-matrices if $n$ is even for the $%
\mathcal{C}_{n}^{(1)}$ models.

\subsection{The $\mathcal{D}_{n}^{(1)}$ Diagonal $K$-Matrices}

We begin by pointing out that these models are symmetric under interchange
of indices $k_{n,n}(u)\leftrightarrow k_{n+1,n+1}(u)$.

The $\mathcal{D}_{2}^{(1)}$ diagonal $K$-matrices exhibit a special
structure. Besides the identity, we have two solutions $\mathbb{K}_{\alpha
\beta }$ with two free parameters related each other by exchanging $%
k_{n,n}(u)\leftrightarrow k_{n+1,n+1}(u)$: 
\begin{equation}
\mathbb{K}_{\alpha \beta }=\left( 
\begin{array}{cccc}
1 & 0 & 0 & 0 \\ 
0 & \frac{\alpha (\mathrm{e}^{u}-1)-2}{\alpha (\mathrm{e}^{-u}-1)-2} & 0 & 0
\\ 
0 & 0 & \frac{\beta (\mathrm{e}^{u}-1)-2}{\beta (\mathrm{e}^{-u}-1)-2} & 0
\\ 
0 & 0 & 0 & \frac{\alpha (\mathrm{e}^{u}-1)-2}{\alpha (\mathrm{e}^{-u}-1)-2}%
\frac{\beta (\mathrm{e}^{u}-1)-2}{\beta (\mathrm{e}^{-u}-1)-2}%
\end{array}%
\right) ,  \label{D(n)diag1}
\end{equation}%
where $\alpha $ and $\beta $ are the free parameters. We remark that the
isotropic limit of this solution has been presented in \cite{AADF}.

For $n>2$ we have the identity and seven $1$-parameter solutions $\mathbb{K}%
_{\beta }^{[i]}$, $i=1,2,...,7$, listed below.

$\bullet $ The $\mathbb{K}_{\beta }^{[1]}$-matrix has the following entries:%
\begin{eqnarray}
&&k_{11}(u)=1,  \notag \\
&&k_{22}(u)=k_{33}(u)=...=k_{2n-1,2n-1}(u)=\frac{\beta (\mathrm{e}^{u}-1)-2}{%
\beta (\mathrm{e}^{-u}-1)-2},  \notag \\
&&k_{2n,2n}(u)=\frac{\beta (\mathrm{e}^{u}-1)-2}{\beta (\mathrm{e}^{-u}-1)-2}%
\frac{\beta (q^{2n-4}\mathrm{e}^{u}-1)-2}{\beta (q^{2n-4}\mathrm{e}^{-u}-1)-2%
}.  \label{D(n)diag2}
\end{eqnarray}

$\bullet $ The $\mathbb{K}_{\beta }^{[2]}$-matrix has the following entries:%
\begin{eqnarray}
&&k_{11}(u)=k_{22}(u)=...=k_{n,n}(u)=1,  \notag \\
&&k_{n+1,n+1}(u)=k_{n+2,n+2}(u)=...=k_{2n,2n}(u)=\frac{\beta (\mathrm{e}%
^{u}-1)-2}{\beta (\mathrm{e}^{-u}-1)-2}.  \notag \\
&&  \label{D(n)diag3}
\end{eqnarray}

$\bullet $ The $\mathbb{K}_{\beta }^{[3]}$-matrix has the following entries:%
\begin{eqnarray}
&&k_{11}(u)=k_{22}(u)=...=k_{n-1,n-1}(u)=k_{n+1,n+1}(u)=1,  \notag \\
&&k_{n,n}(u)=k_{n+2,n+2}(u)=...=k_{2n,2n}(u)=\frac{\beta (\mathrm{e}^{u}-1)-2%
}{\beta (\mathrm{e}^{-u}-1)-2}.  \notag \\
&&  \label{D(n)diag4}
\end{eqnarray}

$\bullet $ The $\mathbb{K}_{\beta }^{[4]}$-matrix has the following entries:%
\begin{eqnarray}
&&k_{11}(u)=1,  \notag \\
&&k_{22}(u)=k_{33}(u)=...=k_{n,n}(u)=\frac{\beta (\mathrm{e}^{u}-1)-2}{\beta
(\mathrm{e}^{-u}-1)-2},  \notag \\
&&k_{n+1,n+1}(u)=k_{n+2,n+2}(u)=...=k_{2n-1,2n-1}(u)=\mathrm{e}^{2u},  \notag
\\
&&k_{2n,2n}(u)=\mathrm{e}^{2u}\frac{\beta (\mathrm{e}^{u}-1)-2}{\beta (%
\mathrm{e}^{-u}-1)-2}.  \label{D(n)diag5}
\end{eqnarray}

$\bullet $ The $\mathbb{K}_{\beta }^{[5]}$-matrix has the following entries:%
\begin{eqnarray}
&&k_{11}(u)=1,  \notag \\
&&k_{22}(u)=k_{33}(u)=...=k_{n-1,n-1}(u)=k_{n+1,n+1}(u)=\frac{\beta (\mathrm{%
e}^{u}-1)-2}{\beta (\mathrm{e}^{-u}-1)-2},  \notag \\
&&k_{n,n}(u)=k_{n+2,n+2}(u)=...=k_{2n-1,2n-1}(u)=\mathrm{e}^{2u},  \notag \\
&&k_{2n,2n}(u)=\mathrm{e}^{2u}\frac{\beta (\mathrm{e}^{u}-1)-2}{\beta (%
\mathrm{e}^{-u}-1)-2}.  \label{D(n)diag6}
\end{eqnarray}

$\bullet $ The $\mathbb{K}_{\beta }^{[6]}$-matrix has the following entries:%
\begin{eqnarray}
&&k_{11}(u)=k_{22}(u)=...=k_{n-1,n-1}(u)=1,  \notag \\
&&k_{n,n}(u)=\mathrm{e}^{2u}\frac{\beta (\mathrm{e}^{-u}-q^{2n-4})-2q^{2n-4}%
}{\beta (\mathrm{e}^{u}-q^{2n-4})-2q^{2n-4}},  \notag \\
&&k_{n+1,n+1}(u)=\frac{\beta (\mathrm{e}^{u}-1)-2}{\beta (\mathrm{e}%
^{-u}-1)-2},  \notag \\
&&k_{n+2,n+2}(u)=k_{n+3,n+3}(u)=...=k_{2n,2n}(u)=\mathrm{e}^{2u}.
\label{D(n)diag7}
\end{eqnarray}

$\bullet $ The $\mathbb{K}_{\beta }^{[7]}$-matrix has the following entries:%
\begin{eqnarray}
&&k_{11}(u)=k_{22}(u)=...=k_{n-1,n-1}(u)=1,  \notag \\
&&k_{n,n}(u)=\frac{\beta (\mathrm{e}^{u}-1)-2}{\beta (\mathrm{e}^{-u}-1)-2},
\notag \\
&&k_{n+1,n+1}(u)=\mathrm{e}^{2u}\frac{\beta (\mathrm{e}%
^{-u}-q^{2n-4})-2q^{2n-4}}{\beta (\mathrm{e}^{u}-q^{2n-4})-2q^{2n-4}}, 
\notag \\
&&k_{n+2,n+2}(u)=k_{n+3,n+3}(u)=...=k_{2n,2n}(u)=\mathrm{e}^{2u}.
\label{D(n)diag8}
\end{eqnarray}

Moreover, for $n>3$ we get $n-3$ solutions $\mathbb{K}^{[p]},$ $%
p=3,4,...,n-1,$ which have no free parameter given by the following matrix
elements%
\begin{eqnarray}
&&k_{11}(u)=k_{22}(u)=...=k_{p-1,p-1}(u)=1,  \notag \\
&&k_{p,p}(u)=k_{p+1,p+1}(u)=...=k_{2n-p+1,2n-p+1}(u)=\mathrm{e}^{2u}\frac{%
\mathrm{e}^{-u}+\epsilon _{p}q^{2p-n-2}}{\mathrm{e}^{u}+\epsilon
_{p}q^{2p-n-2}},  \notag \\
&&k_{2n-p+2,2n-p+2}(u)=k_{2n-p+3,2n-p+3}(u)=...=k_{2n,2n}(u)=\mathrm{e}^{2u},
\notag \\
&&  \label{D(n)diag9}
\end{eqnarray}%
where $\epsilon _{p}=\pm 1$ for $2p\neq n+2$ and $\epsilon _{p}=1$ for $%
2p=n+2$. Therefore, for $n\geq 3$, we have found $2n+1$ regular diagonal $K$%
-matrices if $n$ is odd, and $2n$ regular diagonal $K$-matrices if $n$ is
even for the $\mathcal{D}_{n}^{(1)}$ models.

We observe that the cases $\mathbb{K}^{[p=n]}$ are not computed because they
are reductions of the cases $\mathbb{K}_{\beta }^{[6]}$ and $\mathbb{K}%
_{\beta }^{[7]}$ due to an appropriate choice of the free parameter $\beta $
in such a way that $k_{n,n}(u)=k_{n+1,n+1}(u)$. We also note that the cases $%
\mathbb{K}^{[p=2]}$ are reductions of $\mathbb{K}_{\beta }^{[1]}$ by
choosing $\beta $ such that $k_{2n,2n}(u)=\mathrm{e}^{2u}$.

\subsection{The $\mathcal{A}_{2n}^{(2)}$ Diagonal $K$-Matrices}

For $n\geq 1$ we have $2n$ solutions $\mathbb{K}^{[p]}$, $p=1,2,...,n$,
which have no free parameter as follows%
\begin{eqnarray}
&&k_{11}(u)=k_{22}(u)=...=k_{p,p}(u)=\mathrm{e}^{-u},  \notag \\
&&k_{p+1,p+1}(u)=k_{p+2,p+2}(u)=...=k_{2n-p+1,2n-p+1}(u)=\frac{q^{2p-n-1/2}%
\mathrm{e}^{u}\pm \frac{i}{q}}{q^{2p-n-1/2}\pm \frac{i}{q}\mathrm{e}^{u}}, 
\notag \\
&&k_{2n-p+2,2n-p+2}(u)=k_{2n-p+3,2n-p+3}(u)=...=k_{2n+1,2n+1}(u)=\mathrm{e}%
^{u}.  \notag \\
&&  \label{A(2n)diag1}
\end{eqnarray}

We also get the trivial solution which is multiple of the identity. Thus, we
have found $2n+1$ regular diagonal $K$-matrices for the $\mathcal{A}%
_{2n}^{(2)}$ models. We remark that the $\mathcal{A}_{2}^{(2)}$ diagonal $K$%
-matrices have been obtained by Mezincescu, Nepomechie and Rittenberg in 
\cite{Ne4}.

\subsection{The $\mathcal{A}_{2n-1}^{(2)}$ Diagonal $K$-Matrices}

Here we have two $1$-parameter solutions $\mathbb{K}_{\beta }$ with the
following normalized entries%
\begin{eqnarray}
&&k_{11}(u)=k_{22}(u)=...=k_{n-1,n-1}(u)=1,  \notag \\
&&k_{n,n}(u)=\left( \frac{\beta (\mathrm{e}^{u}-1)-2}{\beta (\mathrm{e}%
^{-u}-1)-2}\right) ,  \notag \\
&&k_{n+1,n+1}(u)=\mathrm{e}^{2u}\left( \frac{\beta (\mathrm{e}%
^{-u}+q^{2n-2})+2q^{2n-2}}{\beta (\mathrm{e}^{u}+q^{2n-2})+2q^{2n-2}}\right)
,  \notag \\
&&k_{2n,2n}(u)=k_{2n-1,2n-1}(u)=...=k_{n+2,n+2}(u)=\mathrm{e}^{2u},  \notag
\\
&&  \label{A(2n-1)diag1}
\end{eqnarray}%
where $\beta $ is the free parameter and the second solution is obtained
from (\ref{A(2n-1)diag1}) by applying the symmetry of interchangeable
indices $k_{n,n}(u)\leftrightarrow k_{n+1,n+1}(u)$.

Moreover, for $n>2$ we get $2n-4$ solutions $\mathbb{K}^{[p]},$ $%
p=2,3,...,n-1,$ which have no free parameter given by%
\begin{eqnarray}
&&k_{11}(u)=k_{22}(u)=...=k_{p-1,p-1}(u)=1,  \notag \\
&&k_{p,p}(u)=k_{p+1,p+1}(u)=...=k_{2n-p+1,2n-p+1}(u)=\mathrm{e}^{2u}\frac{%
\mathrm{e}^{-u}\pm iq^{2p-n-1}}{\mathrm{e}^{u}\pm iq^{2p-n-1}},  \notag \\
&&k_{2n,2n}(u)=k_{2n-1,2n-1}(u)=...=k_{2n-p+2,2n-p+2}(u)=\mathrm{e}^{2u}. 
\notag \\
&&  \label{A(2n-1)diag2}
\end{eqnarray}

We point out that the cases $p=n$ are not computed since these solutions $%
\mathbb{K}^{[p=n]}$ can be obtained from the solutions $\mathbb{K}_{\beta }$
by assigning a special value to the free parameter $\beta $ such that $%
k_{n,n}(u)=k_{n+1,n+1}(u)$.

Additionally, we also have a trivial solution which is proportional to the
identity. Therefore, we have found $2n-1$ regular diagonal $K$-matrices for
the $\mathcal{A}_{2n-1}^{(2)}$ models.

\subsection{The $\mathcal{D}_{n+1}^{(2)}$ Diagonal $K$-Matrices}

We have the \textquotedblleft almost unity\textquotedblright\ regular
diagonal $K$-matrices for the $\mathcal{D}_{n+1}^{(2)}$ models given by%
\begin{eqnarray}
&&k_{11}(u)=\mathrm{e}^{-2u},  \notag \\
&&k_{22}(u)=k_{33}(u)=...=k_{2n+1,2n+1}(u)=1,  \notag \\
&&k_{2n+2,2n+2}(u)=\mathrm{e}^{2u},  \label{D(n+1)diag1}
\end{eqnarray}%
which exist only for even $n$ and have no free parameter. This result
features the $U(1)\otimes U(1)$ symmetries of the models with an even number
of $U(1)$ conserved charges, and has been presented by Martins and Guan in 
\cite{Ma2}.

\section{Special Cases}

We now focus on the $K$-matrices which are ruled out of our classification
scheme, namely the cases $\mathcal{B}_{n}^{(1)},$ $\mathcal{C}_{n}^{(1)},$ $%
\mathcal{D}_{n}^{(1)},$ $\mathcal{A}_{2n}^{(2)},$ $\mathcal{A}_{2n-1}^{(2)},$
and $\mathcal{D}_{n+1}^{(2)}$ for $n=1$, $\mathcal{A}_{n-1}^{(1)},$ $%
\mathcal{C}_{n}^{(1)},$ $\mathcal{D}_{n}^{(1)},$ $\mathcal{A}_{2n-1}^{(2)}$
for $n=2$, the $\mathcal{A}_{2}^{(1)},\mathcal{A}_{3}^{(1)},$ and $\mathcal{A%
}_{4}^{(1)}$ models as well as the $\mathcal{D}_{3}^{(1)}$ $K$-matrix with $%
\mathcal{G}^{(-)}(u)$ and the $\mathcal{C}_{3}^{(1)}$ $K$-matrix with $%
\mathcal{G}^{(+)}(u)$. We regard these solutions as special because they do
not exhibit all the properties featured and shared by most of the reflection 
$K$-matrices.

\subsection{The $\mathcal{A}_{1}^{(1)}$ Case}

We have one general solution which is a very special case among the $%
\mathcal{A}_{n-1}^{(1)}$ models, given by%
\begin{equation}
\mathcal{K}_{12}^{I}=\left( 
\begin{array}{cc}
\beta _{11}(\mathrm{e}^{u}-1)+1 & \frac{1}{2}\beta _{12}(\mathrm{e}^{2u}-1)
\\ 
\frac{1}{2}\beta _{21}(\mathrm{e}^{2u}-1) & \mathrm{e}^{2u}\beta _{11}(%
\mathrm{e}^{-u}-1)+1%
\end{array}%
\right) .  \label{A(1)gen1}
\end{equation}%
The above $K$-matrix may be recognized either from the solution of type $I$
itself or from the solution of type $II${\small a}. Although there is no
constraint equation in this case, the regular condition (\ref{K-matrixReg})
has yielded three free parameters, $\beta _{11}$, $\beta _{12}$, $\beta
_{21} $, in accordance with all reflection $K$-matrices of type $I$. We note
that there is only one general solution containing four non-null matrix
elements \cite{VGR1,GZ}.

\subsection{The $\mathcal{A}_{2}^{(1)}$ Case}

In this special case, all the solutions of type $II$ are indeed solutions of
type $I$, namely $\mathcal{K}_{12}^{I},\mathcal{K}_{13}^{I},\mathcal{K}%
_{23}^{I}$. From (\ref{A(n)gen15}) we get $\mathcal{K}_{12}^{I}$ as follows 
\begin{eqnarray}
\mathcal{K}_{12}^{I} &=&f_{11}(u)E_{11}+\mathrm{e}%
^{2u}f_{11}(-u)E_{22}+h_{12}(u)E_{12}+h_{21}(u)E_{21}+\mathcal{X}%
_{3}(u)E_{33}  \notag \\
&=&\left( 
\begin{array}{ccc}
f_{11}(u) & h_{12}(u) & 0 \\ 
h_{21}(u) & \mathrm{e}^{2u}f_{11}(-u) & 0 \\ 
0 & 0 & \mathcal{X}_{3}(u)%
\end{array}%
\right) ,  \label{A(2)gen1a}
\end{eqnarray}%
with four parameters, $\beta _{11}$, $\beta _{12}$, $\beta _{21}$, $\beta
_{33}$, satisfying the constraint equation%
\begin{equation}
\beta _{12}\beta _{21}=(\beta _{33}-\beta _{11}-2)(\beta _{33}+\beta
_{11}-2).  \label{A(2)gen2a}
\end{equation}%
Due to this constraint equation, we can derive from (\ref{A(2)gen1a}) two
diagonal solutions given by 
\begin{eqnarray}
&&\lim_{\beta _{33}\rightarrow -\beta _{11}+2}\mathcal{X}_{3}(u)=\mathrm{e}%
^{2u}f_{11}(-u),  \notag \\
&\Rightarrow &D_{1}=\text{diag}\left( f_{11}(u),\mathrm{e}^{2u}f_{11}(-u),%
\mathrm{e}^{2u}f_{11}(-u)\right)  \label{A(2)gen3a}
\end{eqnarray}%
and%
\begin{eqnarray}
&&\lim_{\beta _{33}\rightarrow \beta _{11}+2}\mathcal{X}_{3}(u)=\mathrm{e}%
^{2u}f_{11}(u),  \notag \\
&\Rightarrow &D_{2}=\text{diag}\left( f_{11}(u),\mathrm{e}^{2u}f_{11}(-u),%
\mathrm{e}^{2u}f_{11}(u)\right) .  \label{A(2)gen4a}
\end{eqnarray}%
The matrix $\mathcal{K}_{13}^{I}$ is also given by (\ref{A(n)gen15})%
\begin{eqnarray}
\mathcal{K}_{13}^{I} &=&f_{11}(u)E_{11}+\mathrm{e}%
^{2u}f_{11}(-u)E_{33}+h_{13}(u)E_{13}+h_{31}(u)E_{31}+\mathcal{Y}%
_{2}^{(1)}(u)E_{22}  \notag \\
&=&\left( 
\begin{array}{ccc}
f_{11}(u) & 0 & h_{13}(u) \\ 
0 & \mathcal{Y}_{2}^{(1)}(u) & 0 \\ 
h_{31}(u) & 0 & \mathrm{e}^{2u}f_{11}(-u)%
\end{array}%
\right) ,  \label{A(2)gen5a}
\end{eqnarray}%
but now the constraint equation is%
\begin{equation}
\beta _{13}\beta _{31}=(\beta _{22}+\beta _{11}-2)(\beta _{22}-\beta _{11}),
\label{A(2)gen6a}
\end{equation}%
and the corresponding diagonal reductions are%
\begin{eqnarray}
&&\lim_{\beta _{22}\rightarrow -\beta _{11}+2}\mathcal{Y}_{2}^{(1)}(u)=%
\mathrm{e}^{2u}f_{11}(-u),  \notag \\
&\Rightarrow &D_{3}=\text{diag}\left( f_{11}(u),\mathrm{e}^{2u}f_{11}(-u),%
\mathrm{e}^{2u}f_{11}(-u)\right)  \label{A(2)gen7a}
\end{eqnarray}%
and%
\begin{eqnarray}
&&\lim_{\beta _{22}\rightarrow \beta _{11}}\mathcal{Y}%
_{2}^{(1)}(u)=f_{11}(u),  \notag \\
&\Rightarrow &D_{4}=\text{diag}\left( f_{11}(u),f_{11}(u),\mathrm{e}%
^{2u}f_{11}(-u)\right) .  \label{A(2)gen8a}
\end{eqnarray}%
Here we recall (\ref{A(n)gen12}) with $i=2$ and $j=3$ in order to get the
matrix $\mathcal{K}_{23}^{I}$, given by 
\begin{eqnarray}
\mathcal{K}_{23}^{I} &=&f_{22}(u)E_{22}+\mathrm{e}%
^{2u}f_{22}(-u)E_{33}+h_{23}(u)E_{23}+h_{32}(u)E_{32}+\mathcal{Z}%
_{2}(u)E_{11}  \notag \\
&=&\left( 
\begin{array}{ccc}
\mathcal{Z}_{2}(u) & 0 & 0 \\ 
0 & f_{22}(u) & h_{23}(u) \\ 
0 & h_{32}(u) & \mathrm{e}^{2u}f_{22}(-u)%
\end{array}%
\right) ,  \label{A(2)gen9a}
\end{eqnarray}%
with the constraint equation%
\begin{equation}
\beta _{23}\beta _{32}=(\beta _{11}+\beta _{22})(\beta _{11}-\beta _{22}),
\label{A(2)gen10a}
\end{equation}%
and the following diagonal solutions 
\begin{eqnarray}
&&\lim_{\beta _{22}\rightarrow -\beta _{11}}\mathcal{Z}_{2}(u)=f_{22}(-u), 
\notag \\
&\Rightarrow &D_{5}=\text{diag}\left( f_{22}(-u),f_{22}(u),\mathrm{e}%
^{2u}f_{22}(-u)\right)  \label{A(2)gen11a}
\end{eqnarray}%
and%
\begin{eqnarray}
&&\lim_{\beta _{22}\rightarrow \beta _{11}}\mathcal{Z}_{2}(u)=f_{22}(u), 
\notag \\
&\Rightarrow &D_{6}=\text{diag}\left( f_{22}(u),f_{22}(u),\mathrm{e}%
^{2u}f_{22}(-u)\right) .  \label{A(2)gen12a}
\end{eqnarray}

The $K$-matrices $\mathcal{K}_{12}^{I},\mathcal{K}_{13}^{I},\mathcal{K}%
_{23}^{I}$ have only three free parameters, while their corresponding
diagonal solutions have just one free parameter due to the existence of
constraint equations. We observe that only four diagonal solutions are
independent since $D_{1}=D_{3}$ and $D_{4}=D_{6}$. We also note that the
solutions $D_{1}$ and $D_{4}$ have been derived by de Vega and Gonz\'{a}%
lez-Ruiz in \cite{VGR1}, and the non-diagonal solution $\mathcal{K}_{13}^{I}$
has been derived by Abad and Rios in \cite{Ab1}.

\subsection{The $\mathcal{A}_{3}^{(1)}$ Case}

For this model, the structure of the general solution begins to appear, but
it is still particular because half of the solutions of type $II$ turns out
to be solutions of type $I$.

Let us first write the $K$-matrices of type $I$ given by (\ref{A(n)gen15}).
The matrix $\mathcal{K}_{12}^{I}$ is given by 
\begin{equation}
\mathcal{K}_{12}^{I}=\left( 
\begin{array}{cccc}
f_{11}(u) & h_{12}(u) & 0 & 0 \\ 
h_{21}(u) & \mathrm{e}^{2u}f_{11}(-u) & 0 & 0 \\ 
0 & 0 & \mathcal{X}_{3}(u) & 0 \\ 
0 & 0 & 0 & \mathcal{X}_{3}(u)%
\end{array}%
\right) ,  \label{A(3)gen1a}
\end{equation}%
with the constraint equation%
\begin{equation}
\beta _{12}\beta _{21}=(\beta _{33}+\beta _{11}-2)(\beta _{33}-\beta
_{11}-2).  \label{A(3)gen2a}
\end{equation}%
We get $\mathcal{K}_{13}^{I}$ as follows%
\begin{equation}
\mathcal{K}_{13}^{I}=\left( 
\begin{array}{cccc}
f_{11}(u) & 0 & h_{13}(u) & 0 \\ 
0 & \mathcal{Y}_{2}^{(1)}(u) & 0 & 0 \\ 
h_{31}(u) & 0 & \mathrm{e}^{2u}f_{11}(-u) & 0 \\ 
0 & 0 & 0 & \mathcal{X}_{4}(u)%
\end{array}%
\right) ,  \label{A(3)gen3a}
\end{equation}%
where the constraint equation is%
\begin{eqnarray}
\beta _{13}\beta _{31} &=&(\beta _{44}+\beta _{11}-2)(\beta _{44}-\beta
_{11}-2)=(\beta _{22}+\beta _{11}-2)(\beta _{22}-\beta _{11}).  \notag \\
&&  \label{A(3)gen4a}
\end{eqnarray}%
The matrix $\mathcal{K}_{14}^{I}$ is%
\begin{equation}
\mathcal{K}_{14}^{I}=\left( 
\begin{array}{cccc}
f_{11}(u) & 0 & 0 & h_{14}(u) \\ 
0 & \mathcal{Y}_{2}^{(1)}(u) & 0 & 0 \\ 
0 & 0 & \mathcal{Y}_{2}^{(1)}(u) & 0 \\ 
h_{41}(u) & 0 & 0 & \mathrm{e}^{2u}f_{11}(-u)%
\end{array}%
\right) ,  \label{A(3)gen5a}
\end{equation}%
with%
\begin{equation}
\beta _{14}\beta _{41}=(\beta _{22}+\beta _{11}-2)(\beta _{22}-\beta _{11}).
\label{A(3)gen6a}
\end{equation}%
The remaining $K$-matrices of type $I$ are given by (\ref{A(n)gen12}). We
obtain $\mathcal{K}_{23}^{I}$ as follows 
\begin{equation}
\mathcal{K}_{23}^{I}=\left( 
\begin{array}{cccc}
\mathcal{Z}_{2}(u) & 0 & 0 & 0 \\ 
0 & f_{22}(u) & h_{23}(u) & 0 \\ 
0 & h_{32}(u) & \mathrm{e}^{2u}f_{22}(-u) & 0 \\ 
0 & 0 & 0 & \mathrm{e}^{2u}\mathcal{Z}_{2}(u)%
\end{array}%
\right) ,  \label{A(3)gen7a}
\end{equation}%
with the constraint equation%
\begin{equation}
\beta _{23}\beta _{32}=(\beta _{11}+\beta _{22})(\beta _{11}-\beta _{22}).
\label{A(3)gen8a}
\end{equation}%
The matrix $\mathcal{K}_{24}^{I}$ is%
\begin{equation}
\mathcal{K}_{24}^{I}=\left( 
\begin{array}{cccc}
\mathcal{Z}_{2}(u) & 0 & 0 & 0 \\ 
0 & f_{22}(u) & 0 & h_{24}(u) \\ 
0 & 0 & \mathcal{Y}_{3}^{(2)}(u) & 0 \\ 
0 & h_{42}(u) & 0 & \mathrm{e}^{2u}f_{22}(-u)%
\end{array}%
\right) ,  \label{A(3)gen9a}
\end{equation}%
where the constraint equation is%
\begin{equation}
\beta _{24}\beta _{42}=(\beta _{11}+\beta _{22})(\beta _{11}-\beta
_{22})=(\beta _{33}+\beta _{22}-2)(\beta _{33}-\beta _{22}),
\label{A(3)gen10a}
\end{equation}%
and $\mathcal{K}_{34}^{I}$ is given by%
\begin{equation}
\mathcal{K}_{34}^{I}=\left( 
\begin{array}{cccc}
\mathcal{Z}_{3}(u) & 0 & 0 & 0 \\ 
0 & \mathcal{Z}_{3}(u) & 0 & 0 \\ 
0 & 0 & f_{33}(u) & h_{34}(u) \\ 
0 & 0 & h_{43}(u) & \mathrm{e}^{2u}f_{33}(-u)%
\end{array}%
\right) ,  \label{A(3)gen11a}
\end{equation}%
with%
\begin{equation}
\beta _{34}\beta _{43}=(\beta _{11}+\beta _{33})(\beta _{11}-\beta _{33}).
\label{A(3)gen12a}
\end{equation}

From (\ref{A(n)gen29}) we get two solutions of type $II${\small a} given by%
\begin{equation}
\mathcal{K}_{12}^{II}=\left( 
\begin{array}{cccc}
f_{11}(u) & h_{12}(u) & 0 & 0 \\ 
h_{21}(u) & \mathrm{e}^{2u}f_{11}(-u) & 0 & 0 \\ 
0 & 0 & \mathrm{e}^{2u}f_{11}(-u) & \mathrm{e}^{u}h_{34}(u) \\ 
0 & 0 & \mathrm{e}^{u}h_{43}(u) & \mathrm{e}^{2u}f_{11}(u)%
\end{array}%
\right) ,  \label{A(3)gen13a}
\end{equation}%
where the non-diagonal matrix elements satisfy the following constraint
equation%
\begin{equation}
\beta _{12}\beta _{21}=\beta _{34}\beta _{43},  \label{A(3)gen14a}
\end{equation}%
and%
\begin{equation}
\mathcal{K}_{14}^{II}=\left( 
\begin{array}{cccc}
f_{11}(u) & 0 & 0 & h_{14}(u) \\ 
0 & f_{11}(u) & h_{23}(u) & 0 \\ 
0 & h_{32}(u) & \mathrm{e}^{2u}f_{11}(-u) & 0 \\ 
h_{41}(u) & 0 & 0 & \mathrm{e}^{2u}f_{11}(-u)%
\end{array}%
\right) ,  \label{A(3)gen15a}
\end{equation}%
with the constraint equation%
\begin{equation}
\beta _{14}\beta _{41}=\beta _{23}\beta _{32}.  \label{A(3)gen16a}
\end{equation}%
Note that both $K$-matrices of type $II${\small a} (\ref{A(3)gen13a}) and (%
\ref{A(3)gen15a}) have four free parameters.

Next we solve these constraint equations and derive eighteen diagonal
solutions. Using the following reductions for the scalar functions $\mathcal{%
X}_{j+1}(u)$, $\mathcal{Y}_{l}^{(i)}(u)$, $\mathcal{Z}_{i}(u)$, namely 
\begin{eqnarray*}
\lim_{\beta _{j+1,j+1}\rightarrow -\beta _{11}+2}\mathcal{X}_{j+1}(u) &=&%
\mathrm{e}^{2u}f_{11}(-u), \\
\lim_{\beta _{j+1,j+1}\rightarrow \beta _{11}+2}\mathcal{X}_{j+1}(u) &=&%
\mathrm{e}^{2u}f_{11}(u),
\end{eqnarray*}%
\begin{eqnarray}
\lim_{\beta _{l,l}\rightarrow -\beta _{i,i}+2}\mathcal{Y}_{l}^{(i)}(u) &=&%
\mathrm{e}^{2u}f_{i,i}(-u),\text{ \ \ \ }\lim_{\beta _{l,l}\rightarrow \beta
_{i,i}}\mathcal{Y}_{l}^{(i)}(u)=f_{i,i}(u),  \notag \\
\lim_{\beta _{11}\rightarrow -\beta _{i,i}}\mathcal{Z}_{i}(u) &=&f_{i,i}(-u),%
\text{ \ \ \ \ \ \ \ \ \ }\lim_{\beta _{11}\rightarrow \beta _{i,i}}\mathcal{%
Z}_{i}(u)=f_{i,i}(u),  \notag \\
&&  \label{A(3)gen17a}
\end{eqnarray}%
we can realize that only half of these diagonal solutions are independent:%
\begin{eqnarray}
D_{1} &=&\text{diag}\left( f(u),\mathrm{e}^{2u}f(-u),\mathrm{e}^{2u}f(-u),%
\mathrm{e}^{2u}f(-u)\right) ,  \notag \\
D_{2} &=&\text{diag}\left( f(u),\mathrm{e}^{2u}f(-u),\mathrm{e}^{2u}f(u),%
\mathrm{e}^{2u}f(u)\right) ,  \notag \\
D_{3} &=&\text{diag}\left( f(u),f(u),\mathrm{e}^{2u}f(-u),\mathrm{e}%
^{2u}f(-u)\right) ,  \notag \\
D_{4} &=&\text{diag}\left( f(u),\mathrm{e}^{2u}f(-u),\mathrm{e}^{2u}f(-u),%
\mathrm{e}^{2u}f(u)\right) ,  \notag \\
D_{5} &=&\text{diag}\left( f(u),f(u),\mathrm{e}^{2u}f(-u),\mathrm{e}%
^{2u}f(u)\right) ,  \notag \\
D_{6} &=&\text{diag}\left( f(u),f(u),f(u),\mathrm{e}^{2u}f(-u)\right) , 
\notag \\
D_{7} &=&\text{diag}\left( f(-u),f(u),\mathrm{e}^{2u}f(-u),\mathrm{e}%
^{2u}f(-u)\right) ,  \notag \\
D_{8} &=&\text{diag}\left( f(-u),f(u),f(u),\mathrm{e}^{2u}f(-u)\right) , 
\notag \\
D_{9} &=&\text{diag}\left( f(-u),f(-u),f(u),\mathrm{e}^{2u}f(-u)\right) ,
\label{A(3)gen18a}
\end{eqnarray}%
where we have used a compact notation for the functions $f_{i,i}(u)$, 
\begin{equation}
f_{i,i}(u)\equiv f(u)=\beta (\mathrm{e}^{u}-1)+1  \label{A(3)gen19a}
\end{equation}%
and $\beta $ is the free parameter.

\subsection{The $\mathcal{A}_{4}^{(1)}$ Case}

Considering that this model starts to reveal all the properties featured by
most of the $\mathcal{A}_{n-1}^{(1)}$ reflection $K$-matrices, we will only
quote five solutions of type $II$ and their corresponding constraint
equations found in this case. They have nine non-null matrix elements and
four free parameters:%
\begin{equation*}
\mathcal{K}_{12}^{II}=\left( 
\begin{array}{ccccc}
f_{11}(u) & h_{12}(u) & 0 & 0 & 0 \\ 
h_{21}(u) & \mathrm{e}^{2u}f_{11}(-u) & 0 & 0 & 0 \\ 
0 & 0 & \mathrm{e}^{2u}f_{11}(-u) & 0 & \mathrm{e}^{u}h_{35}(u) \\ 
0 & 0 & 0 & \mathcal{X}_{4}(u) & 0 \\ 
0 & 0 & \mathrm{e}^{u}h_{53}(u) & 0 & \mathrm{e}^{2u}f_{11}(u)%
\end{array}%
\right) ,
\end{equation*}%
\begin{equation}
\beta _{12}\beta _{21}=\beta _{35}\beta _{53}=(\beta _{44}+\beta
_{11}-2)(\beta _{44}-\beta _{11}-2),  \label{A(4)gen1}
\end{equation}%
\begin{equation*}
\mathcal{K}_{13}^{II}=\left( 
\begin{array}{ccccc}
f_{11}(u) & 0 & h_{13}(u) & 0 & 0 \\ 
0 & \mathcal{Y}_{2}^{(1)}(u) & 0 & 0 & 0 \\ 
h_{31}(u) & 0 & \mathrm{e}^{2u}f_{11}(-u) & 0 & 0 \\ 
0 & 0 & 0 & \mathrm{e}^{2u}f_{11}(-u) & \mathrm{e}^{u}h_{45}(u) \\ 
0 & 0 & 0 & \mathrm{e}^{u}h_{54}(u) & \mathrm{e}^{2u}f_{11}(u)%
\end{array}%
\right) ,
\end{equation*}%
\begin{equation}
\beta _{13}\beta _{31}=\beta _{45}\beta _{54}=(\beta _{22}+\beta
_{11}-2)(\beta _{22}-\beta _{11}),  \label{A(4)gen2}
\end{equation}%
\begin{equation*}
\mathcal{K}_{14}^{II}=\left( 
\begin{array}{ccccc}
f_{11}(u) & 0 & 0 & h_{14}(u) & 0 \\ 
0 & f_{11}(u) & h_{23}(u) & 0 & 0 \\ 
0 & h_{32}(u) & \mathrm{e}^{2u}f_{11}(-u) & 0 & 0 \\ 
h_{41}(u) & 0 & 0 & \mathrm{e}^{2u}f_{11}(-u) & 0 \\ 
0 & 0 & 0 & 0 & \mathcal{X}_{5}(u)%
\end{array}%
\right) ,
\end{equation*}%
\begin{equation}
\beta _{14}\beta _{41}=\beta _{23}\beta _{32}=(\beta _{55}+\beta
_{11}-2)(\beta _{55}-\beta _{11}-2),  \label{A(4)gen3}
\end{equation}%
\begin{equation*}
\mathcal{K}_{15}^{II}=\left( 
\begin{array}{ccccc}
f_{11}(u) & 0 & 0 & 0 & h_{15}(u) \\ 
0 & f_{11}(u) & 0 & h_{24}(u) & 0 \\ 
0 & 0 & \mathcal{Y}_{3}^{(1)}(u) & 0 & 0 \\ 
0 & h_{42}(u) & 0 & \mathrm{e}^{2u}f_{11}(-u) & 0 \\ 
h_{51}(u) & 0 & 0 & 0 & \mathrm{e}^{2u}f_{11}(-u)%
\end{array}%
\right) ,
\end{equation*}%
\begin{equation}
\beta _{15}\beta _{51}=\beta _{24}\beta _{42}=(\beta _{33}+\beta
_{11}-2)(\beta _{33}-\beta _{11}),  \label{A(4)gen4}
\end{equation}%
\begin{equation*}
\mathcal{K}_{25}^{II}=\left( 
\begin{array}{ccccc}
\mathcal{Z}_{2}(u) & 0 & 0 & 0 & 0 \\ 
0 & f_{22}(u) & 0 & 0 & h_{25}(u) \\ 
0 & 0 & f_{22}(u) & h_{34}(u) & 0 \\ 
0 & 0 & h_{43}(u) & \mathrm{e}^{2u}f_{22}(-u) & 0 \\ 
0 & h_{52}(u) & 0 & 0 & \mathrm{e}^{2u}f_{22}(-u)%
\end{array}%
\right) ,
\end{equation*}%
\begin{equation}
\beta _{25}\beta _{52}=\beta _{34}\beta _{43}=(\beta _{11}+\beta
_{22})(\beta _{11}-\beta _{22}).  \label{A(4)gen5}
\end{equation}

The corresponding diagonal solutions have one free parameter.

\subsection{The $\mathcal{B}_{1}^{(1)}$ Case}

We have one general solution with three free parameters, $\beta _{12}$, $%
\beta _{13}$, $\beta _{23}$. The $\mathcal{B}_{1}^{(1)}$ $K$-matrix takes
the form%
\begin{equation}
K^{-}(u)=\left( 
\begin{array}{ccc}
k_{11} & k_{12} & k_{13} \\ 
k_{21} & k_{22} & k_{23} \\ 
k_{31} & k_{32} & k_{33}%
\end{array}%
\right) ,  \label{B(1)gen1}
\end{equation}%
and by analyzing the functional equations we notice that the relations (\ref%
{B(n),A(2n)gen7}) vanish in this case. Thus we do not have the simplified
structure for non-diagonal matrix elements in terms of the function $%
\mathcal{G}^{(\pm )}(u)$ given by (\ref{B(n),A(2n)gen8}). After solving the
functional equations, we derived the following non-diagonal entries%
\begin{eqnarray}
k_{21}(u) &=&\frac{\beta _{21}}{\beta _{12}}k_{12}(u),\text{ \ \ }%
k_{12}(u)=\left( \frac{\sqrt{q}\beta _{23}(\mathrm{e}^{u}-1)+\beta _{12}(q%
\mathrm{e}^{u}-1)}{q\mathrm{e}^{2u}-1}\right) \frac{k_{13}(u)}{\beta _{13}},
\notag \\
k_{32}(u) &=&\frac{\beta _{21}}{\beta _{12}}k_{23}(u),\text{ \ \ }%
k_{23}(u)=\left( \frac{\sqrt{q}\beta _{12}(\mathrm{e}^{u}-1)+\beta _{23}(q%
\mathrm{e}^{u}-1)}{q\mathrm{e}^{2u}-1}\right) \frac{\mathrm{e}^{u}k_{13}(u)}{%
\beta _{13}},  \notag \\
k_{31}(u) &=&\left( \frac{\beta _{21}}{\beta _{12}}\right) ^{2}k_{13}(u),
\label{B(1)gen2}
\end{eqnarray}%
where%
\begin{equation}
\beta _{21}=-\sqrt{q}\left( \frac{(q-1)\beta _{12}\beta _{23}-2(q+1)\beta
_{13}}{q^{2}-1}\right) \frac{\beta _{12}}{\beta _{13}^{2}}.  \label{B(1)gen3}
\end{equation}%
Similarly, we have no recurrent relation like (\ref{B(n),A(2n)gen18}) for
the diagonal matrix elements. However, by performing a direct computation we
are able to identify the following diagonal entries which read%
\begin{eqnarray}
k_{11}(u) &=&2\frac{k_{13}(u)}{\beta _{13}(\mathrm{e}^{2u}-1)}  \notag \\
&&-\left( \frac{\sqrt{q}[(q\mathrm{e}^{u}-1)\beta _{12}^{2}+(\mathrm{e}%
^{u}-q)\beta _{23}^{2}]+(q+1)(q\mathrm{e}^{u}-1)\beta _{12}\beta _{23}}{%
(q+1)(q\mathrm{e}^{2u}-1)(\mathrm{e}^{u}+1)}\right)  \notag \\
&&\times \frac{k_{13}(u)}{\beta _{13}^{2}},  \notag \\
k_{22}(u) &=&-2\frac{(\mathrm{e}^{2u}-q)k_{13}(u)}{\beta _{13}(q-1)(\mathrm{e%
}^{2u}-1)}  \notag \\
&&+\frac{1}{(q+1)(q\mathrm{e}^{2u}-1)(\mathrm{e}^{u}+1)}\{\sqrt{q}\mathrm{e}%
^{u}(q\mathrm{e}^{u}-1)(\beta _{12}^{2}+\beta _{23}^{2})  \notag \\
&&+[(\mathrm{e}^{u}+q)(q\mathrm{e}^{2u}-1)+2q\mathrm{e}^{u}(\mathrm{e}%
^{u}-1)]\beta _{12}\beta _{23}\}\frac{k_{13}(u)}{\beta _{13}^{2}},  \notag \\
k_{33}(u) &=&2\frac{\mathrm{e}^{2u}k_{13}(u)}{\beta _{13}(\mathrm{e}^{2u}-1)}
\notag \\
&&-\left( \frac{\sqrt{q}[(q\mathrm{e}^{u}-1)\beta _{23}^{2}+(\mathrm{e}%
^{u}-q)\beta _{12}^{2}]+(q+1)(q\mathrm{e}^{u}-1)\beta _{12}\beta _{23}}{%
(q+1)(q\mathrm{e}^{2u}-1)(\mathrm{e}^{u}+1)}\right)  \notag \\
&&\times \frac{\mathrm{e}^{2u}k_{13}(u)}{\beta _{13}^{2}}.  \label{B(1)gen4}
\end{eqnarray}%
This is the $3$-parameter general reflection $K$-matrix for the $\mathcal{B}%
_{1}^{(1)}$ model, also known as Zamolodchikov-Fateev model, which has been
revealed by Inami et al. in \cite{IOZ}. The corresponding diagonal solution
is given by (\ref{B(n)diag1}).

The $\mathcal{B}_{1}^{(1)}$ $K$-matrix may further lead us to two $2$%
-parameter solutions by taking the limit $\beta _{23}=\pm \beta
_{12}\Leftrightarrow \beta _{32}=\pm \beta _{21}$, which will satisfy the
procedure used previously to find the solutions for the $\mathcal{B}%
_{n}^{(1)}$ series.

\subsection{The $\mathcal{A}_{2}^{(2)}$ Case}

Here we get two complex conjugate general solutions with two free
parameters, $\beta _{12}$, $\beta _{13}$, for the Izergin-Korepin model. Let
us begin by first pointing out that the relations (\ref{B(n),A(2n)gen7})
still hold such that%
\begin{equation}
\beta _{23}=\pm \frac{i}{q}\beta _{12},\text{ \ \ \ }\beta _{32}=\pm \frac{i%
}{q}\beta _{21}.  \label{A(2)gen1}
\end{equation}%
In what follows we will consider the case $+\frac{i}{q}$. The non-diagonal
matrix elements can be read from (\ref{B(n),A(2n)gen8}),%
\begin{eqnarray}
k_{12}(u) &=&\beta _{12}\mathcal{G}(u),\text{ \ \ \ }k_{21}(u)=\beta _{21}%
\mathcal{G}(u),\text{ \ \ \ }k_{23}(u)=\beta _{23}\mathrm{e}^{u}\mathcal{G}%
(u),  \notag \\
k_{31}(u) &=&\beta _{31}\mathcal{G}(u),\text{ \ \ \ }k_{32}(u)=\beta _{32}%
\mathcal{G}(u),  \label{A(2)gen2}
\end{eqnarray}%
with%
\begin{equation}
\mathcal{G}(u)=\frac{1}{\beta _{13}}\left( \frac{\sqrt{q}+i}{\sqrt{q}+i%
\mathrm{e}^{u}}\right) k_{13}(u)  \label{A(2)gen3}
\end{equation}%
and $\beta _{31}$ is the last non-diagonal parameter we fix before looking
at the diagonal entries,%
\begin{equation}
\beta _{31}=\left( \frac{\beta _{21}}{\beta _{12}}\right) ^{2}\beta _{13}.
\label{A(2)gen4}
\end{equation}

The special recurrent relations (\ref{B(n),A(2n)gen19}) and (\ref%
{B(n),A(2n)gen20}) are also valid for the diagonal terms%
\begin{eqnarray}
k_{22}(u) &=&k_{11}(u)+(\beta _{22}-\beta _{11})\mathcal{G}(u)-\frac{i}{q}%
\mathcal{L}(u),  \notag \\
k_{33}(u) &=&k_{22}(u)+(\beta _{33}-\beta _{22})\mathrm{e}^{u}\mathcal{G}(u)-%
\sqrt{q}\mathcal{L}(u),  \label{A(2)gen5}
\end{eqnarray}%
where we have defined a new scalar function $\mathcal{L}(u)$ a bit different
from $\mathcal{J}^{(\pm )}(u)$ (\ref{B(n),A(2n)gen20a}) and $\mathcal{F}%
^{(\pm )}(u)$ (\ref{B(n),A(2n)gen21}), given by%
\begin{equation}
\mathcal{L}(u)=\frac{\sqrt{q}\beta _{13}\beta _{21}}{\beta _{12}}\left( 
\frac{\mathrm{e}^{u}-1}{\sqrt{q}+i}\right) \mathcal{G}(u).  \label{A(2)gen6}
\end{equation}%
Now, the functional equations selected from block $B[4,6]$ close the above
recurrent relations and determine $k_{33}(u)$, namely%
\begin{equation}
k_{33}(u)=\mathrm{e}^{2u}k_{11}(u)+(\beta _{33}-\beta _{11}-2)\mathrm{e}^{u}%
\mathcal{G}(u)\left( \frac{\sqrt{q}+i\mathrm{e}^{u}}{\sqrt{q}+i}\right) .
\label{A(2)gen7}
\end{equation}%
Thus, we obtain the matrix element $k_{11}(u)$ as follows%
\begin{eqnarray}
k_{11}(u) &=&\left( \frac{2\mathrm{e}^{u}-(\beta _{22}-\beta _{11})(\mathrm{e%
}^{u}-1)}{\mathrm{e}^{2u}-1}\right) \mathcal{G}(u)-\left( \frac{\sqrt{q}+%
\frac{i}{q}}{\mathrm{e}^{2u}-1}\right) \mathcal{L}(u)  \notag \\
&&-i\left( \frac{\beta _{33}-\beta _{11}-2}{\sqrt{q}+i}\right) \frac{\mathrm{%
e}^{u}}{\mathrm{e}^{u}+1}\mathcal{G}(u)  \label{A(2)gen8}
\end{eqnarray}%
depending on the diagonal parameters $\beta _{22}$ and $\beta _{33}$ given by%
\begin{eqnarray}
\beta _{22} &=&\beta _{11}+\left( \frac{2q^{3/2}}{q^{3/2}-i}\right) -\left( 
\frac{1+q-iq^{3/2}}{\sqrt{q}}\right) \frac{\beta _{13}\beta _{21}}{\beta
_{12}},  \notag \\
\beta _{33} &=&\beta _{11}+2+i\left( \frac{q^{2}+1}{q}\right) \frac{\beta
_{13}\beta _{21}}{\beta _{12}},  \label{A(2)gen9}
\end{eqnarray}%
with%
\begin{equation}
\beta _{21}=\left( \frac{2iq}{(\sqrt{q}+i)(q^{3/2}-i)}\right) \frac{\beta
_{12}}{\beta _{13}}-\left( \frac{i}{\sqrt{q}(q+1)}\right) \frac{\beta
_{12}^{2}}{\beta _{13}^{2}}.  \label{A(2)gen10}
\end{equation}

We observe that the major differences between the cases $\mathcal{A}%
_{2n}^{(2)}$ $(n>1)$ and $\mathcal{A}_{2}^{(2)}$ are due to our previous
choice of the free parameters. It means that we cannot take the limit $%
n\rightarrow 1$ into the $\mathcal{A}_{2n}^{(2)}$ general solutions in order
to get the $\mathcal{A}_{2}^{(2)}$ $K$-matrices. The $\mathcal{A}_{2}^{(2)}$
diagonal solutions are given by (\ref{A(2n)diag1}) and a second type of the $%
\mathcal{A}_{2}^{(2)}$ solution\ with two free parameters is exhibited in
Section 4.1 (\ref{A(2n)red3}).

Finally, we remark that although there is an apparent simplification in
these calculations compared to those developed in \cite{Li1}, after
substituting the fixed parameters the final form of this solution still
remained cumbersome. However, there is an equivalent solution for this model
derived by Nepomechie in \cite{Ne5} which looks simpler than our one.

\subsection{The $\mathcal{C}_{1}^{(1)}$, $\mathcal{D}_{1}^{(1)},$ and $%
\mathcal{A}_{1}^{(2)}$ Cases}

These models share the same general $K$-matrix with three free parameters, $%
\beta _{11}$, $\beta _{12}$, $\beta _{21}$, given by%
\begin{equation}
K^{-}(u)=\left( 
\begin{array}{cc}
1+\beta _{11}(\mathrm{e}^{u}-1) & \frac{1}{2}\beta _{12}(\mathrm{e}^{2u}-1)
\\ 
\frac{1}{2}\beta _{21}(\mathrm{e}^{2u}-1) & \mathrm{e}^{2u}-\beta _{11}%
\mathrm{e}^{u}(\mathrm{e}^{u}-1)%
\end{array}%
\right)  \label{C(1),D(1),A(1)gen}
\end{equation}%
and also share two diagonal solutions, namely the identity and the one
obtained by setting $\beta _{12}=\beta _{21}=0$ in (\ref{C(1),D(1),A(1)gen}).

\subsection{The $\mathcal{C}_{2}^{(1)}$ Case}

We have one general solution with three free parameters, $\beta _{12}$, $%
\beta _{13}$, $\beta _{14}$. The $\mathcal{C}_{2}^{(1)}$ $K$-matrix takes
the form%
\begin{equation}
K^{-}(u)=\left( 
\begin{array}{cccc}
k_{11} & k_{12} & k_{13} & k_{14} \\ 
k_{21} & k_{22} & k_{23} & \frac{\mathrm{e}^{u}}{q^{2}}k_{13} \\ 
k_{31} & k_{32} & \mathrm{e}^{u}k_{11} & -\mathrm{e}^{u}k_{12} \\ 
k_{41} & \frac{\mathrm{e}^{u}}{q^{2}}k_{31} & -\mathrm{e}^{u}k_{21} & 
\mathrm{e}^{u}k_{22}%
\end{array}%
\right) ,  \label{C(2)gen1}
\end{equation}%
where the following remaining diagonal entries are given by%
\begin{equation}
k_{11}(u)=1-\frac{\beta _{12}\beta _{13}}{q^{2}\beta _{14}}f(u),\text{ \ \ \ 
}k_{22}(u)=1+\frac{\beta _{12}\beta _{13}}{\beta _{14}}f(u),
\label{C(2)gen2}
\end{equation}%
and the non-diagonal matrix elements are%
\begin{eqnarray}
k_{12}(u) &=&\beta _{12}f(u),\text{ \ \ \ }k_{13}(u)=\beta _{13}f(u),\text{
\ \ \ }k_{14}(u)=\beta _{14}(\mathrm{e}^{u}-1),  \notag \\
k_{21}(u) &=&-\frac{\beta _{13}}{q^{2}\beta _{14}}\Gamma f(u),\text{ \ \ \ }%
k_{23}(u)=\frac{\beta _{13}}{q^{2}\beta _{12}}\Gamma (\mathrm{e}^{u}-1), 
\notag \\
k_{31}(u) &=&\frac{\beta _{12}}{\beta _{14}}\Gamma f(u),\text{ \ \ \ }%
k_{32}(u)=-\frac{\beta _{12}}{\beta _{13}}\Gamma (\mathrm{e}^{u}-1),  \notag
\\
k_{41}(u) &=&-\frac{\Gamma ^{2}}{q^{2}\beta _{14}}(\mathrm{e}^{u}-1).
\label{C(2)gen3}
\end{eqnarray}%
Here we have defined a new scalar function $f(u)$ as%
\begin{equation}
f(u)=\left( \frac{q^{2}+1}{q^{2}+\mathrm{e}^{u}}\right) (\mathrm{e}^{u}-1)%
\text{ \ and\ \ }\Gamma =\frac{\beta _{12}\beta _{13}}{\beta _{14}}-\frac{%
q^{2}}{q^{2}+1}.  \label{C(2)gen4}
\end{equation}%
This solution can be identified with the $2$-parameter general reflection $K$%
-matrix with $\mathcal{G}^{(+)}(u)$ after an appropriate choice of $\beta
_{12}$. In addition, by applying the reduction procedure we get the $1$%
-parameter solution given by (\ref{C(n),D(n),A(2n-1)red30}) as well as the $%
1 $-parameter diagonal matrix $\mathbb{K}_{\beta }$ (\ref{C(n)diag1}).

\subsection{The $\mathcal{D}_{2}^{(1)}$ Case}

The structure of the $\mathcal{D}_{2}^{(1)}$ general $K$-matrix shows that
this is a very special case since we have found no solution which possesses
all matrix elements different from zero. We have one $1$-parameter solution
given by%
\begin{equation}
K^{-}(u)=\left( 
\begin{array}{cccc}
\mathrm{e}^{-u}\frac{q^{2}-\mathrm{e}^{2u}}{q^{2}-1} & 0 & 0 & 0 \\ 
0 & \mathrm{e}^{u} & \frac{1}{2}\beta (\mathrm{e}^{2u}-1) & 0 \\ 
0 & \frac{2q^{2}(\mathrm{e}^{2u}-1)}{\beta (q^{2}-1)^{2}} & \mathrm{e}^{u} & 
0 \\ 
0 & 0 & 0 & \mathrm{e}^{u}\frac{q^{2}-\mathrm{e}^{2u}}{q^{2}-1}%
\end{array}%
\right) ,  \label{D(2)gen}
\end{equation}%
where $\beta $ is the free parameter, and one $1$-parameter reduced solution
previously presented (\ref{C(n),D(n),A(2n-1)red29}). Here we also get the
identity and two $2$-parameter diagonal matrices $\mathbb{K}_{\alpha \beta }$
(\ref{D(n)diag1}).

\subsection{The $\mathcal{A}_{3}^{(2)}$ Case}

We have one general solution with four free parameters, $\beta _{12}$, $%
\beta _{13}$, $\beta _{14}$ and $\beta _{24}$. The $\mathcal{A}_{3}^{(2)}$ $%
K $-matrix takes the form%
\begin{equation}
K^{-}(u)=\left( 
\begin{array}{cccc}
k_{11} & k_{12} & k_{13} & k_{14} \\ 
k_{21} & k_{22} & k_{23} & k_{24} \\ 
k_{31} & k_{32} & k_{33} & k_{34} \\ 
k_{41} & k_{42} & k_{43} & k_{44}%
\end{array}%
\right) ,  \label{A(3)gen1}
\end{equation}%
where the normalized diagonal entries are given by%
\begin{eqnarray}
k_{22}(u) &=&\mathrm{e}^{u}+\frac{\beta _{12}\mathrm{e}^{u}(\mathrm{e}^{u}-1)%
}{2q^{2}\beta _{14}\beta _{24}(q^{2}+\mathrm{e}^{2u})}\{\beta _{24}\left[
q^{2}\left( \beta _{13}+\beta _{24}\right) -\beta _{13}\right] (\mathrm{e}%
^{u}+q^{2})  \notag \\
&&+\beta _{13}^{2}(q^{2}\mathrm{e}^{u}+1)\},  \notag \\
k_{33}(u) &=&\mathrm{e}^{u}-\frac{\beta _{12}\mathrm{e}^{u}(\mathrm{e}^{u}-1)%
}{2q^{2}\beta _{14}\beta _{24}(q^{2}+\mathrm{e}^{2u})}\{\beta _{13}\left[
\beta _{13}+\beta _{24}-q^{2}\beta _{24}\right] (\mathrm{e}^{u}+q^{2}) 
\notag \\
&&+\beta _{24}^{2}q^{2}(q^{2}\mathrm{e}^{u}+1)\},  \notag \\
k_{44}(u) &=&\mathrm{e}^{u}+\frac{\beta _{12}\mathrm{e}^{u}(\mathrm{e}^{u}-1)%
}{2q^{2}\beta _{14}\beta _{24}(q^{2}+\mathrm{e}^{2u})}\{\beta _{13}\beta
_{24}\left[ (\mathrm{e}^{u}+q^{2})^{2}-q^{2}(\mathrm{e}^{2u}-1)\right] 
\notag \\
&&+(\beta _{13}^{2}-q^{2}\beta _{24}^{2})\mathrm{e}^{u}(\mathrm{e}%
^{u}+q^{2})\},  \label{A(3)gen2}
\end{eqnarray}%
and the non-diagonal matrix elements are%
\begin{eqnarray}
k_{12}(u) &=&\frac{\beta _{12}}{2\beta _{24}}f(u),\text{ \ \ \ }k_{13}(u)=%
\frac{1}{2}g(u),\text{ \ \ \ }k_{14}(u)=\frac{1}{2}\beta _{14}(\mathrm{e}%
^{2u}-1),  \notag \\
k_{21}(u) &=&-\frac{1}{2}\Omega f(u),\text{ \ \ \ }k_{23}(u)=\frac{\beta
_{14}\beta _{24}}{2\beta _{12}}\Omega (\mathrm{e}^{2u}-1),\text{ \ \ \ }%
k_{24}(u)=\frac{1}{2}\mathrm{e}^{u}f(u),  \notag \\
k_{31}(u) &=&\frac{\beta _{12}}{2q^{2}\beta _{24}}\Omega g(u),\text{ \ \ \ }%
k_{32}(u)=-\frac{\beta _{12}\beta _{14}}{2q^{2}\beta _{24}}\Omega (\mathrm{e}%
^{2u}-1),  \notag \\
k_{34}(u) &=&-\frac{\beta _{12}}{2q^{2}\beta _{24}}\mathrm{e}^{u}g(u), 
\notag \\
k_{41}(u) &=&-\frac{\beta _{14}}{2q^{2}}\Omega ^{2}(\mathrm{e}^{2u}-1),\text{
\ \ \ }k_{42}(u)=\frac{\beta _{12}}{2q^{2}\beta _{24}}\Omega \mathrm{e}%
^{u}f(u),  \notag \\
k_{43}(u) &=&\frac{1}{2q^{2}}\Omega \mathrm{e}^{u}g(u).  \label{A(3)gen3}
\end{eqnarray}%
Here we have defined two scalar functions $f(u)$ and $g(u)$ which are
different from $\mathcal{G}^{(\pm )}(u)$:%
\begin{eqnarray}
f(u) &=&[\beta _{13}(\mathrm{e}^{u}-1)+\beta _{24}(\mathrm{e}%
^{u}+q^{2})]\left( \frac{\mathrm{e}^{2u}-1}{\mathrm{e}^{2u}+q^{2}}\right) , 
\notag \\
g(u) &=&[\beta _{13}(\mathrm{e}^{u}+q^{2})-q^{2}\beta _{24}(\mathrm{e}%
^{u}-1)]\left( \frac{\mathrm{e}^{2u}-1}{\mathrm{e}^{2u}+q^{2}}\right) ,
\label{A(3)gen4}
\end{eqnarray}%
and%
\begin{equation}
\Omega =\frac{\beta _{12}\beta _{13}(q^{2}+1)-2q^{2}\beta _{14}}{\beta
_{14}^{2}(q^{2}+1)}.  \label{A(3)gen5}
\end{equation}

The above solution can be regarded as the most general reflection $K$-matrix
because the $2$-parameter solutions with $\mathcal{G}^{(\pm )}(u)$ turn out
to be obtained by assigning specific values to $\beta _{12}$ and $\beta
_{24} $. Furthermore, by applying the reduction procedure we get one $1$%
-parameter solution (\ref{C(n),D(n),A(2n-1)red29}) as well as two $1$%
-parameter diagonal matrices $\mathbb{K}_{\beta }$ given by (\ref%
{A(2n-1)diag1}).

\subsection{The $\mathcal{C}_{3}^{(1)}$ Case with $\mathcal{G}^{(+)}(u)$}

In Section 3.2, we have found one $3$-parameter general solution with $%
\mathcal{G}^{(-)}(u)$ for this model. The reduction procedure gives us
another solution with $\mathcal{G}^{(+)}(u)$ which also has three free
parameters. The corresponding $K$-matrix takes the form%
\begin{equation}
K^{-}(u)=\left( 
\begin{array}{cccccc}
k_{11} & k_{12} & 0 & 0 & k_{15} & k_{16} \\ 
k_{21} & k_{22} & 0 & 0 & k_{25} & k_{26} \\ 
0 & 0 & k_{33} & 0 & 0 & 0 \\ 
0 & 0 & 0 & k_{44} & 0 & 0 \\ 
k_{51} & k_{52} & 0 & 0 & k_{55} & k_{56} \\ 
k_{61} & k_{62} & 0 & 0 & k_{65} & k_{66}%
\end{array}%
\right)  \label{C(3)gen1}
\end{equation}%
with the non-normalized diagonal matrix elements given by%
\begin{eqnarray}
k_{11}(u) &=&\frac{2\mathrm{e}^{u}\mathcal{G}^{(+)}(u)}{\mathrm{e}^{2u}-1} 
\notag \\
&&-\left( \frac{\beta _{12}\beta _{15}}{\beta _{16}}\frac{1+q^{3}}{q^{3}}-%
\frac{q\beta _{21}\beta _{16}[(1+q^{2})\mathrm{e}^{u}-q^{2}(1+q^{4})]}{\beta
_{15}(1+q^{3})}\right) \frac{\mathcal{G}^{(+)}(u)}{\mathrm{e}^{u}+1},  \notag
\\
k_{22}(u) &=&\frac{2\mathrm{e}^{u}\mathcal{G}^{(+)}(u)}{\mathrm{e}^{2u}-1} 
\notag \\
&&+\left( \frac{\beta _{12}\beta _{15}}{\beta _{16}}\frac{(1+q^{3})\mathrm{e}%
^{u}}{q^{3}}+\frac{q\beta _{21}\beta _{16}[(1+q^{2})\mathrm{e}%
^{u}-q^{2}(1+q^{4})]}{\beta _{15}(1+q^{3})}\right)  \notag \\
&&\times \frac{\mathcal{G}^{(+)}(u)}{\mathrm{e}^{u}+1},  \notag \\
k_{33}(u) &=&k_{44}(u)=\frac{2\mathrm{e}^{u}\mathcal{G}^{(+)}(u)}{\mathrm{e}%
^{2u}-1}  \notag \\
&&+\left( \frac{\beta _{12}\beta _{15}}{\beta _{16}}\frac{(1+q^{3})\mathrm{e}%
^{u}}{q^{3}}+\frac{\beta _{21}\beta _{16}[(1+q)(\mathrm{e}^{u}-q^{6})+(%
\mathrm{e}^{u}+q^{3})^{2}]}{\beta _{15}(1+q^{3})}\right)  \notag \\
&&\times \frac{\mathcal{G}^{(+)}(u)}{\mathrm{e}^{u}+1},  \notag \\
k_{55}(u) &=&\frac{2\mathrm{e}^{u}\mathcal{G}^{(+)}(u)}{\mathrm{e}^{2u}-1} 
\notag \\
&&+\left( \frac{\beta _{12}\beta _{15}}{\beta _{16}}\frac{1+q^{3}}{q^{3}}+%
\frac{q\beta _{21}\beta _{16}[(1+q^{4})(\mathrm{e}^{u}+q^{3})+(1+q)(1+q^{3})]%
}{\beta _{15}(1+q^{3})}\right)  \notag \\
&&\times \frac{\mathrm{e}^{u}\mathcal{G}^{(+)}(u)}{\mathrm{e}^{u}+1},  \notag
\\
k_{66}(u) &=&\frac{2\mathrm{e}^{3u}\mathcal{G}^{(+)}(u)}{\mathrm{e}^{2u}-1}%
-\left( \frac{\beta _{12}\beta _{15}}{\beta _{16}}\frac{(1+q^{3})\mathrm{e}%
^{u}}{q^{3}}\right.  \notag \\
&&\left. +\frac{q\beta _{21}\beta _{16}[(1+q^{2})(\mathrm{e}%
^{u}+q^{3})+q^{2}(1+q)(1+q^{3})\mathrm{e}^{u}]}{\beta _{15}(1+q^{3})}\right) 
\frac{\mathrm{e}^{u}\mathcal{G}^{(+)}(u)}{\mathrm{e}^{u}+1},  \notag \\
&&  \label{C(3)gen2}
\end{eqnarray}%
and the non-diagonal entries are%
\begin{eqnarray}
k_{12}(u) &=&\beta _{12}\mathcal{G}^{(+)}(u),\text{ \ \ }k_{15}(u)=\beta
_{15}\mathcal{G}^{(+)}(u),\text{ \ \ }k_{16}(u)=\beta _{16}\frac{\mathrm{e}%
^{u}+q^{3}}{1+q^{3}}\mathcal{G}^{(+)}(u),  \notag \\
k_{21}(u) &=&\beta _{21}\mathcal{G}^{(+)}(u),\text{\ \ \ }k_{25}(u)=-\beta
_{21}\frac{\beta _{16}}{\beta _{12}}\frac{\mathrm{e}^{u}+q^{3}}{1+q^{3}}%
\mathcal{G}^{(+)}(u),  \notag \\
k_{26}(u) &=&\frac{\beta _{15}}{q^{3}}\mathrm{e}^{u}\mathcal{G}^{(+)}(u), 
\notag \\
k_{51}(u) &=&-\beta _{21}\frac{q^{4}\beta _{12}}{\beta _{15}}\mathcal{G}%
^{(+)}(u),\text{\ \ \ }k_{52}(u)=\beta _{21}\frac{q^{4}\beta _{12}\beta _{16}%
}{\beta _{15}^{2}}\frac{\mathrm{e}^{u}+q^{3}}{1+q^{3}}\mathcal{G}^{(+)}(u), 
\notag \\
k_{56}(u) &=&-q\beta _{12}\mathrm{e}^{u}\mathcal{G}^{(+)}(u),  \notag \\
k_{61}(u) &=&-\beta _{21}^{2}\frac{q^{4}\beta _{16}}{\beta _{15}^{2}}\frac{%
\mathrm{e}^{u}+q^{3}}{1+q^{3}}\mathcal{G}^{(+)}(u),\text{ \ \ }%
k_{62}(u)=-\beta _{21}\frac{q\beta _{12}}{\beta _{15}}\mathrm{e}^{u}\mathcal{%
G}^{(+)}(u),  \notag \\
k_{65}(u) &=&-q\beta _{21}\mathrm{e}^{u}\mathcal{G}^{(+)}(u),
\label{C(3)gen3}
\end{eqnarray}%
where%
\begin{equation}
\mathcal{G}^{(+)}(u)=\frac{1}{\beta _{16}}\left( \frac{1+q^{3}}{\mathrm{e}%
^{u}+q^{3}}\right) k_{16}(u)\text{ \ and \ }\beta _{21}=-\frac{\beta
_{12}\beta _{15}^{2}}{q^{3}\beta _{16}^{2}}+\frac{2}{(q+1)(q^{3}+1)}\frac{%
\beta _{15}}{\beta _{16}}.  \label{C(3)gen4}
\end{equation}

For $n>3$, this type of solution follows the classification scheme presented
in Section 4.2.

\subsection{The $\mathcal{D}_{3}^{(1)}$ Case with $\mathcal{G}^{(-)}(u)$}

The corresponding $\mathcal{D}_{3}^{(1)}$ general solution is given in terms
of $\mathcal{G}^{(+)}(u)$ and has three free parameters. Here we have
another $3$-parameter solution in terms of $\mathcal{G}^{(-)}(u)$ possessing
the same form given by (\ref{C(3)gen1}) with the following non-normalized
diagonal entries%
\begin{eqnarray}
k_{11}(u) &=&\left( \frac{2(\mathrm{e}^{u}-q)}{(1-q)(\mathrm{e}^{u}-1)}+%
\frac{(1+q^{2})\beta _{12}\beta _{15}}{q\beta _{16}}\right) \frac{\mathcal{G}%
^{(-)}(u)}{\mathrm{e}^{u}+1},  \notag \\
k_{22}(u) &=&\left( \frac{2(\mathrm{e}^{u}-q)}{(1-q)(\mathrm{e}^{u}-1)}+%
\frac{[\mathrm{e}^{u}(q-1)+q(1+q)]\beta _{12}\beta _{15}}{q\beta _{16}}%
\right) \frac{\mathcal{G}^{(-)}(u)}{\mathrm{e}^{u}+1},  \notag \\
k_{33}(u) &=&k_{44}(u)=\left( \frac{2(\mathrm{e}^{u}-q)^{2}}{(1-q^{2})(%
\mathrm{e}^{u}-1)}+\frac{(\mathrm{e}^{u}-q^{2})\beta _{12}\beta _{15}}{%
q\beta _{16}}\right) \frac{\mathrm{e}^{u}+q}{1-q}\frac{\mathcal{G}^{(-)}(u)}{%
\mathrm{e}^{u}+1},  \notag \\
k_{55}(u) &=&\left( \frac{2(\mathrm{e}^{u}-q)\mathrm{e}^{u}}{(1-q)(\mathrm{e}%
^{u}-1)}+\frac{[\mathrm{e}^{u}(q+1)+q(1-q)]\beta _{12}\beta _{15}}{q\beta
_{16}}\right) \frac{\mathrm{e}^{u}\mathcal{G}^{(-)}(u)}{\mathrm{e}^{u}+1}, 
\notag \\
k_{66}(u) &=&\mathrm{e}^{2u}k_{11}(u),  \label{D(3)gen1}
\end{eqnarray}%
and the non-diagonal terms are given by%
\begin{eqnarray}
k_{12}(u) &=&\beta _{12}\mathcal{G}^{(-)}(u),\text{ \ \ }k_{15}(u)=\beta
_{15}\mathcal{G}^{(-)}(u),\text{ \ \ }k_{16}(u)=\beta _{16}\frac{\mathrm{e}%
^{u}-q}{1-q}\mathcal{G}^{(-)}(u),  \notag \\
k_{21}(u) &=&\beta _{21}\mathcal{G}^{(-)}(u),\text{\ \ \ }k_{25}(u)=-\beta
_{21}\frac{\beta _{16}}{\beta _{12}}\frac{\mathrm{e}^{u}-q}{1-q}\mathcal{G}%
^{(-)}(u),  \notag \\
k_{26}(u) &=&-\frac{\beta _{15}}{q}\mathrm{e}^{u}\mathcal{G}^{(-)}(u), 
\notag \\
k_{51}(u) &=&\beta _{21}\frac{q^{2}\beta _{12}}{\beta _{15}}\mathcal{G}%
^{(-)}(u),\text{\ \ \ }k_{52}(u)=-\beta _{21}\frac{q^{2}\beta _{12}\beta
_{16}}{\beta _{15}^{2}}\frac{\mathrm{e}^{u}-q}{1-q}\mathcal{G}^{(-)}(u), 
\notag \\
k_{56}(u) &=&-q\beta _{12}\mathrm{e}^{u}\mathcal{G}^{(-)}(u),  \notag \\
k_{61}(u) &=&\beta _{21}^{2}\frac{q^{2}\beta _{16}}{\beta _{15}^{2}}\frac{%
\mathrm{e}^{u}-q}{1-q}\mathcal{G}^{(-)}(u),\text{ \ \ }k_{62}(u)=-\beta _{21}%
\frac{q\beta _{12}}{\beta _{15}}\mathrm{e}^{u}\mathcal{G}^{(-)}(u),  \notag
\\
k_{65}(u) &=&-q\beta _{21}\mathrm{e}^{u}\mathcal{G}^{(-)}(u),
\label{D(3)gen2}
\end{eqnarray}%
where%
\begin{equation}
\mathcal{G}^{(-)}(u)=\frac{1}{\beta _{16}}\left( \frac{1-q}{\mathrm{e}^{u}-q}%
\right) k_{16}(u)\text{ \ and \ }\beta _{21}=\frac{\beta _{12}\beta _{15}^{2}%
}{q\beta _{16}^{2}}-\frac{2}{q^{2}-1}\frac{\beta _{15}}{\beta _{16}}.
\label{D(3)gen3}
\end{equation}%
We point out that this type of solution has $n-1$ free parameters for $n>3$
and follows the classification scheme presented in Section 4.2.

\subsection{The $\mathcal{D}_{2}^{(2)}$ Case}

We get one $3$-parameter general $K$-matrix for the $\mathcal{D}_{2}^{(2)}$
model,%
\begin{equation}
K^{-}(u)=\left( 
\begin{array}{cccc}
k_{11} & k_{12} & k_{13} & k_{14} \\ 
k_{21} & k_{22} & k_{23} & k_{24} \\ 
k_{31} & k_{32} & k_{33} & k_{34} \\ 
k_{41} & k_{42} & k_{43} & k_{44}%
\end{array}%
\right) ,  \label{D(2)gen1}
\end{equation}%
whose entries $k_{11}$, $k_{44}$ and $k_{22}=k_{33}$, $k_{23}$, $k_{32}$ are
directly read from the odd $n$ solution by taking $n=1$ into (\ref%
{D(n+1)gen21}), (\ref{D(n+1)gen22}) and (\ref{D(n+1)gen26}), (\ref%
{D(n+1)gen27}), (\ref{D(n+1)gen28}), respectively, and expressed in the
following form%
\begin{eqnarray}
k_{11}(u) &=&\frac{1}{2}\frac{[2(\mathrm{e}^{2u}-q)(q\beta _{-}^{2}+\beta
_{+}^{2})+(q+1)(\mathrm{e}^{2u}+1)(q\beta _{-}^{2}-\beta _{+}^{2})]k_{14}(u)%
}{\beta _{14}^{2}\sqrt{q}(q+1)(\mathrm{e}^{2u}+1)^{2}}  \notag \\
&&+2\frac{k_{14}(u)}{\beta _{14}(\mathrm{e}^{2u}-1)},  \notag \\
k_{44}(u) &=&\frac{1}{2}\frac{[-2(\mathrm{e}^{2u}-q)(q\beta _{-}^{2}+\beta
_{+}^{2})+(q+1)(\mathrm{e}^{2u}+1)(q\beta _{-}^{2}-\beta _{+}^{2})]\mathrm{e}%
^{2u}k_{14}(u)}{\beta _{14}^{2}\sqrt{q}(q+1)(\mathrm{e}^{2u}+1)^{2}}  \notag
\\
&&+2\frac{\mathrm{e}^{2u}k_{14}(u)}{\beta _{14}(\mathrm{e}^{2u}-1)},  \notag
\\
k_{22}(u) &=&k_{33}(u)=-\frac{1}{2}\left( \frac{q\beta _{-}^{2}-\beta
_{+}^{2}}{\sqrt{q}(q+1)}\frac{\mathrm{e}^{2u}+q}{\mathrm{e}^{2u}+1}+\frac{%
4\beta _{14}(\mathrm{e}^{2u}-q)}{(q-1)(\mathrm{e}^{2u}-1)}\right) \frac{%
k_{14}(u)}{\beta _{14}^{2}},  \notag \\
k_{23}(u) &=&\frac{\mathrm{e}^{u}}{\mathrm{e}^{2u}+1}\left( \frac{q\beta
_{-}^{2}+\beta _{+}^{2}}{\sqrt{q}(\mathrm{e}^{2u}+1)}\mathrm{e}^{u}-\frac{2%
\sqrt{q}\beta _{-}\beta _{+}}{q+1}\right) \frac{k_{14}(u)}{\beta _{14}^{2}},
\notag \\
k_{32}(u) &=&\frac{\mathrm{e}^{u}}{\mathrm{e}^{2u}+1}\left( \frac{q\beta
_{-}^{2}+\beta _{+}^{2}}{\sqrt{q}(\mathrm{e}^{2u}+1)}\mathrm{e}^{u}+\frac{2%
\sqrt{q}\beta _{-}\beta _{+}}{q+1}\right) \frac{k_{14}(u)}{\beta _{14}^{2}},
\label{D(2)gen2}
\end{eqnarray}%
where $\beta _{\pm }=\beta _{12}\pm \beta _{13}$.

Due to the indetermination of $\Delta _{l}$ (\ref{D(n+1)gen6}) when $n=1$,
we can replace $\Delta _{l}$ into (\ref{D(n+1)gen4}) and (\ref{D(n+1)gen5})
by $\Delta _{l}^{\prime }$ defined as%
\begin{equation}
\Delta _{l}\rightarrow \Delta _{l}^{\prime }=\frac{q^{2}-1}{q^{2}-\mathrm{e}%
^{2u}}\frac{\mathrm{e}^{2u}}{1+\mathrm{e}^{2u}}\frac{1}{\beta
_{13}(b_{1}^{+}+b_{1}^{-})(b_{4}^{+}+b_{4}^{-})}.  \label{D(2)gen3}
\end{equation}

This replacement implies that the equations (\ref{D(n+1)gen9})-(\ref%
{D(n+1)gen13}) now hold for the $\mathcal{D}_{2}^{(2)}$ model up to a $q$%
-factor. The result is%
\begin{eqnarray}
k_{41}(u) &=&\frac{\beta _{21}^{2}}{\beta _{13}^{2}}k_{14}(u),  \notag \\
k_{12}(u) &=&\frac{\mathrm{e}^{u}\beta _{-}+\beta _{+}}{\beta _{14}(\mathrm{e%
}^{2u}+1)}k_{14}(u),\text{ \ \ \ }k_{13}(u)=\frac{-\mathrm{e}^{u}\beta
_{-}+\beta _{+}}{\beta _{14}(\mathrm{e}^{2u}+1)}k_{14}(u),  \notag \\
k_{21}(u) &=&\frac{\beta _{21}}{\beta _{13}}k_{13}(u),\text{ \ \ \ }%
k_{31}(u)=\frac{\beta _{21}}{\beta _{13}}k_{12}(u),  \notag \\
k_{42}(u) &=&\frac{\beta _{21}}{\beta _{13}}k_{34}(u),\text{ \ \ \ }%
k_{43}(u)=\frac{\beta _{21}}{\beta _{13}}k_{24}(u),  \notag \\
k_{24}(u) &=&\frac{1}{\sqrt{q}}\frac{-q\mathrm{e}^{-u}\beta _{-}+\beta _{+}}{%
\beta _{14}(\mathrm{e}^{2u}+1)}\mathrm{e}^{2u}k_{14}(u),  \notag \\
k_{34}(u) &=&\frac{1}{\sqrt{q}}\frac{q\mathrm{e}^{-u}\beta _{-}+\beta _{+}}{%
\beta _{14}(\mathrm{e}^{2u}+1)}\mathrm{e}^{2u}k_{14}(u),  \label{D(2)gen4}
\end{eqnarray}%
where $\beta _{21}$ is given by%
\begin{equation}
\beta _{21}=\frac{1}{2}\frac{(q-1)(q\beta _{-}^{2}-\beta _{+}^{2})+4\sqrt{q}%
(q+1)\beta _{14}}{(q^{2}-1)\beta _{14}^{2}}\beta _{13}.  \label{D(2)gen5}
\end{equation}%
Assigning the suitable normalization to the matrix element $k_{14}(u)$ as
follows 
\begin{equation}
k_{14}(u)=\frac{1}{2}\beta _{14}(\mathrm{e}^{2u}-1)  \label{D(2)gen6}
\end{equation}%
we can find $\beta _{11}$,%
\begin{equation}
\beta _{11}=-\frac{2\sqrt{q}}{q+1}\frac{\beta _{12}\beta _{13}}{\beta _{14}},
\label{D(2)gen7}
\end{equation}%
in order to obtain a regular solution with three free parameters, $\beta
_{12}$, $\beta _{13}$ and $\beta _{14}$.

\section{Conclusion}

We have provided an unifying presentation of our calculations originally
described in the references \cite{Li2}, \cite{Li3}, \cite{Li4}, and \cite{LM}%
. After accomplishing a detailed study of the boundary Yang-Baxter
equations, we achieved the regular reflection $K$-matrices for the quantum $%
\mathcal{R}$-matrices based on non-exceptional affine Lie algebras $\mathcal{%
A}_{n-1}^{(1)},$ $\mathcal{B}_{n}^{(1)},$ $\mathcal{C}_{n}^{(1)},$ $\mathcal{%
D}_{n}^{(1)},$ $\mathcal{A}_{2n}^{(2)},$ $\mathcal{A}_{2n-1}^{(2)},$ and $%
\mathcal{D}_{n+1}^{(2)}$ in the fundamental representation. A list of the
main results concerning the general and reduced $K$-matrices is given below
(models which we do not set out in the following list have been designated
as special cases):

$\bullet $ For $\mathcal{A}_{n-1}^{(1)}$ models we have found two classes of
general solutions for $n\geq 5$: the first class is given by $\frac{n(n-1)}{2%
}$ $K$-matrices of type $I$ with three free parameters and $n+2$ non-null
matrix elements; the second family depends on whether $n$ is even or odd,
featuring $n$ solutions of type $II$ with $2+\left[ \frac{n}{2}\right] $
free parameters and $2n-1$ non-null entries for odd $n$, $\frac{n}{2}$ $K$%
-matrices of type $II$ with $2+\frac{n}{2}$ free parameters and $2n$
non-null matrix elements for even $n$, and $\frac{n}{2}$ solutions of type $%
II$ with $1+\frac{n}{2}$ free parameters and $2(n-1)$ non-null entries for
even $n$.

$\bullet $ For $\mathcal{B}_{n}^{(1)}$ models we have found two general
solutions with $n+1$ free parameters for $n>1$, and one reduced solution
with $n$ free parameters for $n\geq 1.$

$\bullet $ For $\mathcal{C}_{n}^{(1)}$ models we have found one general
solution with $n$ free parameters for $n>2$, one reduced solution with $n-1$
free parameters and $8n-6$ null entries depending on the parity of $n$ for $%
n>3$, and one reduced solution with one free parameter for $n>3$. Here we
have concentrated on reduced $K$-matrices which turn out to be new solutions
rather than limit reductions of the general solution.

$\bullet $ For $\mathcal{D}_{n}^{(1)}$ models we have found one general
solution with $n$ free parameters for $n>2$, one reduced solution with $n-1$
free parameters and $8n-6$ null entries depending on the parity of $n$ for $%
n>3$, and one reduced solution with one free parameter for $n>3$. Again, the
emphasis in the reduction procedure has been laid on new $K$-matrices rather
than on simple reductions of the general solution.

$\bullet $ For $\mathcal{A}_{2n}^{(2)}$ models we have found two complex
conjugate general solutions with $n+1$ free parameters for $n>1$, and one
reduced solution with $n+1$ free parameters which can be regarded as another
type of general solution for $n\geq 1$.

$\bullet $ For $\mathcal{A}_{2n-1}^{(2)}$ models we have found two complex
conjugate general solutions with $n$ free parameters for $n>2$, and one
reduced solution with one free parameter for $n>3$ featuring a new solution.

$\bullet $ For $\mathcal{D}_{n+1}^{(2)}$ models we have found one general
solution with $n+2$ free parameters for even $n$ $(n>1)$, one general
solution with $n+2$ free parameters for odd $n$ $(n>1)$, and two independent
block diagonal reduced solutions with one free parameter for any value of $n$%
.

The diagonal $K$-matrices and the special cases were presented as well as
discussed case-by-case in their respective sections. The following point is
worth mentioning about the $\mathcal{C}_{n}^{(1)},$ $\mathcal{D}_{n}^{(1)},$
and $\mathcal{A}_{2n-1}^{(2)}$ diagonal solutions: after proceeding to an
appropriate choice of their free parameters, almost all diagonal $K$%
-matrices degenerated into two classes of solutions, which have been
identified by Batchelor et al. in \cite{Bat1}. Moreover, another remarkable
fact to be noted is that the set of $\mathcal{D}_{n}^{(1)}$ diagonal
solutions contains the set of $\mathcal{C}_{n}^{(1)}$ diagonal solutions.

We point out that a closed expression for boundary reflection amplitudes
which is valid for affine Toda field theories related to all simple Lie
algebras has been constructed by Castro-Alvaredo and Fring \cite{AF} in the
form of blocks of hyperbolic functions and using an integral representation
too.

Previously known non-diagonal $K$-matrices for the $\mathcal{U}_{q}(\widehat{%
gl_{n}})$ case have recently been recovered by Doikou through the Hecke
algebraic approach \cite{Dk4}, based on the structural similarity between
the defining relations of the affine Hecke algebra and the reflection
equation, which suggests that representations of the Hecke algebra should
provide solutions to the reflection equation. A natural direction to be
explored could be the identification of the representations of the affine
Hecke algebra that give rise to the general $K$-matrices presented in this
work.

In addition to the regularity property (\ref{K-matrixReg}), the $K$-matrices
satisfy the unitarity condition, i.e. $K^{-}(u)\times K^{-}(-u)\sim \mathbf{1%
}$. The crossing symmetry proposed by Ghoshal and Zamolodchikov \cite{GZ}
and generalized by Hou et al. \cite{HSY} is more elaborate and involves the $%
\mathcal{R}$-matrix as well.

The classification of reflection matrices, which is an interesting subject
of investigation in itself, is quite important in the quest for the Bethe
ansatz for open spin chains \cite{Dk3,LSW,Ne10,AACDF,YSZ,Li5,Li6,LSY,Dk5},
whose construction is possible for special relations among the boundary
parameters, diagonal cases or \textit{q} root of unity once we are equipped
with $K^{\pm }$-matrices. The Bethe ansatz method would allow us to study
the physical properties of open spin chains. We believe our algebraic
approach and results will motivate further progress in the field of
integrable models with open boundaries.

\section{Acknowledgements}

This work was supported in part by Funda\c{c}\~{a}o de Amparo \`{a} Pesquisa
do Estado de S\~{a}o Paulo-\emph{FAPESP}-Brasil, by Conselho Nacional de
Desenvolvimento-\emph{CNPq}-Brasil, and by Coordena\c{c}\~{a}o de Aperfei%
\c{c}oamento de Pessoal de N\'{\i}vel Superior-\emph{CAPES}-Brasil.

\end{document}